\documentclass[aps,twocolumn,showpacs,prb,preprintnumbers]{revtex4}
\usepackage{graphicx}

\newcommand{\etal}      {{\textit et al}.}
\newcommand{\tc}        {$T_{c}$}
\newcommand{\lsco}      {La$_{2-x}$Sr$_x$CuO$_4$}
\newcommand{\ybco}      {YBa$_2$Cu$_3$O$_{6+\delta}$}
\newcommand{\ncco}      {Nd$_{2-x}$Ce$_x$CuO$_4$}
\newcommand{\dxxyy}     {$d_{x^2-y^2}$}
\newcommand{\xxyy}      {$x^2-y^2$}

\newcommand{\zz}        {$z^2$}
\newcommand{\psigma}    {$p_\sigma$}
\newcommand{\pz}        {$p_z$}
\newcommand{\up}        {\uparrow}
\newcommand{\down}      {\downarrow}
\newcommand{\PRL}       {Phys. Rev. Lett.}
\newcommand{\PRB}       {Phys. Rev. B}
\newcommand{\RMP}       {Rev. Mod. Phys.}

\begin{document}

\title{Chiral Plaquette Polaron Theory of Cuprate
Superconductivity}
\author{Jamil Tahir-Kheli}
\email{jamil@wag.caltech.edu}
\homepage{www.wag.caltech.edu/home/jamil}
\author{William A. Goddard III}
\email{wag@wag.caltech.edu}
\affiliation{Materials and Process Simulation Center,
Beckman Institute 139-74,
California Institute of Technology, Pasadena, CA 91125, USA}
\date{June 2, 2006, \PRB\ \textbf{76}, 014514 (2007)}

\begin{abstract}
Ab-initio density functional calculations on explicitly doped
La$_{2-x}$Sr$_x$CuO$_4$
find doping creates localized holes in out-of-plane orbitals.
A model for cuprate superconductivity is developed based on the
assumption that doping leads to the formation of holes on
a four-site Cu plaquette composed of the out-of-plane A$_1$
orbitals apical O $p_z$, planar Cu d$_{3z^2-r^2}$, and planar
O $p_\sigma$.  This is in contrast to the assumption of
hole doping into planar Cu $d_{x^2-y^2}$ and O $p_\sigma$ orbitals as
in the t-J model.  Allowing these holes to interact with the
$d^9$ spin background leads to chiral polarons
with either a clockwise or anti-clockwise charge current.
When the polaron plaquettes percolate through the crystal at
$x\approx0.05$ for La$_{2-x}$Sr$_x$CuO$_4$,
a Cu $d_{x^2-y^2}$ and planar O $p_\sigma$ band is formed.
The computed percolation doping of $x\approx0.05$ equals the observed
transition to the ``metallic" and superconducting
phase for La$_{2-x}$Sr$_x$CuO$_4$.
Spin exchange Coulomb repulsion with chiral polarons leads to
D-wave superconducting pairing.
The equivalent of the Debye energy in phonon superconductivity
is the maximum energy separation between a chiral polaron and its
time-reversed partner.  This energy separation is on the order of the
antiferromagnetic spin coupling energy, $J_{dd}\sim 0.1$ eV, suggesting
a higher critical temperature.
An additive skew-scattering contribution to the Hall effect
is induced by chiral polarons and leads to a temperature dependent
Hall effect that fits the measured values for La$_{2-x}$Sr$_x$CuO$_4$.
The integrated imaginary susceptibility, observed by
neutron spin scattering, satisfies
$\omega/T$ scaling due to chirality and spin-flip scattering
of polarons along with a uniform
distribution of polaron energy splittings.
The derived functional form is compatible with experiments.
The static spin structure factor for chiral spin
coupling of the polarons to the undoped antiferromagnetic
Cu $d^9$ spins is computed for classical spins
on large two dimensional lattices and is found to be
incommensurate with a separation distance from $(\pi/a,\pi/a)$
given by $\delta Q\approx(2\pi/a)x$ where $x$ is the doping.
When the perturbed \xxyy\ band energy in mean-field is included,
incommensurability along the Cu$-$O bond direction is favored.
A resistivity $\sim T^{\mu+1}$ arises when the polaron energy separation
density is of the form $\sim\Delta^\mu$
due to Coulomb scattering of the $x^2-y^2$ band with polarons.
A uniform density leads to linear resistivity.
The coupling of the $x^2-y^2$ band to the undoped Cu $d^9$ spins
leads to the angle resolved photoemission pseudogap and
its qualitative doping and temperature dependence.
The chiral plaquette polaron leads to an explanation of the
evolution of the bi-layer splitting in Bi-2212.
\end{abstract}

\maketitle

\section{Introduction}
\label{sec:intro}
It is generally assumed that the relevant orbitals for understanding
high temperature cuprate superconductivity arise from holes on
planar Cu \dxxyy\ and O \psigma\ orbitals.  The t-J
model\cite{anderson_tj}
and its generalization to the three band Hubbard model\cite{emery} are
believed to be the correct
Hamiltonians for understanding these materials.
Extensive work since the original
discovery\cite{bednorz_muller} has not led to a complete
understanding of the properties of the cuprates,
despite the rich physics contained in such a simple Hamiltonian.

In this paper, we assume doping creates polarons composed of
apical O \pz\ hybridized with Cu $d_{3z^2-r^2}$
and planar O \psigma\ that form localized chiral states
in the vicinity of the dopant (Sr in \lsco, for example).
The polaron orbital is spread over the 4-site Cu plaquette near
the Sr and is stabilized in a chiral state due to
its interaction with the antiferromagnetic $d^9$ spins on the undoped
Cu sites.  This is similar to prior work
\cite{siggia1, siggia2, siggia3, zee1, gooding1}
suggesting chiral spins states arise from doping
except, in our case, the polaron is formed from out-of-plane
orbitals.

As the doping is increased, the chiral polarons eventually
percolate through the crystal.  We assume a
Cu \dxxyy\ and O \psigma\ delocalized band is formed in the percolating
swath.  This leads to our Hamiltonian of a delocalized
Cu \dxxyy\ band
interacting with chiral plaquette polarons and localized
$d^9$ antiferromagnetic spins on the undoped Cu sites.

For low dopings, momentum $\textbf k$ is
not a good quantum number because the \xxyy/\psigma\ band is formed
on the percolating swath.  This leads to broadening observed in
angle-resolved photoemission (ARPES) measurements.  As the doping
is increased, $\textbf k$ becomes a better quantum number. 

With increasing doping the 4-site
chiral polarons crowd together in the crystal
and several changes occur.
First, the apical O and single Cu closest to Sr is doped with \pz\ 
instead of the four Cu's of the plaquette.\cite{ub3lyp_dope}
Second, the reduction of undoped $d^9$ spins decreases the energy
difference between a polaron state and its time-reversed partner.
Third, the number of \xxyy/\psigma\ band electrons increases.

In our model, the superconducting d-wave pairing is due to the
Coulomb spin exchange
interaction of the \xxyy\ band with chiral polarons
where the Debye energy in phonon superconductors is replaced by
the maximum energy difference of a polaron with its time-reversed
partner (the polaron with flipped chirality and spin).  This leads to
an overdoped phase where superconductivity is suppressed.

Calculations in this paper
of the doping values of \lsco\ and \ybco\ for percolation of polaron
plaquettes and the formation of an \xxyy\ band
are $x\approx 0.05$ and $\delta\approx 0.36$.
Percolation of single doped apical O \pz\ and Cu \zz\ 
described above is
$x\approx 0.20$.  These numbers are close to known phase transitions
in \lsco\cite{birgeneau_rmp, cava_ortho} ($x\approx0.05$ for
the spin-glass to superconducting transition and $x\approx0.20$ for
the orthorhombic to tetragonal transition)
and \ybco\cite{ybco_af_sc} ($\delta\approx0.35$ for the
antiferromagnetic to superconducting transition).

Chiral polarons couple to the Cu $d^9$ spins
on the undoped sites and distort the antiferromagnetic order
leading to incommensurate magnetic neutron scattering peaks.
The charge current
of the polaron induces a chiral coupling of the form
$\pm J_{ch}[{\mathbf{S}_z\cdot(\mathbf{S}_{d1}\times\mathbf{S}_{d2})}]$
\cite{zee1, gooding1, gooding_birgeneau} where $\mathbf{S}_z$ is
the polaron spin and the subscripts $d1$ and $d2$ represent
Cu \xxyy\ spins at adjacent sites.
The sign of the interaction
is determined by the chirality of the polaron.
This term is in addition to an antiferromagnetic
coupling between the polaron spin
and a neighboring $d^9$ spin, $J_{dz}$, and the $d^9-d^9$ spin
coupling, $J_{dd}$.

We have performed energy minimizations on large lattices
of classical spins doped with chiral plaquette polarons over a range of
coupling parameters to compute the static spin
structure factor.  These calculations are similar to
previous computations of the correlation length and incommensurability
due to chiral plaquettes\cite{gooding_birgeneau} using the
Grempel algorithm\cite{grempel} to search for a global minimum.
A neutron incommensurability peak consistently
appears on a circle in k-space centered at $(\pi/a,\pi/a)$
with radius $\approx(2\pi/a)x$.  This result is missing the
kinetic energy perturbation of the \dxxyy\ band.
Computing this contribution in mean field selects the
incommensurate peaks along the Cu-O bond directions in accord
with experiments.
\cite{birgeneau_rmp,incommen_metal1, incommen_metal2, incommen_metal3}

If the
energy difference between a chiral state and its time-reversed
partner, where the spin and chirality are flipped,
is uniformly distributed over an
energy range larger than the temperature,
then the dynamical magnetic response of the polarons
satisfies $\omega/T$ scaling.
\cite{neutron_scaling1, neutron_scaling2, neutron_scaling3,
neutron_scaling4, neutron_scaling5, neutron_scaling6,
neutron_scaling7,birgeneau_rmp}
Since the polarons are randomly
distributed throughout the crystal with different
undoped $d^9$ environments, the probability distribution
of the energy separation of these states may be
approximately uniform.

There are four possible orbital state symmetries for a polaron
delocalized over a four Cu plaquette.
They are $S$, $D_{xy}$, and $P_{x'}\pm iP_{y'}$, where the
last two states are chiral.  $x'$, $y'$ refer to axes along
the diagonals.  Coulomb scattering of Cu \xxyy\ band electrons
with polarons leads to a linear resistivity in the case
of a uniform energy distribution of the energies of the four
polaron states.  This may be uniform for the same reasons discussed
above for neutron scaling.
Any non-uniformity of the energy distribution spectrum
makes the resistivity non-linear.

Spin exchange Coulomb scattering of an \xxyy\ Cooper pair
$(\mathbf{k}\up,\mathbf{-k}\down)$ with a chiral polaron
$P_{x'}\pm iP_{y'}$ and spin $s$ into the time-reversed intermediate
state $P_{x'}\mp iP_{y'}$ and spin $-s$
leads to an anisotropic repulsion that is peaked for 
scattering of a Cooper pair with $\mathbf{k}$ near $(\pm\pi,0)$
to $\mathbf{k'}$ near $(0,\pm\pi)$.
There are two
necessary conditions to obtain d-wave superconducting pairing.
First, time-reversal symmetry must be broken such
that $P_{x'}+iP_{y'}$ and
spin $s$ is not degenerate with $P_{x'}-iP_{y'}$ and spin $-s$.
The maximum energy separation of these two polarons replaces the
Debye energy in phonon superconductivity.
Second, the polaron must be
spread out over more than one site so that phase differences in the
initial and final \xxyy\ band states can interfere.
A single site polaron would lead to an isotropic repulsion and
no superconductivity.

In zero magnetic field, there is an equal number of polarons
of each chirality.  A magnetic
field creates more polarons of one chirality than the other.
An \xxyy\ band electron scattering from a chiral polaron is
skew-scattered\cite{fert1, fert2, fert3, coleman1}
due to second-order Coulomb repulsion with a polaron
where the polaron orbital changes in the intermediate state.
This leads to an additive skew-scattering contribution to the Hall
effect proportional to the difference of the number of
``plus'' and ``minus'' polarons.
For high temperatures, the difference is $\sim 1/T$. 

For the hole-doped cuprates, the polarons are holes.  The Coulomb
matrix element is negative, $U<0$, since the change in the
Hamiltonian amounts to the removal of a Coulomb coupling.
Although we have not identified the nature of the polaron in the
electron-doped system \ncco, the same argument makes $U>0$.

The skew-scattering contribution is derived and computed for
reasonable values of the parameters.
It is found the sign change between the
hole-doped and electron-doped cuprates appears due to the sign
change of $U$.  The magnitude of the skew-scattering term is
shown to be large enough to account for the experimental data.
The derived functional form for the temperature dependence is
shown to fit data\cite{cava_hall} on \lsco.
To our knowledge, the only explanation for this
sign difference between the hole and electron-doped cuprates arises
from the additional $(\pi,\pi)$ nesting
of the \ncco\ Fermi surface.\cite{kontani1}

The $d^9$ undoped spins interact with the \xxyy\ band electrons.
They induce a coupling of a state with momentum $k$ to $k\pm Q$ where
$Q\approx(\pi,\pi)$ is the incommensurate peak momentum.  This
leads to an ARPES pseudogap.\cite{norman1, loeser1, marshall1, ding1}
The strength of the $d^9$ antiferromagnetism decreases with
increasing temperature making the pseudogap close with temperature.
At low doping, there are more undoped $d^9$ spins
and the coupling to the \xxyy\ band electrons is larger
than the coupling at higher doping.  The pseudogap increases
with decreasing doping while \tc\ is reduced.
The couplings leading to the
pseudogap and superconductivity are different in our model.

The outline of the paper is as follows.  In Section
\ref{sec:exist_pz}, the
experimental and theoretical arguments for the existence of \pz\ 
holes with doping are examined.  In particular, experiments considered
to establish the validity of the t-J model\cite{brookes1, brookes2}
and preclude any substantial out-of-plane character
are addressed.\cite{xray1} Section
\ref{sec:chiral}
defines the chiral plaquette polarons.  Section
\ref{sec:percolation} calculates the
percolation phase transitions and compares them to the \lsco\ 
and \ybco\ phase diagrams.   Section
\ref{sec:neutron} describes classical spin calculations
of the neutron structure factor including the effect of the
$d^9$ and polaron spin incommensurability on the kinetic energy
of the band \xxyy\ electrons.  Incommensurate peaks
along the Cu$-$O bond direction are obtained.
The polaron magnetic
susceptibility is calculated, assuming a uniform probability
distribution of polaron energy level separations, and is shown to
satisfy $\omega/T$ scaling.
In Section \ref{sec:pairing}, the 
Coulomb interactions of \xxyy\ band states with
chiral plaquette polarons is examined to determine the possible
superconducting pairing channels.  The spin
exchange interaction leads to an anisotropic repulsion
of the form sufficient to create a d-wave gap with nodes.
Section \ref{sec:transport} describes the resistivity
and Hall effect due to Coulomb interactions with chiral polarons.
If the distribution of energy separations of polarons states 
with different symmetries is uniform, then the
resistivity is linear.  A magnetic field produces a difference
in ``up'' and ``down" chiral polarons leading to an additive
skew-scattering
contribution to the ordinary band Hall effect with a temperature
dependence consistent with measurements.  The magnitude and
temperature dependence of the skew-scattering is calculated.
Section \ref{sec:pseudo} describes
our model of the ARPES pseudogap and its doping
and temperature dependence.  Section \ref{sec:split} discusses
the doping and temperature dependence
of the bilayer splitting observed in ARPES on
Bi-2212.
Section \ref{sec:nmr} discusses the NMR data of
Takigawa et al\cite{takigawa1} that is assumed to be strong
evidence for a one-component theory because of the similar
temperature dependencies of the Knight shifts of planar Cu
and O in underdoped YBa$_{2}$Cu$_{3}$O$_{6.63}$.  We argue
qualitatively that these results are compatible with our model.
Section \ref{sec:conclusions} presents our conclusions.

\section{Existence of $\mathbf{A_1}$ holes}
\label{sec:exist_pz}

\subsection{Ab-initio Calculations}
\label{subsec:abinitio}
Becke-3-Lee-Yang-Parr (B3LYP) is a three parameter
hybrid density functional that includes 20\%
exact Hartree-Fock exchange.
\cite{becke1, becke2, becke3, lyp, vwn}
Its success has extended beyond its original domain
of parametrization to include the thermochemistry of compounds
containing transitions metals.\cite{b3lyp_tm1, b3lyp_tm2,
b3lyp_tm3, b3lyp_tm4}

Several years ago,\cite{ub3lyp} we performed ab-initio
periodic band structure computations using the
spin unrestricted B3LYP functional
on undoped La$_2$CuO$_4$ and explicitly doped \lsco.
For the undoped insulator, the antiferromagnetic
insulator with the experimental bandgap of 2.0 eV was obtained.
\cite{lcogap}

Prior to this calculation, the insulating state had been
obtained by extending local spin density (LSD) computations,
which yielded zero gap or a metal, to approximately
incorporate the self-interaction correction not accounted for
in this functional.  Table \ref{table:gap} summarizes
chronologically corrections to the initial LDA results
and their computed bandgaps.

\begin{table}[tbp]
\caption{\label{table:gap}Ab-initio La$_2$CuO$_4$ bandgap results.
LSDA stands for local spin density approximation, SIC is
self-interaction correction, HF is Hartree-Fock, and UB3LYP is
unrestricted spin B3LYP.  The last line is the experimental gap.}
\begin{ruledtabular}
\begin{tabular}{lcl}
Method & Bandgap (eV) & Author \\
\hline
LSDA & 0.0 & Yu \etal\cite{yu} (1987) \\
LSDA & 0.0 & Mattheiss\cite{mattheiss} (1987) \\
LSDA & 0.0 & Pickett\cite{pickett} (1989) \\
SIC-LSDA & 1.0 & Svane\cite{svane} (1992) \\
LSDA+U & 2.3 & Anisimov \etal\cite{anisimov} (1992) \\
SIC-LSDA & 2.1 & Temmerman \etal\cite{temmerman} (1993) \\
LSDA+U & 1.7 & Czyzyk \etal\cite{czyzyk} (1994) \\
HF & 17.0 & Su \etal\cite{su} (1999) \\
UB3LYP & 2.0 & Perry \etal\cite{ub3lyp} (2001) \\
Experiment & 2.0 & Ginder \etal\cite{lcogap} (1988) \\
\end{tabular}
\end{ruledtabular}
\end{table}

Our result showed that an off-the-shelf functional with an
established track record\cite{b3lyp_tm1, b3lyp_tm2, b3lyp_tm3,
b3lyp_tm4} for molecular systems could reproduce the results
of more elaborate LDA corrections. 

In addition, we found
the highest occupied states to have more out-of-plane orbital
character (apical O \pz\ and Cu \zz)
than obtained by LDA. Svane\cite{svane} also
made this observation in his self-interaction corrected (SIC)
computation.

In a second paper,\cite{ub3lyp_dope} we explicitly doped
La$_2$CuO$_4$ with Sr to form supercells of \lsco\ at
special dopings of $x=0.125$, 0.25, and 0.50.
We found an additional hole was formed
for each Sr atom that localized in the vicinity of the dopant
of apical O \pz, Cu \zz, and
an A$_{1g}$ combination of planar O \psigma\ character.
The Cu sites split into undoped and doped sites.  The undoped
sites had a $d^9$ \xxyy\ hole and the doped sites were still
predominantly $d^9$ with a mixture of \xxyy\ and \zz\ hole
character.  There was corresponding hole character on the
neighboring O atoms in and out of the plane with the appropriate
B$_{1g}$ and A$_{1g}$ symmetries.
This led us to argue that out-of-plane hole orbitals
are a generic characteristic of cuprates and must be considered
in developing theories of these materials.

At the time, B3LYP
had an established track record with molecular systems, but
its use for crystal band structures was in its infancy.
This is likely due to the difficulty of including exact
Hartree-Fock exchange into periodic band structure codes.

Since the appearance of our doped Sr work, it has been found that
B3LYP does remarkably well at obtaining the bandgaps of
insulators.\cite{b3lyp1, b3lyp2, b3lyp3, b3lyp4}  Hybrid
functionals appear to
compensate the overestimate of the gap from Hartree-Fock
with the underestimate arising from local density and gradient
corrected functionals.  Thus, we believe density functionals have
established the existence of non-planar hole character in
\lsco.

\begin{figure}[tbp]
\centering  \includegraphics[width=0.8\linewidth]{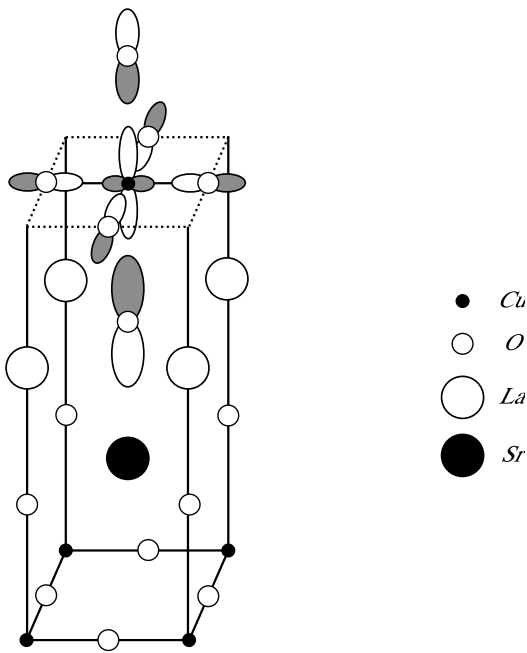}
\caption{La$_2$CuO$_4$ with one doped Sr atom.  Ab-initio
calculations\cite{ub3lyp_dope} find an A$_1$ hole localized above
the Sr with hole character on the apical O \pz, Cu \zz, and planar
O \psigma\ orbitals.  The \pz\ character above the doped Cu is
smaller than the \pz\ below the Cu leading to an A$_1$ state rather
than A$_{1g}$.}
\label{fig:a1_hole}
\end{figure}

For \lsco,
there are five Cu sites in the vicinity of a Sr atom in two
distinct CuO$_2$ planes.  The Sr is centered over four Cu in
a square plaquette.  The fifth Cu couples to the Sr through
the neighboring apical O between them as shown in figure
\ref{fig:a1_hole}.  The hole state composed of
apical O \pz\ , Cu \zz, and planar O \psigma\  as shown in
figure \ref{fig:a1_hole} appeared with Sr doping.

\begin{figure*}[tbp]
\centering \includegraphics[width=\linewidth]{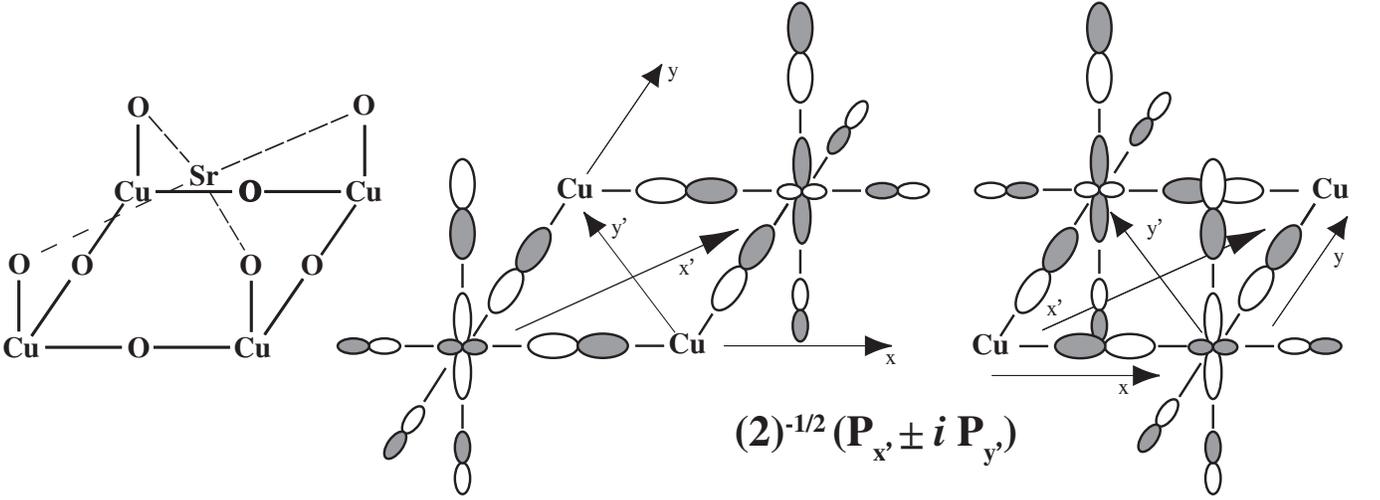}
\caption{Orbital schematic of chiral polarons postulated in
our model.  We assume these chiral hole states are the most stable
due to interactions with the undoped $d^9$ Cu lattice spins.}
\label{fig:chiral}
\end{figure*}

The polaron state with the hole delocalized over two diagonally
opposed Cu in the 
four Cu plaquette is higher in energy in our ab-initio calculation
by 0.57 eV for each Sr
or 0.071 eV for each formula unit La$_{1.875}$Sr$_{0.125}$CuO$_4$.
The value of 0.57 eV is
an upper bound since our geometry optimizations only
allowed the apical O sites to relax.
The polaron localizes on two Cu sites due to spin exchange
coupling with the \xxyy\ hole and the anti-ferromagnetic
spin ordering of the \xxyy\ holes in our periodic supercells.
In this paper,
the hole state in figure \ref{fig:a1_hole} obtained
from our ab-initio calculations is not taken to be the correct polaron.
Instead, we postulate Sr doping leads to chiral polarons over
the four plaquette Cu atoms shown in figure \ref{fig:chiral}.
This is discussed in section \ref{sec:chiral}.

This paper explores the consequences of the
assumptions that Sr doping causes holes to appear
in Cu four-site plaquettes and the most stable configurations
are the chiral states $P_{x'}\pm iP_{y'}$.

From an ab-initio standpoint, our first assumption is
plausible for \lsco, but unproven.  This may be due to the limitation
of the special periodic supercells that were chosen out of necessity to
perform the computation, the restrictive geometry relaxation for the
plaquette polaron, or it may be a limitation of the B3LYP
functional.

For \ybco, we do not have ab-initio proof for doping of four-site
Cu plaquettes in the CuO$_2$ plane either.  In fact,
any polaron plaquettes would likely be in the yz plane where
the Cu$-$O chains are along the y-axis and the z-axis is normal
to the CuO$_2$ planes.
One way in which polaron plaquettes can arise is when two
adjacent Cu$-$O chains each have an occupied O separated
by one lattice spacing along the x-axis (perpendicular to the
direction of the chains).  In other words, the two chain O
reside in neighboring chains with minimum separation
between them.  This may create
four-site polarons on the two CuO$_2$ planes above and below
the two O atoms.  For this paper, the chiral plaquette polarons
in figure \ref{fig:chiral} are assumed.

The second assumption, that the polarons are chiral,
is true for a localized polaron interacting with
an infinite $d^9$ antiferromagnetic lattice in two dimensions
(2D)\cite{siggia1, siggia2, siggia3, zee1, gooding1} by mapping
the 2D Heisenberg antiferromagnet to a continuum model and
analyzing the effective Hamiltonian arising from a path integral
formulation.  These papers
did not specifically consider an out-of-plane hole, but the
analysis is applicable in our scenario.  This is discussed
further in section \ref{sec:chiral}.

\subsection{Experiment}
\label{subsec:experiment}
Resonance circular dichroism photoemission investigating
the spin of the occupied states near the Fermi level
\cite{brookes1, brookes2} find a preponderance of
singlet occupied states just below the Fermi level in CuO and
Bi$_2$Sr$_2$CaCu$_2$O$_{8+\delta}$ (2212).  These results
are considered strong evidence in favor of the correctness of the 
t-J model.  In particular, it is expected that
out-of-plane \pz, \zz, and A$_{1g}$ \psigma\ would lead to
triplet occupied states near E$_{F}$ due to exchange
Coulomb coupling to the orthogonal \xxyy\ orbital.
Since the prima facie evidence is against our
proposal, we review the measurement and its interpretation.

We show our assumption of a delocalized \xxyy\ band on the
percolating out-of-plane polaron doped Cu sites leads
to a null effect for resonance absorption on these sites.  This arises
because a delocalized \xxyy\ band electron spin has
no correlation to the polaron spin.
Thus, the experiment
measures the spin of the highest occupied
states on the undoped Cu d$^9$ sites where it is expected
the first holes would be created in B$_{1g}$ combinations
of ligand planar O \psigma\ orbitals that form a singlet with
the \xxyy\ d$^9$ hole (the Zhang-Rice\cite{zhang_rice} singlet).

The idea behind the dichroism experiment is to use circularly
polarized incident soft x-rays tuned to the Cu L$_3$ ($2p_{3/2}$) white
line energy ($\approx931.5$ eV).  The incident x-rays induce
the photoabsorption transition
$2p^6 3d^9+\hbar\omega\rightarrow 2p^5d^{10}$ that Auger decays
to an ARPES final state $2p^6 d^8 + e$.  The spin-orbit energy
separation
of the core-hole $2p_{1/2}$ and $2p_{3/2}$ states is sufficiently
large ($\sim 20$ eV) to guarantee the intermediate state is
a $j=3/2$ core-hole.

By monitoring the outgoing electron energy and spin along the
incident photon direction for each photon helicity,
$\sigma^+$ and $\sigma^-$, the total spin
of the final Cu d$^8$ is obtained. 

In the Bi-2212 experiment,
\cite{brookes1} the photon is incident normal to the CuO$_2$ planes.
The analysis below is for a normally incident photon.
The transition rates are slightly different for
the CuO case\cite{brookes2} where a poly-crystalline sample
was used.

Consider a Cu initially in the d$^9$ state
$|2p^6;3d_{z^2}\up\down;3d_{x^2-y^2}(A_\up\up+A_\down\down)\rangle$,
where our notation shows the occupied electrons.
The $d_{xy}$, $d_{xz}$, $d_{yz}$ orbitals are always
doubly occupied and are omitted in the wavefunction for convenience.
The $3d_{z^2}$ orbital is doubly occupied
and the single \xxyy\ electron is in a spin state along a
direction that may be different from the incident photon direction.
It is represented as a linear combination of $\up$ and $\down$
along the incident photon direction with $|A_\up|^2+|A_\down|^2=1$.
By summing over all helicities and exiting electron
spin directions, the photoemission becomes
independent of the initial direction of the \xxyy\ electron
as shown below.

Writing $\hat{x}$, $\hat{y}$, and $\hat{z}$ for the angular part
of the Cartesian variables, $x/r$, $y/r$, and $z/r$, the relevant
wave functions and photon polarization operators may be written as
$d_{3z^2-r^2}=C_2(3\hat{z}^2-1)$,
$d_{x^2-y^2}=\sqrt3 C_2(\hat{x}^2-\hat{y}^2)$, and
$Y_{1\pm 1}=C_1(1/\sqrt 2)(\hat{x}\pm\hat{y})$, where
$C_1=\sqrt{3/4\pi}$ and $C_2=\sqrt{5/16\pi}$.  The mod-squared
matrix elements for resonance absorption of
$|2p^6;3d_{z^2}\up\down;3d_{x^2-y^2}(A_\up\up+A_\down\down)\rangle$
to the intermediate $2p_{3/2}$ core-hole states are,
\begin{eqnarray}
\label{eqn:int1}
|2p^6;3d_{z^2}\up\down;3d_{x^2-y^2}(A_\up\up+A_\down\down)\rangle
\nonumber \\
 \nonumber \\
\stackrel{\hbar\omega^+}{\longrightarrow}
\left\{
\begin{array}{llr}
\left|p^5:\frac{3}{2},\frac{3}{2}\right\rangle d^{10}
                      & = & |A_\up|^2 \\
\left|p^5:\frac{3}{2},\frac{1}{2}\right\rangle d^{10}
                      & = & \frac{1}{3}|A_\down|^2 
\end{array}
\right\}
\cdot\frac{1}{6} \left(\frac{C_1}{C_2}\right),
\end{eqnarray}
\begin{eqnarray}
\label{eqn:int2}
|2p^6;3d_{z^2}\up\down;3d_{x^2-y^2}
(A_\up\up+A_\down\down)\rangle \nonumber \\
 \nonumber \\
\stackrel{\hbar\omega^-}{\longrightarrow}
\left\{
\begin{array}{llr}
\left|p^5:\frac{3}{2},-\frac{3}{2}\right\rangle d^{10}
                      & = & |A_\down|^2 \\
\left|p^5:\frac{3}{2},-\frac{1}{2}\right\rangle d^{10}
                      & = & \frac{1}{3}|A_\up|^2 
\end{array}
\right\}
\cdot\frac{1}{6} \left(\frac{C_1}{C_2}\right),
\end{eqnarray}
where $\hbar\omega^\pm$ are positively and negatively
circularly polarized photons.

The Auger scattering rates of the four $2p_{3/2}$ intermediate states
where one \xxyy\ electron fills the $2p_{3/2}$ core hole and
the other is ejected are,
\begin{equation}
\label{eqn:sfin1}
\left|\frac{3}{2},\frac{3}{2}\right\rangle d^{10} \longrightarrow
|2p^6\rangle|d_{z^2}\up\down\rangle + e\up = |V|^2,
\end{equation}
\begin{equation}
\label{eqn:sfin2}
\left|\frac{3}{2},\frac{1}{2}\right\rangle d^{10} \longrightarrow
\left\{
\begin{array}{c}
|2p^6\rangle|d_{z^2}\up\down\rangle + e\up=
\left(\frac{2}{3}\right)|V|^2, \\
|2p^6\rangle|d_{z^2}\up\down\rangle + e\down=
\left(\frac{1}{3}\right)|V|^2, 
\end{array}
\right.
\end{equation}
\begin{equation}
\label{eqn:sfin3}
\left|\frac{3}{2},-\frac{1}{2}\right\rangle d^{10}\longrightarrow
\left\{
\begin{array}{c}
|2p^6\rangle|d_{z^2}\up\down\rangle + e\up =
 \left(\frac{1}{3}\right)|V|^2, \\
|2p^6\rangle|d_{z^2}\up\down\rangle + e\down =
 \left(\frac{2}{3}\right)|V|^2, 
\end{array}
\right.
\end{equation}
\begin{equation}
\label{eqn:sfin4}
\left|\frac{3}{2},-\frac{3}{2}\right\rangle d^{10}\longrightarrow 
|2p^6\rangle|d_{z^2}\up\down\rangle + e\down = |V|^2,
\end{equation}
where $|V|^2$ is the Auger matrix element.  The \zz\ is doubly
occupied making the $d^8$ state a singlet.  There are analogous
matrix elements if the Auger process scatters the two \zz\ electrons
instead of \xxyy and also if the final $d^8$ is composed of
one electron in \zz\ and one in \xxyy\ in a singlet configuration.

The total scattering rate is given by the products through the
various intermediate states.  Using the convention
\cite{brookes1, brookes2} $\sigma^+\up$, $\sigma^-\up$,
$\sigma^+\down$, and $\sigma^-\down$ to represent a positively
circularly polarized photon ejecting an electron with $\up$
spin etc, the scattering leaving a singlet $d^8$ final state is,
\begin{eqnarray}
\label{eqn:sigud_s1}
\sigma^+\up   & = &
\left(|A_\up|^2 + \frac{2}{9}|A_\down|^2\right)|V|^2, \\
\label{eqn:sigud_s2}
\sigma^+\down & = & \frac{1}{9}|A_\down|^2|V|^2, \\
\label{eqn:sigud_s3}
\sigma^-\up   & = & \frac{1}{9}|A_\up|^2|V|^2, \\
\label{eqn:sigud_s4}
\sigma^-\down & = &
\left(\frac{2}{9}|A_\up|^2 + |A_\down|^2\right)|V|^2. 
\end{eqnarray}

The total parallel and anti-parallel scattering is,
\begin{eqnarray}
\label{eqn:pol_s1}
\up\up  \equiv(\sigma^+\up + \sigma^-\down) & = & \frac{11}{9}|V|^2, \\
\label{eqn:pol_s2}
\up\down\equiv(\sigma^+\down + \sigma^-\up) & = & \frac{1}{9}|V|^2,
\end{eqnarray}
where we have neglected the $(C_1/6C_2)$ from equations
\ref{eqn:int1} and \ref{eqn:int2} since it cancels out when we
evaluate the polarization defined below.
These two sums are independent of the starting spin orientation of
the \xxyy\ electron.  The ``polarization,'' defined as a ratio
$(\up\up-\up\down)/(\up\up+\up\down)=5/6$ for pure singlet $d^8$ states.

There are three possible triplet $d^8$ spin states.  There is one
electron in \xxyy\ and \zz.  The scattering from the intermediate
$d^{10}$ state with a $2p_{3/2}$ core-hole to triplet $d^8$
is given by,
\begin{equation}
\label{eqn:tf1}
\left|\frac{3}{2},\frac{3}{2}\right\rangle d^{10}\rightarrow
\left\{
\begin{array}{l}
|2p^6\rangle|\down\down\rangle + e\down = 2|V|^2, \\
|2p^6\rangle|\frac{\up\down+\down\up}{\sqrt2}\rangle + e\up 
= |V|^2, 
\end{array}
\right.
\end{equation}
\begin{equation}
\label{eqn:tf2}
\left|\frac{3}{2},\frac{1}{2}\right\rangle d^{10}\rightarrow
\left\{
\begin{array}{l}
|2p^6\rangle|\down\down\rangle + e\down
 =\left(\frac{2}{3}\right)\cdot2|V|^2, \\
|2p^6\rangle|\frac{\up\down+\down\up}{\sqrt2}\rangle + e\up
 =\left(\frac{2}{3}\right)|V|^2, \\
|2p^6\rangle|\frac{\up\down+\down\up}{\sqrt2}\rangle + e\down
 =\left(\frac{1}{3}\right)|V|^2, \\
|2p^6\rangle|\up\up\rangle + e\up 
 =\left(\frac{1}{3}\right)\cdot2|V|^2,
\end{array}
\right.
\end{equation}
\begin{equation}
\label{eqn:tf3}
\left|\frac{3}{2},-\frac{1}{2}\right\rangle d^{10}\rightarrow
\left\{
\begin{array}{l}
|2p^6\rangle|\down\down\rangle + e\down=
 \left(\frac{1}{3}\right)\cdot2|V|^2, \\
|2p^6\rangle|\frac{\up\down+\down\up}{\sqrt2}\rangle + e\down=
 \left(\frac{2}{3}\right)|V|^2, \\
|2p^6\rangle|\frac{\up\down+\down\up}{\sqrt2}\rangle + e\up=
 \left(\frac{1}{3}\right)|V|^2, \\
|2p^6\rangle|\up\up\rangle + e\up =
 \left(\frac{2}{3}\right)\cdot2|V|^2, 
\end{array}
\right.
\end{equation}
\begin{equation}
\label{eqn:tf4}
\left|\frac{3}{2},-\frac{3}{2}\right\rangle d^{10}\rightarrow
\left\{
\begin{array}{l}
|2p^6\rangle|\up\up\rangle + e\up = 2|V|^2, \\
|2p^6\rangle|\frac{\up\down+\down\up}{\sqrt2}\rangle + e\down =
|V|^2. 
\end{array}
\right. 
\end{equation}
Multiplying by the transition rates to the intermediate
state for all possible photon and electron spin polarizations,
\begin{eqnarray}
\label{eqn:sud_t1}
\sigma^+\up   & = & (|A_\up|^2 + \frac{4}{9}|A_\down|^2)|V|^2, \\
\label{eqn:sud_t2}
\sigma^+\down & = & (2|A_\up|^2+ \frac{5}{9}|A_\down|^2)|V|^2, \\
\label{eqn:sud_t3}
\sigma^-\up   & = & (\frac{5}{9}|A_\up|^2 + 2|A_\down|^2)|V|^2, \\
\label{eqn:sud_t4}
\sigma^-\down & = & (\frac{4}{9}|A_\up|^2 + |A_\down|^2)|V|^2. 
\end{eqnarray}

The total parallel and anti-parallel scattering is,
\begin{eqnarray}
\label{eqn:pol_t1}
\up\up  \equiv(\sigma^+\up + \sigma^-\down) & = & \frac{13}{9}|V|^2, \\
\label{eqn:pol_t2}
\up\down\equiv(\sigma^+\down + \sigma^-\up) & = & \frac{23}{9}|V|^2,
\end{eqnarray}
leading to polarization
$(\up\up-\up\down)/(\up\up+\up\down)=-(1/3)(5/6)$
for pure triplet $d^8$ states.

The measured value of the polarization for each photoelectron
energy gives an estimate of
the amount of singlet and triplet character in the occupied states
below E$_F$.  The experiments\cite{brookes1,brookes2} find singlet
character just below E$_F$ consistent with the t-J model and
in contradiction to A$_1$ holes that
would Hund's rule triplet couple to the \xxyy\ electron.

In our model, there are two types of Cu sites.  The first is
undoped with a single \xxyy\ hole
in a $d^9$ state.  The ejected photoelectron
near the Fermi level comes from the
B$_{1g}$ combination of neighboring \psigma\ orbitals on the
planar O that couples to the \xxyy\ electron in a singlet
as described by Zhang and Rice.\cite{zhang_rice} This is consistent
with experiment and the t-J model.

The second Cu is on a doped site with an out-of-plane polaron
and a delocalized band comprised of \xxyy\ and \psigma\ 
in our model.  In this case, the final Cu $d^8$
state has one \zz\ and one \xxyy\ hole with
no spin correlation between them.  Thus, $\up\up=\up\down$
and the ``polarization'' arising from resonance scattering
on doped Cu sites is zero.  The only polarization
observed arises from the undoped sites with singlet holes
near the Fermi energy.

The second experiment we consider is polarized xray absorption
on \lsco\ for $x=0.04-0.30$.\cite{xray1, pellegrin1}
A substantial O absorption with z-axis polarized xrays indicates
there are holes in apical O \pz. 
In addition, xray absorption fine structure (XAFS)
\cite{haskel_Sr1, haskel_Sr2} measurements
observe displacement
of the apical O away from the Sr towards Cu consistent with hole
formation in O \pz.  Since the \pz\ hole
character is compatible with our out-of-plane polaron assumption,
we focus on the Cu result.

The Cu absorption finds a few percent \zz\ character on the Cu sites.
Our ab-initio calculations find the \zz\ hole character to be
approximately 85\% of the \xxyy\ hole character.  It is too
large compared with experiment.
One could argue that the many-body response to the formation of a
Cu $2p$ core-hole is different for an undoped Cu versus
a doped Cu where the delocalized \xxyy\ band may suppress
the white line due to the orthogonality catastrophe or more
strongly screen the core-hole potential.  We
are not convinced this is the sole reason for
the small amount of \zz\ hole character observed in the white
line.

A possible explanation is that the chiral polaron,
spread out over
four Cu sites as in figure \ref{fig:chiral}, has more
\psigma\ and \pz\ character at the expense of \zz\ from
delocalization compared to the polaron centered around a
single Cu site in figure \ref{fig:a1_hole}.
A recent neutron pair distribution analysis\cite{billinge}
is more compatible with a chiral plaquette polaron.
In this case, extracting a very small
signal from a bulk average of bond distances and then using the
measured bond distances to infer orbital occupations is very model
dependent.

\section{Chiral Polarons}
\label{sec:chiral}

\begin{figure}[tbp]
\centering  \includegraphics[width=\linewidth]{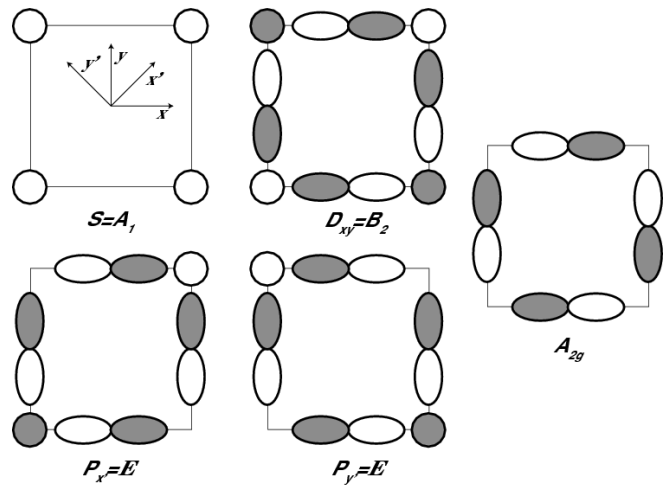}
\caption{Projection onto CuO$_2$ of four polaron states and their
symmetries.  The higher energy anti-bonding states are shown.
The fifth state of A$_{2g}$ symmetry is composed entirely
of \psigma\ orbitals and is not a polaron state.  This state becomes
part of the delocalized \xxyy\ band at the unoccupied $k$ state
$(\pi,\pi)$ and the occupied bonding \xxyy, \psigma\ band.
We assume interaction with the undoped $d^9$ spin
background, as shown in prior work,
\cite{siggia1, siggia2, siggia3, zee1, gooding1}
makes the chiral combinations $P_{x'}\pm iP_{y'}$ the
most unstable electronic states (most stable hole states).
When these chiral states percolate through
the crystal, we further assume the electronic states
composed of \xxyy\ and \psigma\ delocalize over these doped sites.
There are a total of eight states.  The lowest three have E
(P$_{x'}$ and P$_{y'}$) and B$_2$ (D$_{xy}$) symmetry and are bonding
combinations of the states in the figure.  They are always occupied.
The energies of the states are listed in
table \ref{table:polaron_energies}}
\label{fig:polaron_states}
\end{figure}

The higher energy anti-bonding electronic
states with apical \pz, \zz, and \psigma\ over
a four Cu doped plaquette are shown in figure \ref{fig:polaron_states}.
The $P_{x'}$ and $P_{y'}$ are degenerate.  For simplicity, we
have taken the two apical O \pz\ above and below each Cu and
the Cu \zz\ and 4s to be one A$_1$ orbital.  Thus, there are
a total of eight states.  The figure does not show the lower
energy three bonding states (E and B$_2$) since they are occupied.

Table \ref{table:polaron_energies} lists the energies of the eight
polaron states for the case where the orbital energy of the
``effective'' A$_1$ composed of \pz\ and \zz,
is taken to be equal to the \psigma\ orbital energy,
$\epsilon_z=\epsilon_p=0$.
There is an effective hopping matrix element, $t_{pz}$, from \pz\ to
\psigma.  $t_{pp}$ is the diagonal \psigma\ matrix element.
It is expected $0<t_{pz}<t_{pp}$.

\begin{table}[tbp]
\caption{\label{table:polaron_energies}Polaron symmetries and energies
from highest (most unstable electronic states) to lowest.  The
effective \pz\ and \psigma\ orbital energies are taken to be 0.}
\begin{ruledtabular}
\begin{tabular}{ccc}
State & Symmetry & Energy \\
\hline
P$_{x'}$, P$_{y'}$ & E & $+\sqrt2 t_{pz}$ \\
D$_{xy}$ & B$_2$ & $-t_{pp}+\sqrt{t_{pp}^2+4t_{pz}^2}$ \\
S & A$_1$ & 0 \\
P$_{x'}$, P$_{y'}$ & E & $-\sqrt2 t_{pz}$ \\
D$_{xy}$ & B$_2$ & $-t_{pp}-\sqrt{t_{pp}^2+4t_{pz}^2}$ \\
Not Polaron & A$_2$ & Coupled to \xxyy\ \\
\end{tabular}
\end{ruledtabular}
\end{table}

Antiferromagnetic (AF) interaction of the polaron spin
with the undoped Cu $d^9$ lattice renormalizes these
couplings, but we expect P$_{x'}$ and P$_{y'}$ to remain the
most unstable electronic states.

The effect of the undoped $d^9$ spin background is seen
in mean-field where the $d^9$ AF spins surrounding a plaquette
are frozen with an $\uparrow$ spin on one sublattice and
a $\downarrow$ spin on the other sublattice.  The additional
energy of an $S$ or $D_{xy}$ polaron with spin $\sigma$
due to AF coupling of $\sigma$ with the $d^9$ spins is zero since the
average $d^9$ spin seen by the polaron is zero.  For $P$ states,
the polaron spin couples to one sublattice.  $\sigma$ can
be aligned with the sublattice spin leading to a further
destabilization of the $P$ state.

$P$ hole states were found in the exact
diagonalization of Gooding\cite{gooding1} for a t-J
model on a $4\times4$ lattice with an additional hole allowed to
delocalize on the interior $2\times2$ lattice.
This is in accordance with theoretical predictions.
\cite{siggia1, siggia2, siggia3, zee1}

Based on the energies in table \ref{table:polaron_energies}, the
mean-field description of the $d^9$ spins, and exact results on a
$4\times4$ lattice,\cite{gooding1}
we assume the polaron hole has $P$ symmetry.

For a single hole delocalized in a small region of
an AF spin background,
it has been shown\cite{siggia1, siggia2, siggia3, zee1} the chiral
states
$1/\sqrt2(P_{x'}\pm iP_{y'})$
are the correct spontaneous
symmetry breaking states for the hole rather than P$_{x'}$ and P$_{y'}$,
because the complex linear combinations are compatible with the
long range twisting of the AF lattice spins into a stable
configuration topologically distinct from the AF
ground state.\cite{polyakov}

In this paper, we assume doping
introduces hole character in out-of-plane orbitals
that can delocalize over a small number of sites in the vicinity
of the dopant.  The most favorable configuration for the polaron
is taken to be the chiral state.
If there was a single dopant in an
infinite $d^9$ crystal, then the chiral states,
$1/\sqrt2(P_{x'}\pm iP_{y'})$, would be degenerate.  These two
states are time-reversed partners.

In a finitely
doped system, the environment of each polaron is different and
the two chiral states may have different energies.  We
assume, in a doped cuprate, the chiral states are the
correct polaron eigenstates, but the energies of the two states
may be different.  This leads to a model of the polarons where
the splitting between the chiral states along with all the
other states represented in table \ref{table:polaron_energies}
and figure \ref{fig:polaron_states} are distributed differently
for each plaquette.  The assumption of a completely uniform
probability distribution of different polaron state energies
throughout the crystal leads to neutron $\omega/T$ scaling
as shown in section \ref{subsec:scaling}.
A linear resistivity,
derived in section \ref{sec:transport}, arising from the
Coulomb scattering of \xxyy\ band electrons with
the polarons is also obtained with a uniform energy distribution.

This model of non-degenerate chiral polarons
implies time-reversal symmetry is broken.  At any instant, the
number of ``up'' chiral polarons should equal the number of
``down'' chiral polarons and macroscopically the cuprate is
time-reversal invariant.  There is recent experimental evidence for
local time-reversal symmetry breaking in neutron scattering.
\cite{neutron_TR}

\section{Percolation}
\label{sec:percolation}
There are three basic assumptions of our model.  First, doping leads
to additional holes in out-of-plane orbitals that form chiral
states as shown in figures \ref{fig:chiral} and \ref{fig:polaron_states}.

Second, when these polaron plaquettes percolate through the crystal, a
band is formed with the \xxyy\ and \psigma\ orbitals on the percolating
swath.  This metallic band interacts with the \xxyy\ hole $d^9$ spins
on the undoped Cu sites and the plaquette polarons.
The random distribution of impurities leads to a distribution
of the energy separations of polaron states shown in figure
\ref{fig:polaron_states}.

Third, this energy distribution
is uniform.  The linear resistivity arises from this assumption
as shown in section \ref{sec:transport}.
Since the resistivity is non-linear for certain dopings and
temperature ranges, this assumption is not always valid.

The transition from spin-glass to superconductor in \lsco\ at
$x\approx0.05$\cite{birgeneau_rmp}
and from an antiferromagnet to superconductor at
$\delta\approx0.35$\cite{ybco_af_sc} in \ybco\ occur 
at the doping when the polarons percolate through the crystal.

In this section, the site percolation doping values are computed for
\lsco\ and \ybco.  Reasonable assumptions for the
distribution of plaquettes are used to approximately simulate the
repulsion of the dopants.  The computed
values are close to known phase transitions in these materials.

We also computed the percolation for two additional systems.
The first is \lsco\ where each Sr dopes exactly
one Cu site as shown in figure \ref{fig:a1_hole}
and the second is a 2D square lattice with plaquette doping.
The computed
\lsco\ 1-Cu percolation value of $x\approx0.20$ is associated
with the observed orthorhombic to tetragonal phase transition.
\cite{birgeneau_rmp}

For the 2D square
lattice with four Cu plaquette doping, percolation occurs
at $x\approx0.15$.  We believe the 2D percolation of the
plaquettes should be associated with the transition from
insulator to metal at $x\approx0.15$ found
by low temperature resistivity
measurements in large pulsed magnetic fields.\cite{boeb1}
This is further discussed in section \ref{sec:transport}.

All percolation computations described here were performed using
the linear scaling algorithm of Newman and Ziff.\cite{ziff}

In all these calculations, we simplify the problem by using
Cu sites only.  For \lsco, we take each Cu to have four neighbors
in the plane at vectors $(\pm a,0,0)$, $(0,\pm b,0)$ and 
eight neighbors out of the plane at $(\pm a/2,\pm b/2,\pm c/2)$
where $a$, $b$, and $c$ are the cell dimensions.  Thus, each Cu has
a total of 12 neighbors in the site percolation calculations.

\begin{figure}[tbp]
\centering  \includegraphics[width=0.9\linewidth]{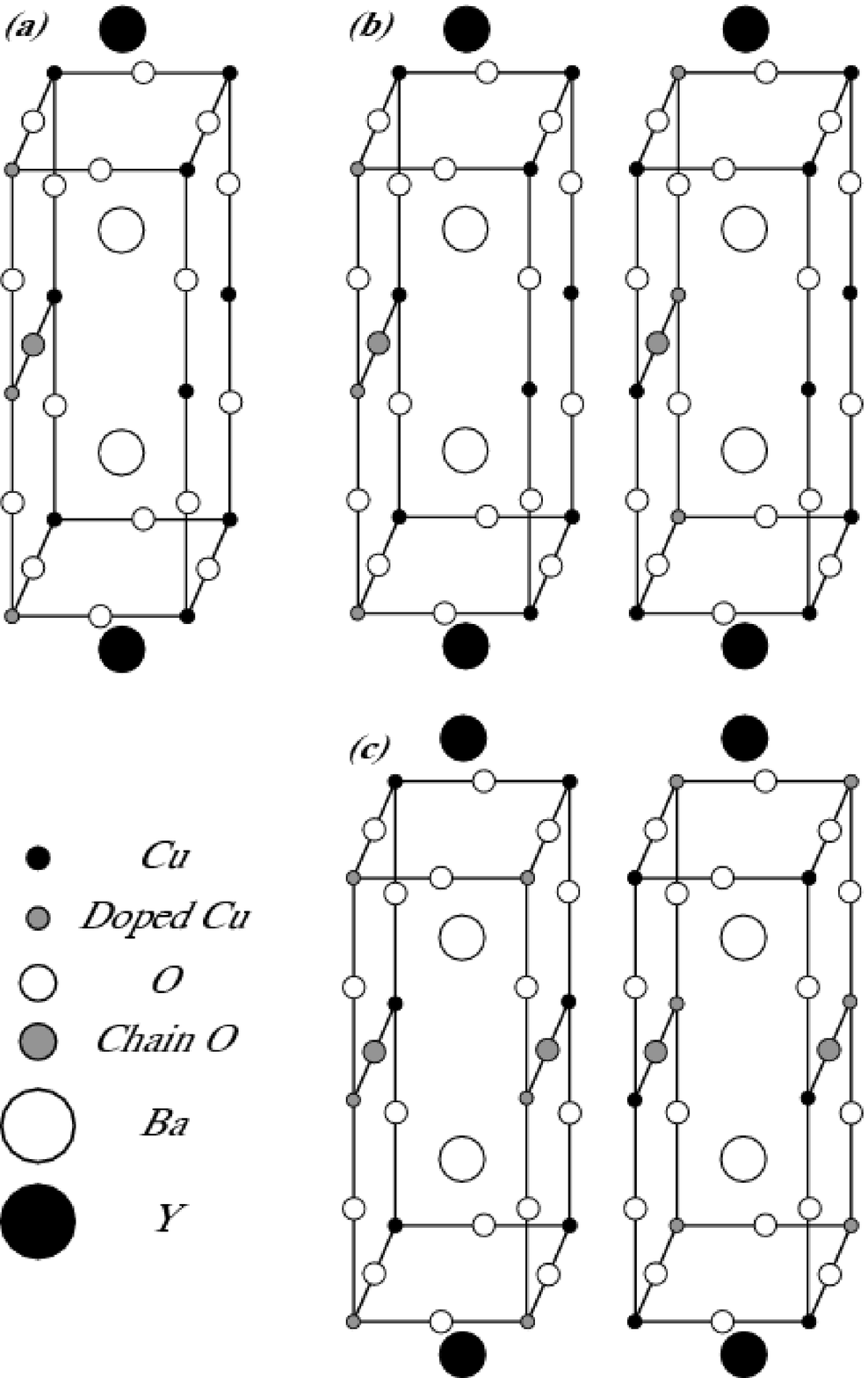}
\caption{Doping constraints used for \ybco\ calculations
shown in table \ref{table:percolation}.  The chain O and its
three doped Cu sites are shaded gray.  Three different doping
scenarios are shown.  (a) the chain O dopes a fixed Cu triple.
(b) the chain O randomly dopes one of the two possible Cu triples.
No triple may be doped by two adjacent chain O.  (c) If two
chain O are in the same cell, the two Cu doping configurations
where the triples are closest to each other are not permitted.
}
\label{fig:ybco_con}
\end{figure}

For all \ybco\ calculations, we take each planar Cu to be
connected to a total of six Cu atoms.
There are four nearest neighbors in the same
CuO$_2$ plane, one neighboring Cu on the adjacent
CuO$_2$ across the intervening Y atom and one Cu on the neighboring
chain.  The chain Cu is connected to the two Cu atoms in
the CuO$_2$ planes above and below itself.
We assume a chain O dopes
three Cu atoms, two in CuO$_2$ planes and the corresponding
chain Cu as shown in the constraints of figure \ref{fig:ybco_con}.

Table \ref{table:percolation} lists the computed percolation values for
\lsco, a 2D square lattice, and \ybco\ for various types of
doping and doping constraints.
These constraints were chosen to simulate
the repulsion of the dopants and are approximations
to the actual distribution of dopants in the cuprates.

The \lsco\ and \ybco\ calculations are on
$200\times200\times200$ lattices with $2,500$ different dopings.  The
square lattice size is $2,000\times2,000$ with $5,000$
different dopings.

\begin{table}[tbp]
\caption{\label{table:percolation}Percolation values for various
structures and doping scenarios.  The constraints column
references the figures describing the applied constraint.
All \lsco\ and \ybco\ results are obtained for a
$200\times200\times200$ lattice
with $2,500$ doped ensembles.  The percolation value is the computed
critical $x$ in \lsco\ and $\delta$ in \ybco.
The square lattice results are
for a $2,000\times2,000$ lattice with $5,000$ ensembles.
The digit in parentheses is the error in the last digit.}
\begin{ruledtabular}
\begin{tabular}{cccc}
Structure & Dopant Type & Constraints & Percolation \\
\hline
LSCO & 1-Cu & none & $0.19617(2)$ \\
LSCO & 4-Cu & none & $0.05164(1)$ \\
LSCO & 4-Cu & fig. \ref{fig:lsco_con}a & $0.05097(1)$ \\
LSCO & 4-Cu & fig. \ref{fig:lsco_con}a,b & $0.04834(1)$ \\
LSCO & 4-Cu & fig. \ref{fig:lsco_con}a,b,c & $0.04880(1)$ \\
LSCO & 4-Cu & fig. \ref{fig:lsco_con}a,b,c,e & $0.04904(1)$ \\
LSCO & 4-Cu & fig. \ref{fig:lsco_con}a,b,e & $0.04862(1)$ \\
LSCO & 4-Cu & fig. \ref{fig:lsco_con}a,b,c,d & $0.04926(1)$ \\
LSCO & 4-Cu & fig. \ref{fig:lsco_con}a,b,c,d,e & $0.04943(1)$ \\
Square & 4-Cu & fig. \ref{fig:lsco_con}a,b & $0.15053(1)$ \\
YBCO & 3-Cu & fig. \ref{fig:ybco_con}a & $0.31162(2)$ \\
YBCO & 3-Cu & fig. \ref{fig:ybco_con}b & $0.32890(2)$ \\
YBCO & 3-Cu & fig. \ref{fig:ybco_con}b,c & $0.36098(2)$ \\
\end{tabular}
\end{ruledtabular}
\end{table}

\begin{figure}[tbp]
\centering  \includegraphics[width=0.8\linewidth]{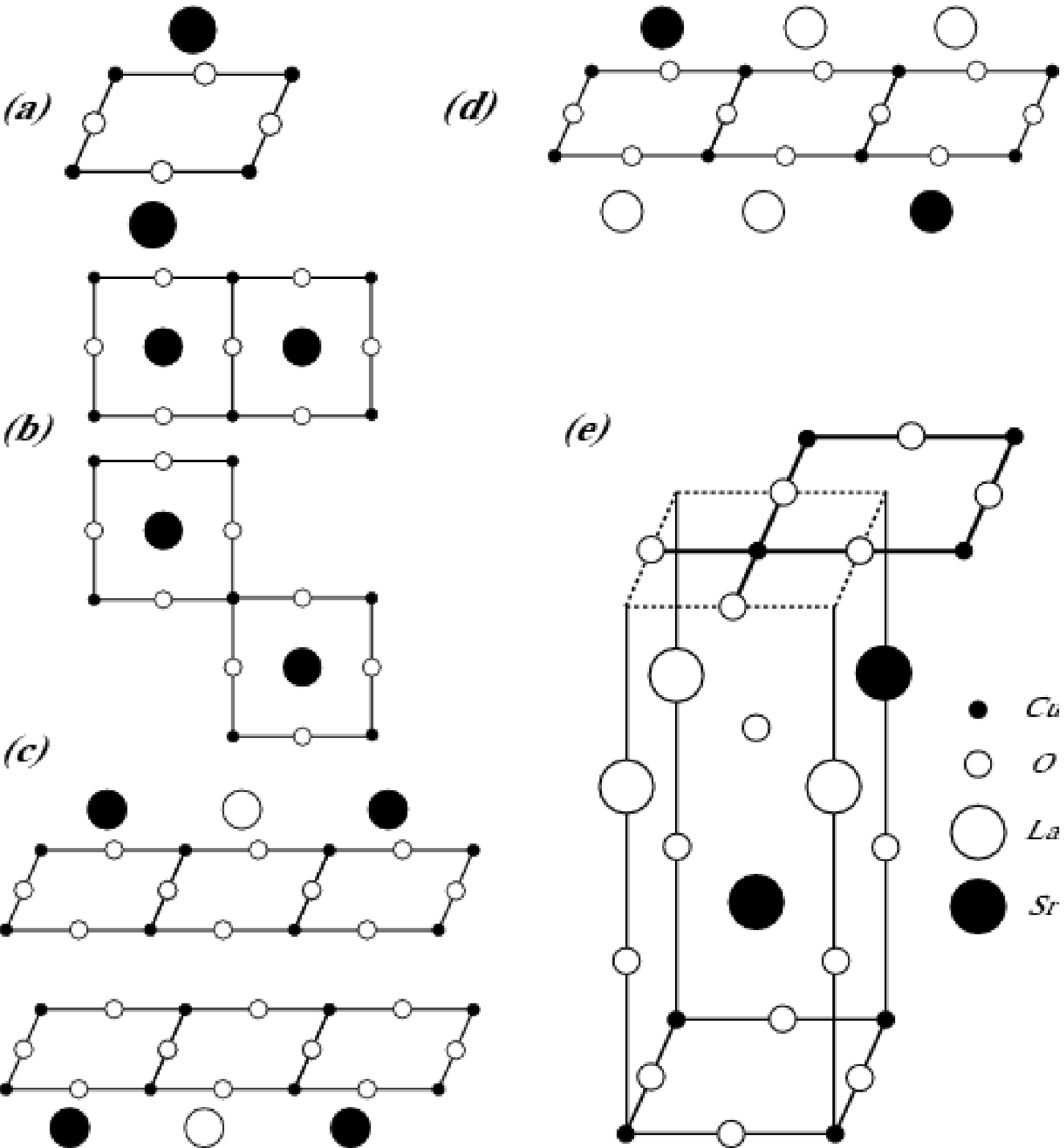}
\caption{Applied Sr doping constraints used for the \lsco\ 
and 2D square lattice plaquette percolation
calculations shown in table \ref{table:percolation}.  For each
four Cu plaquette in figures \ref{fig:chiral} and
\ref{fig:polaron_states},
two Sr atoms, one above and one below the plaquette (upper and
lower Sr), can dope the Cu's.  Each figure shows the disallowed
configuration of Sr doping.  It is assumed all $\pm90^\circ$ and
$180^\circ$ rotated configurations are equivalent to the figure
and also disallowed.
(a) an upper and lower Sr doping the same plaquette.
(b) a Cu atom in a plaquette doped by two different
Sr atoms.  This figure includes the three cases of two upper Sr,
two lower Sr, and one upper and one lower Sr.  This is the no
overlap constraint.
(c) adjacent plaquettes doped by two upper Sr or two lower Sr.
(d) adjacent plaquettes doped by one upper Sr and one lower Sr.
(e) nearest neighbor upper Sr and lower Sr in different LaO planes.
}
\label{fig:lsco_con}
\end{figure}

The first \lsco\ calculation is the critical doping for percolation
of doped Cu where each Sr dopes the single Cu shown in
figure \ref{fig:a1_hole} instead of the four Cu plaquette of
figure \ref{fig:chiral}.
Although we assume plaquettes are created at low dopings, once the
doping is large enough, there is crowding of the
plaquettes.  Single Cu polarons are formed.
This single Cu percolation calculation
is an approximate measure of the doping for the
transition from predominantly doped plaquettes to single
site polarons.  A phase transition at this crossover doping
is expected.
The computed percolation of $x\approx0.20$ matches the
orthorhombic to tetragonal transition\cite{birgeneau_rmp} doping.

From the table, the critical doping for 3D plaquette percolation
in \lsco\ 
is $x\approx0.05$ regardless of the applied doping
constraints and matches the spin-glass to superconductor
transition.\cite{birgeneau_rmp}

This is because the plaquette percolation values are
approximately $1/4$ of the single Cu percolation result of
$x\approx 0.20$.

For \ybco, the more realistic doping constraints are the second and
third cases where $\delta\approx0.33$ and $\delta\approx0.36$ since
O chains should not have a preference for which Cu triple to dope.
Experiment\cite{ybco_af_sc} finds $\delta\approx0.35$.

From these results, we conclude the plaquette
polaron model with percolation can obtain known insulator
to metal phase transitions in \lsco\ and \ybco.

\section{Neutron Scaling and Incommensurability}
\label{sec:neutron}

\subsection{Scaling}
\label{subsec:scaling}
Neutron spin scattering measures the imaginary part of the magnetic
susceptibility, $\chi(\mathbf{q},\omega)$.

The integral of the imaginary part of the spin susceptibility
$\int{\mathbf{d^2q}}\ \chi''({\mathbf q},\omega)$
over the Brillouin zone where $\chi=\chi'+i\chi''$
has been found
\cite{neutron_scaling1, neutron_scaling2, neutron_scaling3,
neutron_scaling4, neutron_scaling5, neutron_scaling6,
neutron_scaling7} to be a function of $\omega/T$.
The integral is the on-site magnetic spin susceptibility.

The $\omega/T$
scaling is unusual because $\chi''\sim\omega/J_{dd}$ for
an antiferromagnet and $\chi''\sim\omega/E_F$ for a band
where $J_{dd}$ is the $d^9$ AF spin coupling and $E_F$ is the
\xxyy\ band Fermi energy.

In this section,
we show that the single polaron susceptibility is
a function of $\omega/T$ when the
energy difference between polaron chiral states with opposite spins
and chiralities is uniformly distributed.

The $\mathbf{q}$ dependence of the polaron susceptibility,
$\chi_p''(\mathbf{q},\omega)$, is peaked at $\mathbf{q}=0$ if
spin-flip polaron scattering dominates at low energy.
$\chi_p''(\mathbf{q},\omega)$ is peaked at $\mathbf{q}=(\pi,\pi)$
if the polaron spin and chirality flip at low energy.  The
latter scatters the polaron to its time-reversed state.  The
time-reversed chiral polarons are the low energy excitations, as
shown in section \ref{subsec:incomm}.  The $\mathbf{q}$ dependence
is broad because the polaron is localized over a four-site
plaquette.  Since the total susceptibility is dominated by
$\mathbf{q}$ near $(\pi,\pi)$, the
polaron susceptibility is approximately momentum independent,
$\chi_p''(\mathbf{q},\omega)\approx\chi_p''(\omega)$.

The imaginary part of the polaron susceptibility is found to
be of the form
$\chi_p''(\omega)\sim\tanh(\beta\omega/2)$ and satisfies scaling.
The coupling of the polaron spin and chiral
orbital state to the undoped $d^9$
spins causes the total susceptibility
to become incommensurate.  This is shown in the next subsection
\ref{subsec:incomm} where we compute the static spin structure
factor for classical spins perturbed by chiral polarons.

In this section, we show that coupling to the undoped $d^9$ spins
leads to a dynamic susceptibility consistent with the
measured form
\cite{neutron_scaling1, birgeneau_rmp} in equation \ref{eqn:chi0}.

\begin{figure}[tbp]
\centering  \includegraphics[width=0.8\linewidth]{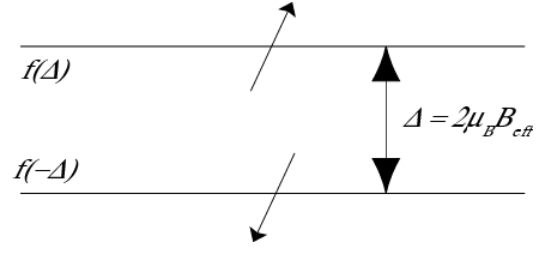}
\caption{Energy difference $\Delta$ between an up and down spin
polaron.  $f(-\Delta)$ and $f(\Delta)$ are the equilibrium occupations
of the spins states where $f(\epsilon)$ is the Fermi-Dirac function,
$f(\epsilon)=1/(e^{\beta\epsilon}+1)$.  B$_{\mathrm{eff}}$ is the
effective magnetic field that splits the spin energies by $\Delta$.
The chiralities of the two states are different for low energies
leading to a broad peak in $\chi_p''(\mathbf{q},\omega)$
at $(\pi,\pi)$.
}
\label{fig:pol_eng}
\end{figure}

Consider a polaron as in figure \ref{fig:pol_eng} with energy
$\Delta$ separating the down spin state from the up spin and
occupations $n_\down$ and $n_\up$ in thermal equilibrium.  Since
the polaron is always singly occupied, $n_\up+n_\down=1$ and
$n_\down=e^{\beta\Delta}n_\up$.  Solving for $n_\down$ and
$n_\up$,
\begin{equation}
\label{eqn:occ}
n_\down=f(-\Delta),\ \ n_\up=f(\Delta), 
\end{equation}
where $f(\epsilon)$ is the Fermi-Dirac function,
\begin{equation}
\label{eqn:fd}
f(\epsilon)=\frac{1}{e^{\beta\epsilon}+1}.
\end{equation}

To compute the dynamical polaron spin susceptibility, consider
an applied magnetic field $Be^{i\omega t}$ normal to the spin
quantization axis.  The alternating field
induces transitions between the two states if $\Delta=\omega$.

Let $W=W_{\up\down}=W_{\down\up}$ be the induced transition rate 
between the two spin states.  The total absorption rate is,
\begin{equation}
\label{eqn:p1}
\left<P(\Delta)\right>=\hbar\omega W(n_\down - n_\up).
\end{equation}
Using equation \ref{eqn:occ},
\begin{equation}
\label{eqn:p2}
\left<P(\Delta)\right>=
\hbar\omega W\left[f(-|\Delta|) - f(|\Delta|)\right],
\end{equation}
\begin{equation}
\label{eqn:p3}
\left<P(\Delta)\right>=\hbar\omega W\tanh(\beta|\Delta|/2).
\end{equation}
The absolute values of $\Delta$ are used above because the absorption
rate is independent of which spin state is lower in energy.
The transition rate $W$ is,
\begin{equation}
\label{eqn:rate}
W(\Delta)=\frac{2\pi}{\hbar}\mu_B^2 B^2\delta(\omega-|\Delta|).
\end{equation}
Averaging $W$ over all spin quantization directions multiplies
equation \ref{eqn:rate} by $2/3$,
\begin{equation}
\label{eqn:rate1}
\langle W(\Delta)\rangle
=
\frac{4\pi}{3\hbar}\mu_B^2 B^2\delta(\omega-|\Delta|).
\end{equation}
Let $\rho(\Delta)$
to be the probability distribution of energy differences
for spin and chirality flips.
Summing over all polarons, the total absorption rate is,
\begin{equation}
\label{eqn:p4}
\left<P(\omega)\right>=\hbar\omega 
\int d\Delta\ \rho(\Delta)
\tanh(\beta|\Delta|/2)\langle W(\Delta)\rangle,
\end{equation}
\begin{equation}
\label{eqn:p5}
\left<P(\omega)\right>= \left(\frac{4\pi}{3}\right)\omega\mu_B^2 B^2
\left[\rho(\omega)+\rho(-\omega)\right]
\tanh\left(\frac{\beta\omega}{2}\right),
\end{equation}
\begin{equation}
\label{eqn:p6}
\left<P(\omega)\right>= \left(\frac{8\pi}{3}\right)\omega\mu_B^2 B^2
\rho(\omega)
\tanh\left(\frac{\beta\omega}{2}\right),
\end{equation}
where $\rho(-\Delta)=\rho(\Delta)$ because there is an
equal number of polarons with an up spin lower in energy than
a down spin as there are down spins lower in energy than up spins.

The absorption rate can be written in terms of the imaginary
part of the polaron susceptibility as,
\begin{equation}
\label{eqn:p7}
\left<P(\omega)\right> =
\frac{1}{2}\omega\chi_p''(\omega)B^2,
\end{equation}
leading to the imaginary susceptibility per polaron of,
\begin{equation}
\label{eqn:chi}
\chi_p''(\omega)= \left(\frac{16\pi}{3}\right)\mu_B^2 \rho(\omega)
\tanh\left(\frac{\beta\omega}{2}\right).
\end{equation}

The probability density of polaron energy separations
$\rho(\Delta)$
is taken to be uniform with
$\int d\Delta\rho(\Delta)=1$
and of the form,
\begin{equation}
\label{eqn:rho}
\rho(\Delta)=\left\{
\begin{array}{cc}
\frac{1}{2\Delta_{\mathrm{max}}}, &
-\Delta_{\mathrm{max}}<\Delta<\Delta_{\mathrm{max}} \\
0, & |\Delta|>\Delta_{\mathrm{max}}
\end{array}
\right.
\end{equation}
where $\Delta_{\mathrm{max}}$ is doping dependent,
$\Delta_{\mathrm{max}}=\Delta_{\mathrm{max}}(x)$.

Equations \ref{eqn:chi} and \ref{eqn:rho} show that the polaron
susceptibility is a function of $\omega/T$ 
and has the approximate form seen
in experiments.\cite{birgeneau_rmp, neutron_scaling1}
The functional form of $\chi_p''$
increasing from $\chi_p''(0)=0$ and saturating for
$\beta\omega>1$ arises from the thermal occupations of polaron
states with energy splitting $\omega$.
When $\beta\omega\ll 1$,
the two states have almost equal occupation and the absorbed energy
is small from equations \ref{eqn:p1} and \ref{eqn:p2}.
For $\beta\omega\gg 1$, the lower energy state is always occupied
and the higher energy spin state is always unoccupied.  In this
case, the absorption saturates.  Since the polaron spin density
$\rho(\Delta)$ is constant up to $\Delta_{\mathrm{max}}$,
it is $\beta\omega$ that determines
the amount of absorption due to the difference of the two
Fermi-Dirac occupation factors.
Finally, if there are no spin flip energies smaller than
$\Delta_{\mathrm{min}}$, then $\chi_p''(\omega)$ is zero
for $\omega<\Delta_{\mathrm{min}}$. 

The measured susceptibility
for \lsco\ at $x=0.04$ is normalized and fit by the expression
\cite{neutron_scaling1, birgeneau_rmp}
\begin{equation}
\label{eqn:chi0}
\left(\frac{2}{\pi}\right)
\tan^{-1}[a_1(\beta\omega) + a_3(\beta\omega)^3],
\end{equation}
with $a_1=0.43$ and $a_3=10.5$.  This curve rises to the saturating
value of one faster than our expression in equation \ref{eqn:chi}.

The contribution to the susceptibility from the undoped $d^9$ spins
and the metallic \xxyy\ band has not been included.  The band
contribution is on the order of $\omega/E_f$ where $E_f$ is
the Fermi energy and can be neglected.  The imaginary susceptibility
from the undoped $d^9$ spins is on the order\cite{pines_scaling1}
of $\omega/\Gamma$ where $\Gamma$ is several $J_{dd}$ to
$E_f$ and can also be neglected.  The real part of the $d^9$
susceptibility is approximately constant up to the
energy $\omega\sim\Gamma$.  We may therefore take the $d^9$
susceptibility to be real and $\omega$ independent for small $\omega$.
Thus, the
$\omega$ dependence of the total susceptibility arises from
the polaron susceptibility in equation \ref{eqn:chi}.

The $\mathbf{q}$ dependence
of the susceptibility is incommensurate from the
calculations of the next section and of the form,\cite{pines_scaling1}
\begin{equation}
\label{eqn:chi1}
\chi(\mathbf{q})=\frac{\mu_B^2\tilde\chi_0\left(\xi/a\right)^2}
{1+\left(\mathbf{q}-\mathbf{Q}\right)^2
\left(\xi/a\right)^2},
\end{equation}
where $\mathbf{Q}$ is the incommensurability peak vector and
$\xi$ is the correlation length.  From the computations in the
next section, $\mathbf{Q}$ is shifted from $(\pi/a,\pi/a)$
along the Cu$-$O bond direction and agrees with experiment.
$\xi\approx a/\sqrt{x}$ is the mean separation between Sr.

\begin{figure}[tbp]
\centering  \includegraphics[width=\linewidth]{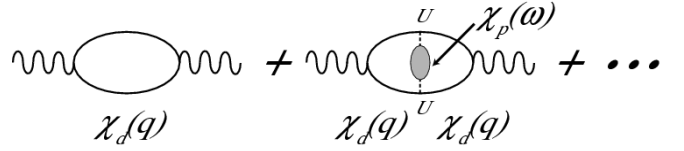}
\caption{Diagrams to sum in the random phase approximation
for the dynamic susceptibility $\chi(\mathbf{q},\omega)$.
The first term is the static susceptibility
$\chi(\mathbf{q})$ in equation \ref{eqn:chi1} and the shaded loop
is the polaron susceptibility $\chi_p(\omega)$ with imaginary
part shown in equation \ref{eqn:chi}.
$U$ is the coupling energy.}
\label{fig:rpa}
\end{figure}

Summing the random phase approximation
diagrams in figure \ref{fig:rpa} leads to,
\begin{equation}
\label{eqn:chi2}
\left[\frac{\chi(\mathbf{q},\omega)}{\mu_B^2}\right]^{-1}=
\left[\frac{\chi(\mathbf{q})}{\mu_B^2}\right]^{-1}-
xU^2\left[\frac{\chi_p(\omega)}{\mu_B^2}\right],
\end{equation}
\begin{equation}
\label{eqn:chi3}
\chi(\mathbf{q},\omega)=\frac{\mu_B^2\tilde\chi_0\left(\xi/a\right)^2}
{1+\left(\mathbf{q}-\mathbf{Q}\right)^2\left(\xi/a\right)^2 -
\tilde\chi_0 U^2\left[\chi_p(\omega)/\mu_B^2\right]},
\end{equation}
where we have used $(\xi/a)^2x=1$ for the $\chi_p$ term.

Using the integral,
\begin{eqnarray}
\label{eqn:chi4}
\mathrm{Im}\int\frac{d^2q}{A+Bq^2-iC} =
\int_0^{+\infty}\frac{\left(2\pi C\right)qdq}
{\left(A+Bq^2\right)^2+C^2}, \nonumber \\
 = 
\left(\frac{\pi}{B}\right)\left[
\frac{\pi}{2}-\tan^{-1}\left(\frac{A}{C}\right)\right]
= \left(\frac{\pi}{B}\right)\tan^{-1}\left(\frac{C}{A}\right),
\end{eqnarray}
and defining $\tilde\chi_p(\omega)=\chi_p(\omega)/\mu_B^2$,
the integrated imaginary susceptibility is,
\begin{equation}
\label{eqn:chi5}
\int\chi''(\mathbf{q},\omega)dq=
\pi\mu_B^2\tilde\chi_0
\tan^{-1}\left[\frac{\tilde\chi_0 U^2\tilde\chi_p''(\omega)}
{1-\tilde\chi_0 U^2\tilde\chi_p'(\omega)}\right].
\end{equation}
In the above expression,
$\tilde\chi_p=\tilde\chi_p'+i\tilde\chi_p''$ has been expanded
into real and imaginary parts.

Using $\chi_p'(-\omega)=\chi_p'(\omega)$ and
$\chi_p''(-\omega)=-\chi_p''(\omega)$, the equation can
fit the experimental curve in equation \ref{eqn:chi0}.

\subsection{Incommensurability}
\label{subsec:incomm}
\begin{figure}[tbp]
\centering  \includegraphics[width=\linewidth]{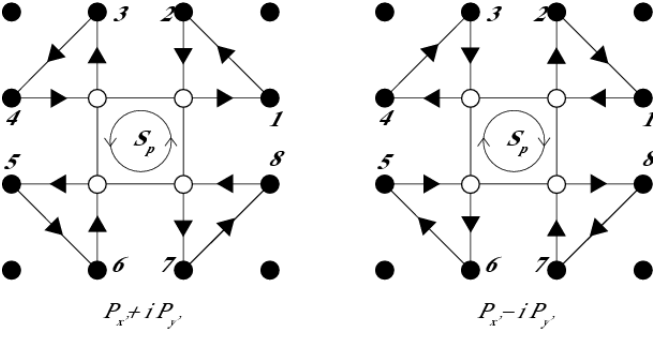}
\caption{Coupling of chiral polarons to neighboring $d^9$ spins.
$\mathbf{S}_p$ is the polaron spin and it couples to four $d^9$ pairs
in the cyclic order shown by the arrows and equation
\ref{eqn:jch}.  The chiral coupling for the $P_{x'}-iP_{y'}$ polaron
reverses the cyclic ordering of the spins as seen in the figure
and leads to the same expression as the $P_{x'}+iP_{y'}$ with
$J_{ch}\rightarrow -J_{ch}$.
}
\label{fig:jch}
\end{figure}

There is an antiferromagnetic coupling, $J_{pd}$, between
the polaron spin $\mathbf{S}_p$ and the neighboring $d^9$ spins.
The polaron is delocalized over four Cu sites.  The probability
of the hole residing on a particular Cu is $1/4$
leading to the estimate $J_{pd}\approx (1/4)J_{dd}$ where
$J_{dd}$ is the undoped $d^9$ AF coupling.
The effective coupling of a chiral polaron to the $d^9$ spin
background is known
\cite{zee1, gooding1, gooding_birgeneau} to induce a twist in
the neighboring spins.  This can be encapsulated in
a topological charge term\cite{polyakov} of the form
$\mp J_{ch}[{\mathbf{S}_p\cdot(\mathbf{S}_{1}\times\mathbf{S}_{2})}]$
where $\mathbf{S}_p$ is the polaron spin and
$\mathbf{S}_{d1}$, $\mathbf{S}_{d2}$ are adjacent $d^9$ spins as
shown in figure \ref{fig:jch}. 

The expectation value
$\langle{\mathbf{S}_p\cdot(\mathbf{S}_{1}
\times\mathbf{S}_{2})}\rangle=0$
for states invariant under time reversal.  Thus,
the expectation value of the topological charge is zero
for the real polaron states $P_{x'}$, $P_{y'}$, $S$, and $D_{xy}$.
The complex linear combinations in the chiral states lead to non-zero
topological charge.  The above chiral coupling term is
the simplest coupling of chiral polarons
to the neighboring spins.

The references\cite{zee1, gooding1, gooding_birgeneau} considered 
holes in both the t-J and three-band Hubbard models
that can delocalize over a four-site plaquette.
In our model, \xxyy\ spins delocalize on the plaquettes forming
a band when the polarons percolate.
Our chiral coupling is between a polaron spin
and the adjacent spin sites that may be undoped $d^9$ or another
polaron spin.  The specific form for the coupling is analogous
to previous work.

The coupling of the \xxyy\ band to the neighboring spins
is smaller than the chiral coupling of the polaron
and $d^9$ spins.
The perturbation arising from the band spin coupling selects
incommensurability along the Cu$-$O bond directions as shown
at the end of this section.

The polaron chiral coupling of $P_{x'}\pm iP_{y'}$ to the
neighboring $d^9$ spins is,
\begin{equation}
\label{eqn:jch}
\begin{array}{ccc}
H_{ch} & = & \mp J_{ch}\left\{
\mathbf{S_p}\cdot(\mathbf{S_1}\wedge\mathbf{S_2}) +
\mathbf{S_p}\cdot(\mathbf{S_3}\wedge\mathbf{S_4}) +
\right. \\
       &   &
\left.
\mathbf{S_p}\cdot(\mathbf{S_5}\wedge\mathbf{S_6}) +
\mathbf{S_p}\cdot(\mathbf{S_7}\wedge\mathbf{S_8})
\right\},
\end{array}
\end{equation}
where $J_{ch}>0$ and the spins are labelled in figure \ref{fig:jch}.
The antiferromagnetic coupling of polaron spins to undoped $d^9$
spins is
\begin{equation}
\label{eqn:jpd}
H_{pd}=J_{pd}\mathbf{S_p}\cdot
\left(\mathbf{S_1}+\dots+\mathbf{S_8}\right).
\end{equation}

Electronic hopping matrix elements are on the order
of $0.5-1.0$ eV.  The chiral coupling,
$J_{ch}$, is estimated to be less than or of the same order.
Gooding \etal\cite{gooding_mailhot1}
obtain $J_{ch}\approx 3J_{dd}$ from numerical
simulations and by computing the effective next-nearest
neighbor antiferromagnetic coupling, $J'_{dd}$, induced by chiral
polarons at very low doping.  $J'_{dd}$ is then
compared with Raman data to obtain $J_{ch}$.

In the spin Hamiltonian of Gooding \etal,\cite{gooding_mailhot1}
the chiral coupling term is squared,
$-J_{ch}[\mathbf{S_1}\cdot(\mathbf{S_2}\wedge\mathbf{S_3})]^2$, in
contrast to our linear terms in equation \ref{eqn:jch}.
The overall energy scale of $J_{ch}$
is similar.  We take $J_{ch}=3J_{dd}$ where $J_{dd}\approx 0.1$ eV
in our computations of the static neutron structure
factor.  We have found our results for the magnitude of the
incommensurability are independent
of the precise values of all of the parameters. 
The only necessary
feature to obtain incommensurability is that the chiral
coupling, $J_{ch}$, is
sufficiently large to break the $(\pi,\pi)$ spin
ordering from the antiferromagnetic spin coupling, $J_{dd}$.

\begin{figure}[tbp]
\centering  \includegraphics[width=\linewidth]{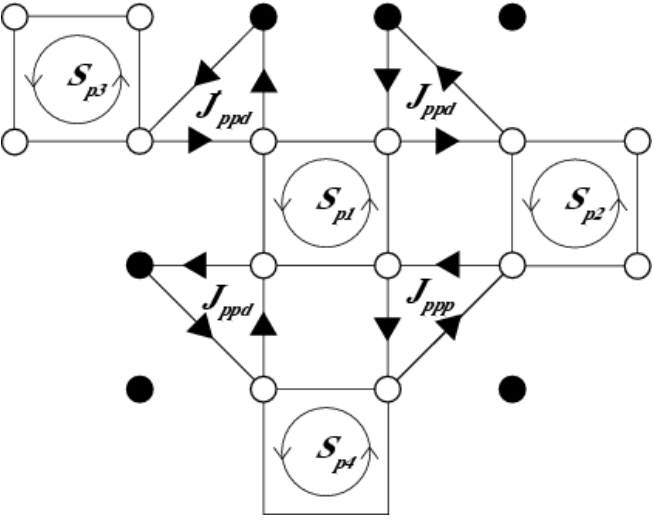}
\caption{Schematic of adjacent polaron configurations.
$J_{ppp}$ is the chiral
coupling between three polarons, $J_{ppd}$ couples
two polarons to a $d^9$ spin, and $J'_{ppd}$ couples two polarons
to a $d^9$ spin when one polaron is shifted by one lattice spacing.
The couplings are shown for the case where all the polarons are
$P_{x'}+iP_{y'}$.  For each opposite chirality polaron,
$P_{x'}-iP_{y'}$, the coupling should be multiplied by $-1$.
The figure does not exhaust all possible couplings.  For
example, there is another term $J'_{ppp}$ if $\mathbf{S_{p4}}$
is shifted to the right by one lattice spacing.
}
\label{fig:jpp}
\end{figure}

There is an antiferromagnetic
spin-spin coupling between the polarons.  $J_{pp}$ is the coupling
between $\mathbf{S_{p1}}$ and $\mathbf{S_{p2}}$ and $J'_{pp}$
is the coupling between
$\mathbf{S_{p1}}$ and $\mathbf{S_{p3}}$ shown in
figure \ref{fig:jpp}.  An estimate
of $J_{pp}$ and $J'_{pp}$ is obtained in a similar manner
to $J_{pd}$.  For $J_{pp}$, the polarons have two adjacent pair
sites.  An antiferromagnetic coupling occurs for every adjacent pair.
This occurs with probability $(1/4)^2=1/16$.  There are two pairs for
$J_{pp}$ and one for $J'_{pp}$ leading to estimates
$J_{pp}\approx(1/8)J_{dd}$ and $J'_{pp}\approx(1/16)J_{dd}$.

Figure \ref{fig:jpp} shows various chiral couplings when polarons
are adjacent to each other.  Using a similar analysis, we
estimate the magnitude of the chiral couplings to be,
$J_{ppp}\approx(1/4)^2J_{ch}$, $J_{ppd}\approx(1/4)J_{ch}$, and
$J'_{ppd}\approx(1/4)J_{ch}$.

The total spin Hamiltonian for the $d^9$ spins and polarons is
\begin{equation}
\label{eqn:htot}
H=H_{dd}+H_{pd}+H_{pp}+H_{ch}^{tot},
\end{equation}
where $H_{dd}$ is the antiferromagnetic $d^9$ spin-spin coupling
with $J_{dd}\approx 0.1$ eV.  $H_{pd}$ is the polaron-$d^9$
coupling and $H_{pp}$ is the polaron-polaron spin coupling.
$H_{ch}^{tot}$ is the total chiral coupling.

The chiral coupling $H_{ch}^{tot}$ is invariant under
polaron time reversal
that flips the chirality of a single polaron
$P_{x'}\pm iP_{y'}\rightarrow P_{x'}\mp iP_{y'}$ or
$(J_{ch}\rightarrow -J_{ch})$ and
the polaron spin, $\mathbf{S_p}\rightarrow\mathbf{-S_p}$.
$H_{pd}\rightarrow -H_{pd}$ is not invariant under time reversal
of the polaron.
When the chiral coupling is much larger than all spin-spin couplings,
$J_{ch}\gg J_{dd}>J_{pd}>J_{pp}>J'_{pp}$, the ground state
energy becomes independent of the polaron chiralities.

This is an important point because it means the energy to
simultaneously flip the chirality and spin of a polaron has an
energy scale $J_{dd}$ while flipping either the chirality or
the spin, but not both, has an energy scale $J_{ch}$.

\begin{figure}[tbp]
\centering  \includegraphics[width=0.7\linewidth]{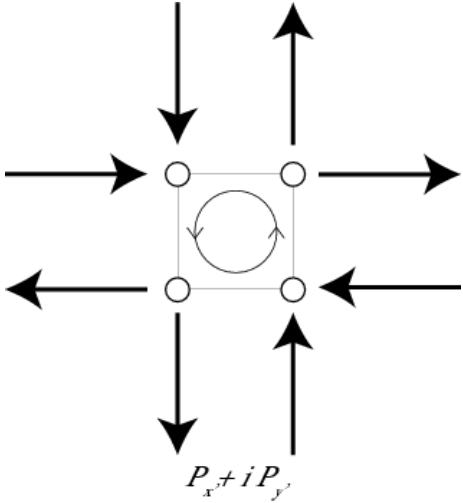}
\caption{Minimized energy configuration for a polaron with
spin pointing out of the page surrounded by 8 $d^9$ spins.
Only $J_{dd}$ and $J_{ch}$ are non-zero.
This represents the regime dominated by $J_{ch}$.
}
\label{fig:4x4}
\end{figure}

Figure \ref{fig:4x4} shows the minimized spin ordering surrounding
a polaron when the chiral coupling, $J_{ch}$, dominates the
polaron spin to $d^9$ coupling, $J_{pd}$.
Increasing $J_{ch}$ further does not change the spin ordering.
This leads to an incommensurability that is weakly parameter dependent.
All the neighboring
$d^9$ spins in the figure are orthogonal to the polaron spin.
The $H_{pd}$ antiferromagnetic energy is zero.  The energy
difference of the time reversed polaron in the same background
is also zero.  The spin-spin couplings, $J_{pd}$ etc, lead to
non-zero energy differences.

The static neutron spin structure factor is computed by minimizing
the energy in
equation \ref{eqn:htot} on a finite 2D lattice with classical
spins $\mathbf{S}_i$ of unit length, $\mathbf{S}_i^2=1$.  Each
undoped $d^9$ site has a spin and every
polaron has an orbital chirality, $P_{x'}\pm iP_{y'}$,
and spin.  All the terms for the classical
Hamiltonian are described above along with the parameters used.
The only constraint on the polarons is that they may not overlap,
but otherwise they are randomly placed in the lattice.  At
this point, the effect of the delocalized \xxyy\ band electrons
is ignored.

One can imagine additional constraints on the
placement of the polarons arising from polaron-polaron
Coulomb repulsion. Also, calculations on 3D lattices
with a small interlayer antiferromagnetic spin coupling
can be done.  The addition of polaron constraints
and the third dimension does not to change the computed
incommensurability or its location in the Brillouin zone.
These effects are ignored in this paper.

Finally, calculations with periodic and non-periodic boundary
conditions were performed to ensure there is no long range twist
in the spins that is frustrated by periodic boundary
conditions.  No major difference was found in the computed structure
factors.  This is likely due to the small energy difference
between a polaron and its time reversed partner.

Computations were done on $256\times 256$ lattices with polaron
dopings of $x=0.075$, $0.10$, and $0.125$.  A random configuration
of polarons was chosen subject to the constraint that no polarons
overlap.  Starting spins and chiralities are randomly generated
and the initial energy is calculated.  The energy is minimized
by performing local minimizations.\cite{walstedt1}

A spin is selected and
the effective magnetic field on the site is computed.  Since the
Hamiltonian in equation \ref{eqn:htot} is linear in the spins,
the energy arising from the chosen spin is minimized by aligning it
with the magnetic field.  If the spin is a polaron spin, then
the effective magnetic field is computed for both orbital
chiralities to determine the chirality and spin that minimizes
the energy.  The chirality is flipped if a lower energy can
be obtained.
The program loops through all the spins, determines
the new energy, and compares it to the previous energy to
decide on convergence.  Calculations were performed on
$5,000$ different polaron configurations for each doping value.

The Grempel algorithm\cite{grempel, gooding_birgeneau} was used
to obtain
the global minimum.  This algorithm is similar to raising the
temperature to allow the energy to climb out of local minima
and then annealing.  Unlike Gooding \etal,\cite{gooding_birgeneau}
we found the Grempel steps lowered the energy minimally and
made no difference to the static neutron structure factor.
It is likely that this is due to our linear chiral coupling in
equation \ref{eqn:jch}.  A squared chiral coupling, used by
Gooding \etal,
makes the minimization more difficult and computationally
expensive.  Thus, we were able to minimize larger lattices and
more ensembles to obtain smaller error bars on the results.
Our calculations constitute a different physical model than
Gooding \etal\cite{gooding_birgeneau} despite the similar
computational methods used.

Finally, we found that including the polaron-polaron spin
and chiral couplings shown in figure \ref{fig:jpp} does not
alter the results.  The dominant couplings in terms of
the minimized spin structure are the $d^9$ spin coupling
$J_{dd}$ and the chiral coupling $J_{ch}$.  The results shown
below exclude any chiral couplings involving more than one
polaron.

\begin{figure}[tbp]
\centering  \includegraphics[width=\linewidth]{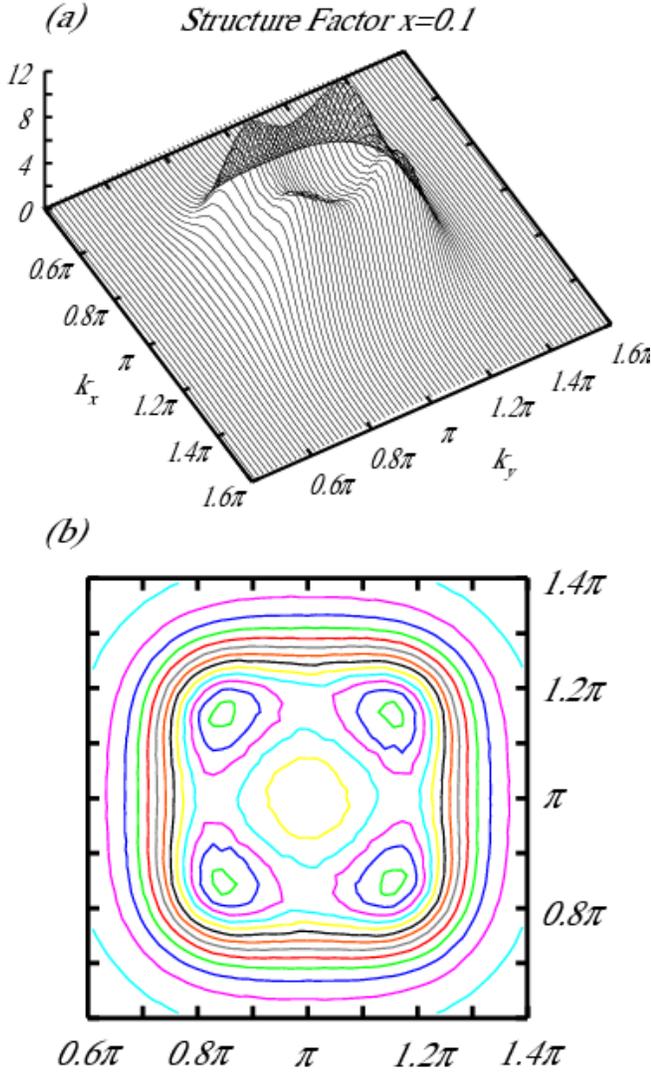}
\caption{Static spin structure factor for $d^9$ spins at doping
$x=0.10$.  The structure factor is incommensurate with a ring
of radius $k\approx(2\pi/a)x$ around $(\pi/a,\pi/a)$.  The total
sum over the Brillouin zone satisfies the normalization,
$N^{-1}\sum_k S(k)=1$ where $N$ is the total number of cells.
(a) 3D plot of structure factor. (b) contour plot centered
at $(\pi/a,\pi/a)$.
}
\label{fig:3d}
\end{figure}

\begin{figure}[tbp]
\centering  \includegraphics[width=0.8\linewidth]{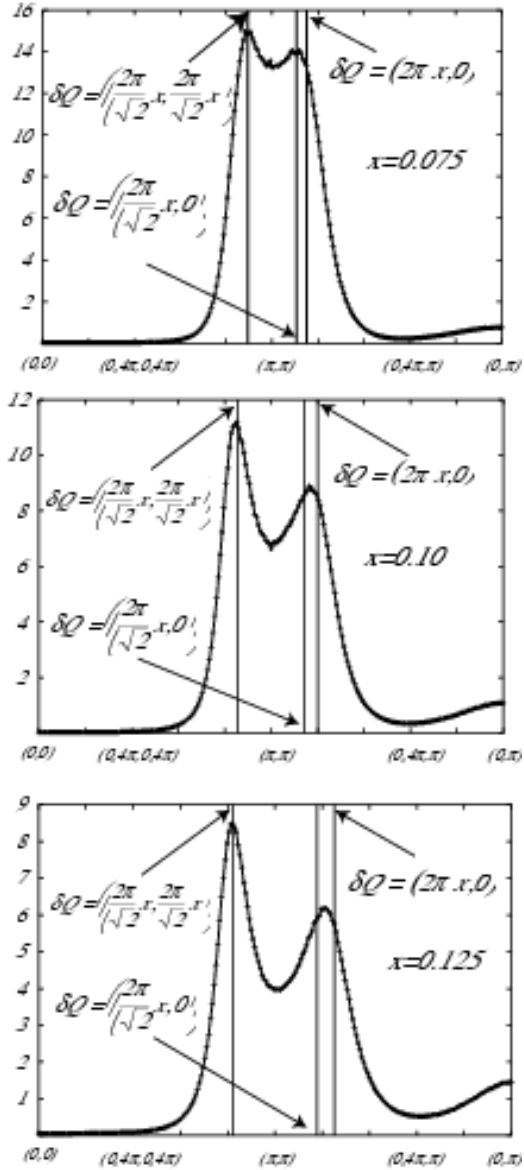}
\caption{Static spin structure factor for $d^9$ spins at dopings
$x=0.075$, $0.10$, and $0.125$ in \lsco.  Each figure plots
$S(k)$ starting from $k=(0,0)$ to
$(\pi/a,\pi/a)$ and then to $(0,\pi/a)$.
The structure factor is incommensurate with a ring
of radius $k\approx(2\pi/a)x$ around $(\pi/a,\pi/a)$. 
The vertical lines are drawn to highlight specific incommensurate
vectors.
If the structure factor was purely derived from the sum of four
Lorenztians along the diagonals a distance $(2\pi/a)x$ from
$(\pi/a,\pi/a)$, then the peak along the Cu$-$O bond direction
would be slightly less than the shift shown at
$\delta Q=(2\pi/\sqrt{2}a,0)$.
The normalization
is the same as figure \ref{fig:3d}.
}
\label{fig:struc}
\end{figure}

\begin{figure*}[tbp]
\centering  \includegraphics[width=0.70\linewidth]{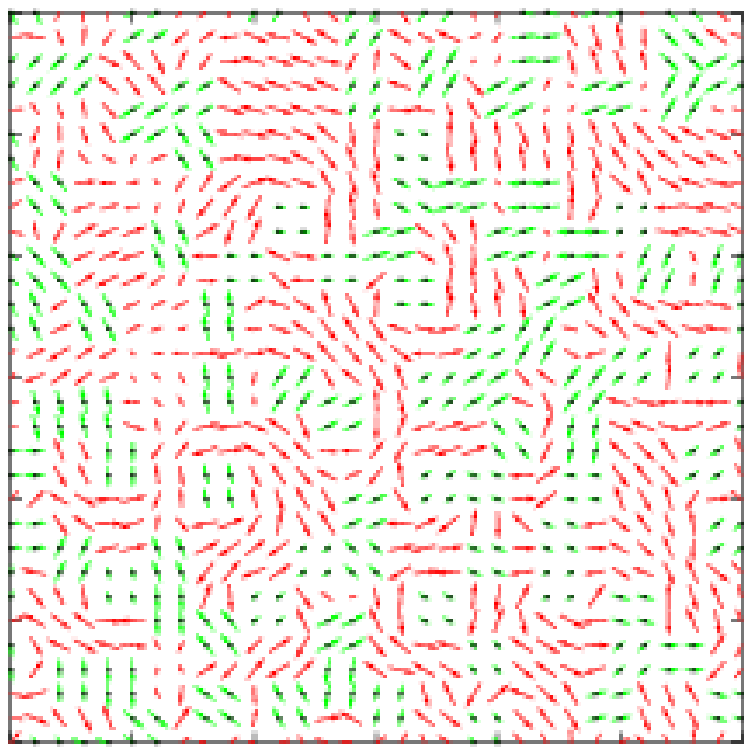}
\caption{Projection of spins onto xy plane for an $x=0.10$ minimized
structure.  The $d^9$ spin staggered magnetization is along the
z-axis out of the plane of the paper and the magnetization of the
$d^9$ spins is along the x-axis.  Only a $30\times 30$ subset
of the $256\times 256$ lattice is shown.  The undoped $d^9$ spins
are shown in red and the polaron spins are green.  The four Cu sites
of each polaron are indicated by black dots.}
\label{fig:spins}
\end{figure*}

Figures \ref{fig:3d} and \ref{fig:struc} show our results for
the spin correlation
at dopings $x=0.075$, $0.10$, and $0.125$ on a $256\times 256$
lattice averaged over an ensemble of $5,000$ configurations for each
doping.  The structure factor is dimensionless and
is normalized such that its integral over the Brillouin zone is
one, $N^{-1}\sum_k S(k)=1$, where $N$ is the total number of
cells.  Figure \ref{fig:spins} shows part of a minimized spin
structure at $x=0.10$.

Due to the large number of ensembles, the
error bars for the plotted values are small.
For $x=0.075$, they are
$\pm 0.12$, $\pm 0.15$, and $\pm 0.12$ for the diagonal peak,
$(\pi,\pi)$, and the Cu$-$O bond peak respectively. 
For $x=0.10$, the error is
$\pm 0.09$, $\pm 0.08$, and $\pm 0.07$.  For $x=0.125$,
the error is $\pm 0.07$, $\pm 0.05$, and $\pm 0.05$.
The error decreases sharply and becomes negligible on the
scale of the figure as $k$ moves past the peaks and farther
from $(\pi,\pi)$.

From figure \ref{fig:struc}, the diagonal incommensurate
peak is shifted from $(\pi/a,\pi/a)$ by
$(2\pi/a)(x/\sqrt{2})(1,1)$ and is of length $(2\pi/a)x$.
The peak along the Cu$-$O bond direction is shifted in the range
$(2\pi/a)x(1/\sqrt{2},0)$ to $(2\pi/a)x(1,0)$.
The Cu$-$O bond direction shift is experimentally
observed for the metallic range $x>0.05$
\cite{incommen_metal3, birgeneau_rmp}
in \lsco\ and the diagonal shift is seen in the spin-glass regime
$0.02<x<0.05$.\cite{incommensurate_sg}

Since the difference
between $(2\pi/a)x(1/\sqrt{2},0)$ and $(2\pi/a)x(1,0)$ 
is small, it is difficult to resolve the precise peak doping value
within our finite size computations.  If the structure factor
derives from broadened Lorenztians centered at the four diagonal
points around $(\pi/a,\pi/a)$, then a peak in the Cu$-$O
bond direction would be expected at
$(2\pi/a)x(1/\sqrt{2},0)$ from the two closest peaks.
The contributions from the remaining two peaks shift
the peak closer to $(\pi/a,\pi/a)$ rather than away from it
as is seen in figure \ref{fig:struc}.  Thus, the best we can
currently say with the calculations is there is a ring of
incommensurate peaks approximately a
distance $k=(2\pi/a)x$ from $(\pi/a,\pi/a)$.

From the widths of the peaks, the correlation length is
approximately the mean separation between the polarons,
$a/\sqrt{x}$.

We present
a heuristic derivation for why the spin structure
factor is incommensurate with a shift from $(\pi/a,\pi/a)$ of
magnitude $(2\pi/a)x$.
Similar to previous work,\cite{gooding_birgeneau} our calculations
find that the minimum spin
configuration consists of undoped patches of $d^9$ spins
aligned antiferromagnetically with the polarons acting to rotate
the direction of the antiferromagnetic alignment of adjacent
patches.  This is seen in figure \ref{fig:spins}.

Consider an area $A$.  The number of polarons in this area
is $N_p=Ax/a^2$.
When the chiral coupling dominates,
the effect of a single polaron on the neighboring $d^9$
spins is shown in figure \ref{fig:4x4}.  The polaron
rotates each adjacent spin on opposite sides of the polaron by
$\pi$, or $2\pi$ total.  If this net $2\pi$ rotation is
rigidly transmitted to an antiferromagnetic patch, then
the polaron rotates a patch by an angle,
$\delta\theta=2\pi/(1/x)=2\pi x$.  Thus, the net rotation per
spin is $N_p\delta\theta$.  If the area $A$ is chosen such that
the net rotation per spin is $2\pi$, or $N_p\delta\theta=2\pi$,
then $A=(a/x)^2$.  A translation by $L=\sqrt{A}=a/x$ returns
to an identical antiferromagnetic patch.  This leads to a shift
of the spin correlation peak from $(\pi/a,\pi/a)$ to
$\delta Q=2\pi/L$ or $\delta Q=(2\pi/a)x$.

\subsection{Kinetic Energy of the \xxyy\ Band}
\label{subsec:ke}
The energy contribution from the delocalized \xxyy\ band electrons
has not been included in the minimization of equation
\ref{eqn:htot}.  A complete minimization would also compute the change
in the band energy to determine the direction to align a
given spin during our sweep through the lattice spins.
This effect is included in mean-field below.

A $d^9$ spin ordering of momentum $\mathbf{q}$ hybridizes
band electrons of momentum $\mathbf{k}$ and $\mathbf{k+q}$
with a coupling energy $V$ on the order of $J_{dd}\sim 0.1$ eV.
This mixing of $\mathbf{k}$ and
$\mathbf{k+q}$ perturbs the band energies and the total
ground state energy.  We calculate the band energy change in
\lsco\ for a ring of $\mathbf{q}$ vectors at the computed
incommensurate length $q=(2\pi/a)x$.
The $\mathbf{q}$ vector producing the lowest band energy is the
observed neutron incommensurability.

The idea that the band kinetic energy change in the $d^9$
spin background determines the final neutron incommensurability
has been suggested by Sushkov \etal\cite{kotov1, kotov2, kotov3}
for the $t-t'-t''-J$ model.  In their model,
the magnitude of the incommensurability arises from the doping $x$,
the antiferromagnetic $d^9$ spin stiffness $\rho_s$,
the hopping matrix element $t$,
and the quasiparticle renormalization $Z$
where $q=(Zt/\rho_s)(2\pi/a)x$.
Their self-consistent Born approximation calculation of
the quasiparticle dispersion in a spin-wave theory background
finds values for the parameters such that $Zt/\rho_s\approx 1$.  
The magnitude of the incommensurability is
less dependent on the detailed parameters for our Hamiltonian
in equation \ref{eqn:htot}.

For a given spin incommensurability vector $\mathbf{q}$ and
$d^9$ to \xxyy\ band coupling, $V\sim 0.1$ eV, the Green's
function satisfies,

\begin{eqnarray}
\label{eqn:g1}
G^{-1}(\mathbf{k},\omega) & = &
G_0^{-1}(\mathbf{k},\omega) - V^2 G_0(\mathbf{k+q},\omega)
\nonumber \\
 & - & V^2 G_0(\mathbf{k-q},\omega).
\end{eqnarray}
The vector $\mathbf{-q}$ must be included with
$\mathbf{q}$ because the coupling Hamiltonian is Hermitean.
Solving for $G(\mathbf{k},\omega)$,
\begin{eqnarray}
\label{eqn:g2}
G(\mathbf{k},\omega) & = & G_0(\mathbf{k},\omega)
\left\{
1-V^2 G_0(\mathbf{k},\omega)\right. \nonumber \\
 & \times & \left.
\left[G_0(\mathbf{k+q},\omega)+G_0(\mathbf{k-q},\omega)\right]
\right\}^{-1}.
\end{eqnarray}
Expanding to order $V^2$,
\begin{eqnarray}
\label{eqn:g3}
G(\mathbf{k},\omega) & = & G_0(\mathbf{k},\omega)
+V^2 G_0(\mathbf{k},\omega)\left\{G_0(\mathbf{k+q},\omega)\right.
\nonumber \\
 & + & \left.G_0(\mathbf{k-q},\omega)\right\}
G_0(\mathbf{k},\omega) \nonumber \\
 & + & O(V^4).
\end{eqnarray}
The number of electrons in the $\mathbf{k}$ state up to energy
$\epsilon$ is,
\begin{equation}
\label{eqn:g4}
n(\mathbf{k},\epsilon)=\int_{-\infty}^{\epsilon}
d\omega\left[-\frac{1}{\pi}\mathrm{Im}\ G(\mathbf{k},\omega)\right],
\end{equation}
\begin{eqnarray}
\label{eqn:g5}
n(\mathbf{k},\epsilon) & = & n_0(\mathbf{k},\epsilon)+
V^2\int_{-\infty}^{\epsilon}d\omega
\left(-\frac{1}{\pi}\right) \nonumber \\
 & \times & \mathrm{Im}
\left\{
G_0(\mathbf{k},\omega)\left[
G_0(\mathbf{k+q},\omega)\right.\right. \nonumber \\
 & + & 
\left.\left.
G_0(\mathbf{k-q},\omega)\right]G_0(\mathbf{k},\omega)\right\},
\end{eqnarray}
where $n_0(\mathbf{k},\epsilon)$ is the unperturbed occupation.
The total density of states per spin at energy $\epsilon$ is,
\begin{equation}
\label{eqn:g5a}
N(\epsilon)=\sum_{\mathbf{k}}
\left(-\frac{1}{\pi}\right)\mathrm{Im}\ G(\mathbf{k},\epsilon)
=\frac{\partial}{\partial\epsilon}
\sum_{\mathbf{k}}n(\mathbf{k},\epsilon),
\end{equation}

A percolating band has Green's function,
\begin{equation}
\label{eqn:g6}
G_0(\mathbf{k},\omega)=\frac{1}
{\omega -\epsilon_k + i\Gamma_k},\ \Gamma_k>0,
\end{equation}
where $\Gamma_k$ is the linewidth.
Using equations \ref{eqn:g4} and \ref{eqn:g6},
\begin{equation}
\label{eqn:g7}
n_0(\mathbf{k},\omega)=\frac{1}{2}+\frac{1}{\pi}
\tan^{-1}\left(\frac{\epsilon_f-\epsilon_k}{\Gamma_k}\right).
\end{equation}

The perturbation shifts the Fermi level to
$\epsilon_f + \delta\epsilon_f$.  The total number of electrons is
conserved, leading to the equation for $\delta\epsilon_f$,
\begin{equation}
\label{eqn:g8}
n_{tot}=
\sum_{\mathbf{k}} n(\mathbf{k},\epsilon_f + \delta\epsilon_f)=
\sum_{\mathbf{k}} n_0(\mathbf{k},\epsilon_f).
\end{equation}
The total energy of the band electrons per spin is given by,
\begin{equation}
\label{eqn:g9}
E_{tot}(\mathbf{q},V)=\int_{-\infty}^{\epsilon_f+\delta\epsilon_f}
d\omega\ \omega\sum_{\mathbf{k}}n(\mathbf{k},\omega).
\end{equation}
In appendix \ref{ap:neut},
the expressions for $\delta\epsilon_f$ and $E_{tot}(\mathbf{q},V)$
to order $V^2$ are derived,
\begin{eqnarray}
\label{eqn:g10}
\delta\epsilon_f
N_0(\epsilon_f) + V^2
\int_{-\infty}^{\epsilon_f}d\omega\sum_{\mathbf{k}}
\left[
R(\mathbf{k},\mathbf{k+q},\omega)\right. \nonumber \\
+ \left.R(\mathbf{k},\mathbf{k-q},\omega)
\right] = 0,
\end{eqnarray}
\begin{eqnarray}
\label{eqn:g11}
E_{tot}(\mathbf{q},V)=E_G + V^2
\int_{-\infty}^{\epsilon_f}d\omega
(\omega-\epsilon_f)
\nonumber \\
\times\ \left[
\sum_{\mathbf{k}}
R(\mathbf{k},\mathbf{k+q},\omega) +
R(\mathbf{k},\mathbf{k-q},\omega)
\right],
\end{eqnarray}
where $N_0(\epsilon_f)$ is the unperturbed density of states per
spin, $E_G$ is the unperturbed $(V=0)$ energy, and
\begin{eqnarray}
\label{eqn:g12}
R(\mathbf{k},\mathbf{p},\omega) & = &
\left(-\frac{1}{\pi}\right)\mathrm{Im}
\left[
\frac{1}{\left(\omega-\epsilon_k+i\Gamma_k\right)^2}
\right. \nonumber \\
 & \times & 
\left.
\frac{1}{\left(\omega-\epsilon_p+i\Gamma_p\right)}
\right].
\end{eqnarray}

The integral
$\int_{-\infty}^{\epsilon_f}R(\mathbf{k},\mathbf{p},\omega)d\omega$
can be evaluated analytically,
thereby allowing us to accurately compute
the small energy change of the Fermi energy $\delta\epsilon_f$
and $E_{tot}(\mathbf{q},V)$.  This is done in appendix
\ref{ap:neut}.

The band energy is given by,
\begin{equation}
\label{eqn:lsco}
\begin{array}{ccc}
\epsilon_k & = & -2t_1(\cos k_x + \cos k_y)
-4t_{11}\cos k_x\cos k_y \\
 &  & -2t_2(\cos 2k_x + \cos 2k_y) + \epsilon_{x^2-y^2}. \\
\end{array}
\end{equation}
We use the band structure parameters\cite{freeman1, freeman2, kontani1}
for \lsco\ given by $t_1=0.25$ eV, $t_{11}=-0.025$ eV,
and $t_2=0.025$ eV.  $t_1$ is the nearest neighbor hopping,
$t_{11}$ is the next-nearest neighbor diagonal term,
and $t_2$ is the hopping along the Cu$-$O bond direction from
two lattices site away.
For $x=0.10$, $\epsilon_{x^2-y^2}=0.133$ eV leads to $\epsilon_f=0$.

We calculated the Fermi energy shift and energy for incommensurability
along the diagonal and Cu$-$O bond direction of magnitude
$q=(2\pi/a)x$ at $x=0.10$ with $V=0.1$ eV.  The electron
linewidth, $\Gamma_k$,
is chosen to be the sum of an s-wave and d-wave term,
\begin{equation}
\label{eqn:gam}
\Gamma_k=\Gamma_S + \Gamma_D(\cos k_x - \cos k_y)^2,
\end{equation}
where $\Gamma_S=0.01$ eV and $\Gamma_D=0.01$ eV.

The addition of a d-wave term to the linewidth arises from the
$\mathbf{k}$ dependence of the Coulomb scattering rate with
polarons discussed in section \ref{resist}.

The energy changes and Fermi level shifts in eV are,
\begin{equation}
\label{eqn:eneut}
\begin{array}{ccc}
\delta E\left(\frac{2\pi x}{\sqrt{2}a},
\frac{2\pi x}{\sqrt{2}a}\right) & = & -0.01456,
\ \ \delta\epsilon_f=0.0046 \\
\delta E\left(\frac{2\pi x}{a},0\right) & = & -0.01478,
\ \ \delta\epsilon_f=-0.0056 \\
\end{array}
\end{equation}
The band energy is lower for incommensurability along
the Cu$-$O bond direction.

The Cu$-$O bond direction
incommensurability is lower in energy due to the additional
Umklapp scattering available for $\mathbf{q}$ on the Brillouin
zone edge rather than inside the zone for diagonal $\mathbf{q}$.

We have shown that chiral coupling of polarons
to $d^9$ spins leads to a ring of incommensurability centered at
$(\pi/a,\pi/a)$ of magnitude $(2\pi/a)x$.  The
perturbation to the kinetic energy of the delocalized \xxyy\ 
band electrons selects incommensurability along the
Cu$-$O bond direction due to Umklapp scattering on the
Brillouin zone edge.

For $0.02<x<0.05$,\cite{birgeneau_rmp,incommensurate_sg}
\lsco\ is a spin-glass.  No \xxyy\ band is formed because
the plaquette polarons do not percolate.  The \xxyy\ states
triplet couple to polaron spins.  The spin interactions
in the spin-glass phase are different from the
Hamiltonian in equation \ref{eqn:htot}.

The Cu \xxyy\ cannot delocalize over
an infinite polaron swath in our model.  We do not know if the
\xxyy\ states remain localized on a single Cu or delocalize
over the finite swath of the polaron.  Any delocalization leads
to an effective ferromagnetic coupling between neighboring polaron
spins due to the triplet coupling with the \xxyy\ spin.
In addition, there is an asymmetry in the chiral coupling due to
orthorhombic crystal symmetry arising from the tilt of the
CuO$_6$ octahedra.  The one-dimensional incommensurability
in the spin-glass phase of \lsco\cite{incommensurate_sg} is
not explained in this paper.

\section{Superconducting Pairing}
\label{sec:pairing}
Coulomb scattering of \xxyy\ band electrons with chiral
plaquette polarons
leads to anisotropic Cooper pair repulsion.
The maximum energy difference between a
chiral polaron and its time-reversed partner is
analogous to the Debye energy in BCS superconductors,
As discussed in
section \ref{subsec:incomm}, this energy difference can be
non-zero and on the order of $J_{dd}$.

\begin{figure}[tbp]
\centering  \includegraphics[width=0.8\linewidth]{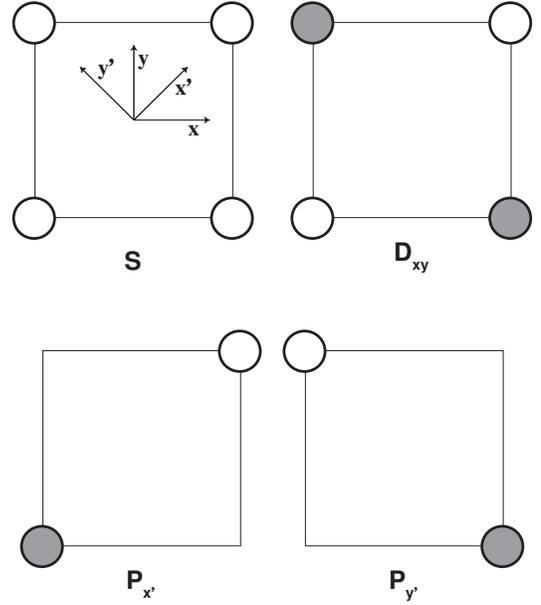}
\caption{Effective polaron orbitals with \psigma\ absorbed
into A$_{1g}$ on the Cu sites.
}
\label{fig:p1}
\end{figure}

We simplify the polaron wavefunctions by absorbing the
\psigma\ orbitals into the A$_{1g}$ orbitals on the Cu sites
for the pairing and transport calculations.  This is
shown in figure \ref{fig:p1}.

The direct and exchange Coulomb terms coupling the \xxyy\ band
with a polaron is,
\begin{equation}
\label{eqn:hc}
H_U = U\sum_{\mathbf{R},\sigma,\sigma'}
d^{\dagger}_{R\sigma'}z^{\dagger}_{R\sigma}
z^{}_{R\sigma}d^{}_{R\sigma'},
\end{equation}
\begin{equation}
\label{eqn:hk}
H_K= -K\sum_{\mathbf{R},\sigma,\sigma'}
d^{\dagger}_{R\sigma'}z^{\dagger}_{R\sigma}
z^{}_{R\sigma'}d^{}_{R\sigma},
\end{equation}
where $U>K>0$.
$d^{\dagger}_{\mathbf{R}}$ creates an \xxyy\ electron at 
$\mathbf{R}$ and $z^{\dagger}_{\mathbf{R}}$ creates an
A$_{1g}$ electron at $\mathbf{R}$.

The \xxyy\ band state with momentum $\mathbf{k}$ is,
\begin{equation}
\label{eqn:k}
d^{\dagger}_{\mathbf{k}\sigma}=N^{-\frac{1}{2}}
\sum_{\mathbf{R}}e^{i\mathbf{k}\cdot\mathbf{R}}
d^{\dagger}_{\mathbf{R}\sigma},
\end{equation}
where $N$ is the total number of Cu sites.
A polaron state of spin $s$ is given by,
\begin{equation}
\label{eqn:z}
z^{\dagger}_s=\sum_{\mathbf{R}}\alpha^{}_{\mathbf{R}}
z^{\dagger}_{\mathbf{R}s},
\end{equation}
where the coefficients $\alpha^{}_R$ determine the type of polaron
in figure \ref{fig:p1}.
The matrix elements for direct and exchange Coulomb scattering
of an \xxyy\ band electron with a polaron electron are,
\begin{equation}
\label{eqn:mu}
\left<k'\sigma',z^{}_{s'}|H_U|k\sigma,z^{}_s\right> =
\left(\frac{U}{N}\right)M(\mathbf{k'-k})
\delta_{\sigma\sigma'}\delta_{ss'}, 
\end{equation}
\begin{equation}
\label{eqn:mk}
\left<k'\sigma',z^{}_{s'}|H_K|k\sigma,z^{}_s\right> =
\left(-\frac{K}{N}\right)M(\mathbf{k'-k})
\delta_{\sigma s'}\delta_{\sigma' s},
\end{equation}
\begin{equation}
\label{eqn:m}
M(\mathbf{q})=
\sum_{\mathbf{R}}
\alpha'^{\ast}_{\mathbf{R}}\alpha^{}_{\mathbf{R}}
e^{-i\mathbf{qR}}.
\end{equation}

\begin{figure}[tbp]
\centering  \includegraphics[width=\linewidth]{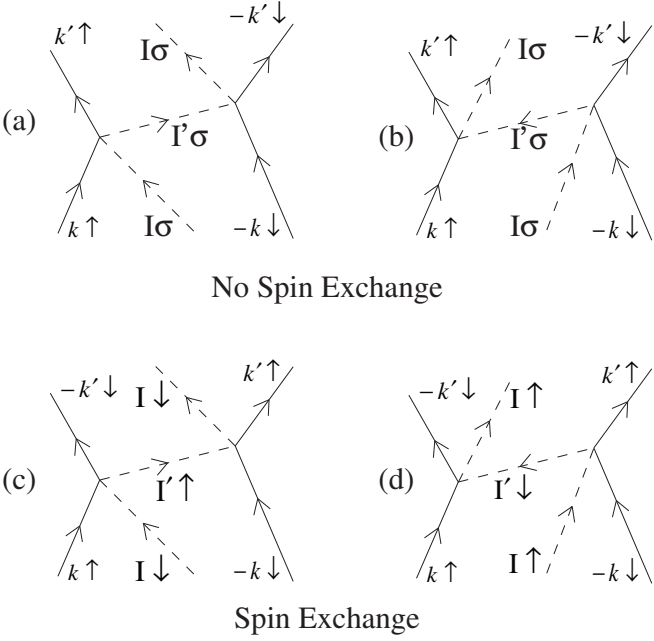}
\caption{Second-order Cooper pairing
processes with polarons.  The Coulomb coupling is represented
by a four-point vertex for simplicity.  The solid lines are
the band electrons and the dashed lines are the polaron states.
$I$ is the initial and final polaron
orbital state. $I'$ is the intermediate state.  (a) and (b) are the
Coulomb couplings with no spin exchange.
(c) and (d) are spin exchange couplings.
In (c) and (d), the final electrons are interchanged compared to
(a) and (b) yielding an extra minus sign in the pairing matrix
element.
}
\label{fig:p2}
\end{figure}

The Cooper pairing matrix element for scattering
arising from figure \ref{fig:p2}(a)
with initial polaron orbital state $I\up$ is,
\begin{eqnarray}
\label{eqn:hpu}
\left\langle
(\mathbf{k'}\up,\mathbf{-k'}\down),I\up\left|H'\right|
(\mathbf{k}\up,\mathbf{-k}\down),I\up
\right\rangle= \nonumber \\
\frac{U(U-K)}{N^2} |M(\mathbf{q})|^2
\sum_{I'}\frac{1}{2}
\left[
\frac{1}{E_f-E_n}+\frac{1}{E_i-E_n}
\right],
\end{eqnarray}
where $\mathbf{q=k'-k}$,
$E_i$ is the total energy of the initial state, $E_f$ is
the final state energy, and $E_n$ is the intermediate state energy.
The factor $U-K$ comes from the left vertex where both direct
and exchange Coulomb matrix elements appear while the factor $U$
arises from the right vertex where there is no exchange.
$E_i$, $E_n$, and $E_f$ are,
\begin{eqnarray}
\label{eqn:eng1}
E_i & = &  2\epsilon_k + E_{I\up}, \\
\label{eqn:eng2}
E_n & = &  \epsilon_k' + \epsilon_k + E_{I'\up}, \\
\label{eqn:eng3}
E_f & = & 2\epsilon_{k'} + E_{I\up}.
\end{eqnarray}
Substituting equations \ref{eqn:eng1}, \ref{eqn:eng2}, and
\ref{eqn:eng3}
into \ref{eqn:hpu} leads to,
\begin{eqnarray}
\label{eqn:hpu1}
\left\langle
(\mathbf{k'}\up,\mathbf{-k'}\down),I\up\left|H'\right|
(\mathbf{k}\up,\mathbf{-k}\down),I\up
\right\rangle= \nonumber \\
\frac{U(U-K)}{N^2} |M(\mathbf{q})|^2
\sum_{I'}
\frac{\left(E_{I'\up}-E_{I\up}\right)}
{\left(\epsilon_{k'}-\epsilon_k\right)^2-
\left(E_{I'\up}-E_{I\up}\right)^2}.
\end{eqnarray}

The pairing matrix element is the probability weighted
average of equation \ref{eqn:hpu1} over
all initial polaron states $I$.  States with
$E_{I'\sigma}-E_{I\sigma}>0$
dominate the average because the polaron is primarily in
its lowest energy state.
Since $U>0$ and $U-K>0$, there is an attractive pairing
with momentum anisotropy determined by $|M(\mathbf{q})|^2$ for pairs
close to the Fermi level.  The coupling is attractive for Cooper
pairs near the Fermi level within the polaron energy separation,
$|\epsilon_k|$ and $|\epsilon_{k'}| <
\left(E_{I'\sigma}-E_{I\sigma}\right)$.  There are
similar terms for $I\down$ and figure \ref{fig:p2}(b).

For the spin exchange processes shown in figures \ref{fig:p2}(c)
and (d), the matrix element picks up an extra minus sign
compared to figures \ref{fig:p2}(a) and (b) due
to the exchange of $\mathbf{k'}\up$ and $\mathbf{-k'}\down$
in the final state,
\begin{eqnarray}
\label{eqn:hpk1}
\left\langle
(\mathbf{k'}\up,\mathbf{-k'}\down),I\sigma\left|H''\right|
(\mathbf{k}\up,\mathbf{-k}\down),I\sigma
\right\rangle= \nonumber \\
(-)\left(\frac{-K}{N}\right)^2 |M(\mathbf{k'+k})|^2
\sum_{I'}
\frac{\Delta}
{\left(\epsilon_{k'}-\epsilon_k\right)^2-
\Delta^2},
\end{eqnarray}
where $\Delta=E_{I'\up}-E_{I\down}$.  This leads to an overall
repulsive interaction for pairs near the Fermi level.

The direct on-site isotropic Coulomb
repulsion between Cooper pairs must be included in the
net pairing interaction in addition to the second-order pairings
in figure \ref{fig:p2}.

For hole-doped cuprates, one chiral polaron state is
singly occupied and the remaining polaron states shown in figure
\ref{fig:polaron_states} are doubly occupied.
The intermediate polaron state $I'$ must
be a chiral state, $P_{x'}\pm iP_{y'}$.  For electron-doped
cuprates, the initial state $I$ is a chiral polaron.
The pairing is identical
because the matrix element between $I$ and $I'$ is
mod-squared.  We focus on the hole-doped derivation here.

The momentum dependence of the pairing is determined by
$|M(\mathbf{q})|^2$ where $\mathbf{q=k'-k}$ for non-spin
exchange attraction and $\mathbf{q=k'+k}$ for
spin exchange repulsion.
$|M(\mathbf{q})|^2$ is the same for $I=S$ and $D_{xy}$.  From
figure \ref{fig:p1} and equation \ref{eqn:m},
\begin{equation}
\label{eqn:m1}
|M(\mathbf{q})|^2_{S,D}\sim
\sin^2\frac{1}{2}\left(q_x+q_y\right)+
\sin^2\frac{1}{2}\left(-q_x+q_y\right),
\end{equation}
where we have dropped all constants and wavefunction
normalizations in the expression because they do not affect the
momentum dependence.  For $I=P_{x'}\pm iP_{y'}$ and
$I'=P_{x'}\mp iP_{y'}$,
\begin{equation}
\label{eqn:m2}
|M(\mathbf{q})|^2_{+-}\sim
\left[\cos\frac{1}{2}\left(q_x+q_y\right)-
\cos\frac{1}{2}\left(-q_x+q_y\right)\right]^2.
\end{equation}
For spin exchange repulsion, there is an additional possibility
where $I'=I=P_{x'}\pm iP_{y'}$.  $|M(\mathbf{q})|^2_{++}$ is,
\begin{equation}
\label{eqn:m3}
|M(\mathbf{q})|^2_{++}\sim
\left[\cos\frac{1}{2}\left(q_x+q_y\right)+
\cos\frac{1}{2}\left(-q_x+q_y\right)\right]^2.
\end{equation}
Table \ref{table:m} shows the value of $|M(\mathbf{q})|^2$ for
various vectors $\mathbf{q}$.
There are several pairing possibilities.

First, consider non-spin exchange attractive pairing through
figures \ref{fig:p2}(a) and (b) added to direct isotropic
repulsion.
In this case, $\mathbf{q=k'-k}$.  $|M(\mathbf{q})|^2_{S,D}$
leads to a net attraction for $\mathbf{q}=(\pi,0)$.
An s-wave gap cannot occur because the isotropic repulsion dominates.
A gap of any other symmetry has a net repulsion
for $\mathbf{q}=(0,0)$ and cannot lead to superconductivity.
$|M(\mathbf{q})|^2_{+-}$ also repels at
$\mathbf{q}=(0,0)$ and does not superconduct.
$M_{++}$ does not occur for non-spin exchange.  Thus, there
is no superconductivity through non-spin exchange polaron
coupling.

\begin{table}[tbp]
\caption{\label{table:m}The values of $|M(q)|^2$ at special vectors
in the Brillouin zone.  For non-spin exchange couplings shown in
figure \ref{fig:p2}(a) and (b), the last column does not apply
because the intermediate cannot be identical to the initial
polaron state, $I'\ne I$. $M(q)$ is defined in equations
\ref{eqn:m1}, \ref{eqn:m2}, and \ref{eqn:m3}.
$\mathbf{q=k'\mp k}$ for non-spin exchange and spin exchange
respectively.
}
\begin{ruledtabular}
\begin{tabular}{cccc}
 $(q_x,q_y)$ & $|M(q)|^2_{S,D}$ & $|M(q)|^2_{+-}$ & $|M(q)|^2_{++}$ \\
\hline
 $(0,0)$              & 0 & 0 & 4 \\
 $(\pi,0)$, $(0,\pi)$ & 2 & 0 & 0 \\
 $(\pi,\pi)$          & 0 & 4 & 0 
\end{tabular}
\end{ruledtabular}
\end{table}

For spin exchange repulsion in figures
\ref{fig:p2}(c) and (d) added to direct isotropic repulsion,
$M_{S,D}$ has the largest repulsion for $\mathbf{q=k'+k}=(\pi,0)$,
$M_{+-}$ for $\mathbf{q}=(\pi,\pi)$, and $M_{++}$
for $\mathbf{q}=(0,0)$.  $M_{++}$ cannot be correct for the cuprates
because $\mathbf{q}=0$ when
$\mathbf{k'=-k}$.  This leads to a gap that changes
sign for $\mathbf{k}\rightarrow\mathbf{-k}$ and triplet spin pairing.
For $M_{S,D}$, there is no favorable Fermi surface nesting near
$\mathbf{k'+k}=(\pi,0)$.  

$M_{+-}$ has the time-reversed chiral polaron as the intermediate
state and has a large repulsion for $\mathbf{k'+k}=(\pi,\pi)$.
For $\mathbf{k}\approx(\pi,0)$ and $\mathbf{k'}\approx(0,\pi)$,
there is a large repulsion between pairs close to
the Fermi surface.  Pairing through spin exchange coupling
with time-reversed polarons leads to d-wave superconductivity.

As discussed in section \ref{subsec:incomm} on the
neutron spin incommensurability, the energy
splitting between time-reversed chiral polarons is on the 
scale of $J_{dd}\sim 0.1$ eV.  It is this scale that
is analogous to the phonon Debye energy in conventional
superconductors.

The energy separation between two
time-reversed chiral polarons, $|P_{x'}+iP_{y'},\sigma\rangle$ and
$|P_{x'}-iP_{y'},-\sigma\rangle$, is largest for low dopings
where there are more undoped $d^9$ spins.
The energy splitting decreases with increasing doping
leading to a reduction in the energy range surrounding the
Fermi surface involved in Cooper pairing.
Concurrently, the number of polarons that can induce
Cooper pairing increases. 
The number of band \xxyy\ electrons available to
form the superconducting ground state also increases with increasing
doping.  Finally,
as the number of polarons increases, they crowd into
each other, eventually creating polarons localized on
single Cu sites, as found in UB3LYP calculations\cite{ub3lyp_dope}.
The single Cu state is not chiral and leads to isotropic Cooper
pair repulsion.
The competition between all of these factors leads
to the superconducting phase diagram in cuprates.

\section{Normal State Transport}
\label{sec:transport}
\subsection{Resistivity}
\label{resist}
The temperature dependence of the resistivity arising from
\xxyy\ band electrons Coulomb scattering from polarons is shown to
depend on the density of polaron energy separations, $\rho(\Delta)$.
If the density is constant, then the resistivity is linear in $T$.

The scattering rate of an \xxyy\ electron initially in the
state $\mathbf{k}\sigma$ to Coulomb scatter from a polaron
to another band electron and polaron is the sum,
\begin{eqnarray}
\label{eqn:tau1}
\frac{1}{\tau(\epsilon_{k\sigma})} & = & \frac{2\pi}{\hbar}
\sum_{\stackrel{\mathbf{k'}\sigma'}{Is,I's'}}
|\langle\mathbf{k'\sigma'},I's'|H'|
\mathbf{k\sigma},Is\rangle|^2 P(Is) \nonumber \\
 & & \times\ \delta(\epsilon_{\mathbf{k'\sigma'}}+E_{I's'}-
\epsilon_{\mathbf{k\sigma}}-E_{Is}),
\end{eqnarray}
where $H'=H_U+H_K$ is the total Coulomb interaction of band
electrons with polarons.  $H_U$ and $H_K$ are defined 
in equations \ref{eqn:hc} and \ref{eqn:hk}.
$E_{Is}$ and $E_{I's'}$ are
the initial and final polaron energies respectively.
$P(Is)$ is the probability the polaron is initially in the
state $Is$.  The delta function enforces total energy conservation,
\begin{equation}
\label{eqn:e1}
\epsilon_{\mathbf{k\sigma}}+E_{Is}=
\epsilon_{\mathbf{k'\sigma'}}+E_{I's'}.
\end{equation}

The matrix element in equation \ref{eqn:tau1} is proportional
to $|M(\mathbf{k'-k})|^2$ defined in equations \ref{eqn:m1},
\ref{eqn:m2}, and \ref{eqn:m3}.
To obtain the qualitative temperature dependence,
we simplify equation \ref{eqn:tau1} by assuming the
matrix element is constant for all momenta and
there are only two polaron states at any doped center.
This does not alter the temperature
dependence, but smears out the anisotropy of the scattering
rate.  The $M(\mathbf{k'-k})$
have different momentum dependencies for each polaron
symmetry.  Since the scattering rate
is a sum over all scattering channels, the approximation
of isotropic scattering is likely to be good.

There is a polaron Coulomb scattering process we have neglected.
This is second-order hopping of a $\mathbf{k}$ \xxyy\ electron into
the unoccupied polaron state that then hops off into
a $\mathbf{k'}$ \xxyy\ band state.  Since the polaron is comprised
of orbitals with A$_1$ symmetry, the hopping is zero for
momenta along the zone diagonal and largest
at $(\pm\pi,0)$ and $(0,\pm\pi)$.
This term is approximately temperature independent
and of ``d-wave" symmetry.  The temperature dependence of this term
arises from the displacement of the apical O due to
the addition of an electron into the polaron.
Since the initial and final states in this second-order process have
one hole on the polaron, the temperature
dependence is expected to be weak.  We neglect this
additive temperature independent anisotropic constant term in
the remainder of this section.

The neglected term justifies the addition of a d-wave
lifetime broadening of the band electrons for
the computation of the change in the band energy
from the neutron spin incommensurability in section
\ref{subsec:ke}.

Consider an \xxyy\ band electron with energy $\epsilon_1$
and a polaron with energy $E_i$ scattering to an electron with
energy $\epsilon_2$ and polaron with energy $E_f$.
$\epsilon_1$ and $\epsilon_2$ are measured relative to the Fermi
level, $\epsilon_f=0$.  Conservation of energy is,
\begin{equation}
\label{eqn:e2}
\epsilon_1=\epsilon_2+\Delta,\ \ \Delta\equiv E_f-E_i.
\end{equation}

The thermal occupations of a doped site with two polaron states
of energy $E_i$ and $E_f$ is given by
the Fermi-Dirac functions $f(-\Delta)$ and $f(\Delta)$
defined in equation \ref{eqn:fd}
and depends only on the energy difference $\Delta$.
This is identical to the results in equation \ref{eqn:occ}. 

The total
lifetime of a band state with energy $\epsilon_1$ is given by,
\begin{eqnarray}
\label{eqn:e3}
\frac{1}{\tau(\epsilon_1)} & = &
\int d\epsilon_2\int d\Delta\ \rho(\Delta)
\frac{2\pi}{\hbar}|H'|^2 \nonumber \\
& \times &
\delta(\epsilon_2-\epsilon_1-\Delta)
f(-\Delta)\left[1-f(\epsilon_2)\right],
\end{eqnarray}
where $\rho(\Delta)$ is the density of polaron sites with
an energy splitting $\Delta$ and $|H'|^2$ is taken to be a constant.
The $\rho(\Delta)$ used here is different from the spin and
chirality flipped density in section \ref{subsec:scaling}.
$\rho(\Delta)$ in equation \ref{eqn:e3} includes the energy splittings
of the $S$, $D_{xy}$, and the bonding combinations of the
$P_{x'}$ and $P_{y'}$ polaron states.

Integrating over the final electron energy $\epsilon_2$,
\begin{eqnarray}
\label{eqn:e4}
\frac{1}{\tau(\epsilon_1)} & = & \frac{2\pi}{\hbar}|H'|^2 N(0)
\nonumber \\
& \times & \int d\Delta\ \rho(\Delta)
f(-\Delta)\left[1-f(\epsilon_1-\Delta)\right],
\end{eqnarray}
where $N(0)$ is the band density of states per spin.

For $\epsilon_1$ at the Fermi level, $\epsilon_1=0$, the
integrand becomes $\rho(\Delta)f(\Delta)[1-f(\Delta)]$ leading
to a temperature dependence $T^{\mu+1}$ if 
$\rho(\Delta)\sim \Delta^\mu$.  For a uniform distribution of
polaron energy splittings, $\rho(\Delta)=\rho_0$ for
$|\Delta|\le\Delta_{max}$ and zero otherwise,
the integral in equation \ref{eqn:e4} is,
\begin{eqnarray}
\label{eqn:e5}
\frac{1}{\tau(\epsilon_1)} & = &
\frac{2\pi}{\hbar}|H'|^2 N(0) \rho_0 \nonumber \\
 & \times &
\frac{T}{\left(1-e^{-\beta\epsilon_1}\right)}
\ln\left[
\frac{e^{\beta\Delta_{max}}+e^{-\beta\epsilon_1}}
{e^{\beta(\Delta_{max}-\epsilon_1)}+1}
\right].
\end{eqnarray}

The energy scale for $\Delta_{max}$ is determined by the scale
for the hopping matrix elements and is several tenths of
an eV.  This is much larger than the temperature,
$\beta\Delta_{max}\gg 1$.  For $|\epsilon_1|\ll\Delta_{max}$,
we may take the limit $\Delta_{max}\rightarrow+\infty$ leading to,
\begin{equation}
\label{eqn:e5a}
\frac{1}{\tau(\epsilon_1)} = \frac{2\pi}{\hbar}|H'|^2 N(0) \rho_0
\left(\frac{\epsilon_1}{1-e^{-\beta\epsilon_1}}\right).
\end{equation}

The $\tau_B$ that appears in the Boltzmann transport
equation is related to the linewidth and satisfies,
\begin{equation}
\label{eqn:e6}
\frac{1}{\tau_B(\epsilon_1)}=
-\frac{2}{\hbar}\mathrm{Im}\ \Sigma(\epsilon_1)=
\frac{1}{\tau(\epsilon_1)}\cdot\frac{1}{1-f(\epsilon_1)},
\end{equation}
\begin{equation}
\label{eqn:e7}
\frac{1}{\tau_B(\epsilon)}=
\frac{2\pi}{\hbar}|H'|^2 N(0) \rho_0
\ \epsilon\coth\frac{1}{2}\beta\epsilon.
\end{equation}
At the Fermi level, $\epsilon=0$, the scattering rate is,
\begin{equation}
\label{eqn:e8}
\frac{1}{\tau_B(\epsilon)}=
\frac{2\pi}{\hbar}|H'|^2 N(0) \rho_0 (2T),
\end{equation}
leading to a linear resistivity.

If $\rho(\Delta)\sim\Delta$ for $|\Delta|\le\Delta_{min}$ and
is constant for $\Delta_{min}\le\Delta\le\Delta_{max}$,
then the resistivity
is $\sim T^2$ at low temperature and crosses over to $\sim T$ at
high temperature.  Different forms for
$\rho(\Delta)$ lead to different temperature dependencies
for the resistivity as seen in \lsco\ from $0.05<x<0.35$.
\cite{cava_resist}

The optical spectrum is composed of a Drude scattering term for
finite $\epsilon$ in equations \ref{eqn:e5} and \ref{eqn:e7}
plus polaron to polaron scattering due to photon absorption.
The Drude scattering rate becomes proportional to $\epsilon$ for
$\beta\epsilon\gg 1$ with the crossover from $T$ to
$\epsilon$ occurring when
$\beta\epsilon\approx 2$.  When the energy is larger than
the largest polaron splitting, $\epsilon\gg\Delta_{max}$ the
scattering rate saturates and becomes proportional to $\Delta_{max}$.

Direct optical absorption from
$S$ and $D_{xy}$ polarons to $P_{x'}\pm iP_{y'}$ can occur
for light polarized in the CuO$_2$ planes.
This leads to the excess mid-IR absorption.
\cite{optics1}

Chiral polarons can lead to the logarithmic
resistivity upturn at low temperatures due to a Kondo effect.
\cite{boeb1}  As \lsco\ is doped, polaron islands are formed
composed of exactly one polaron with no adjacent polarons.
Island polarons are surrounded by $d^9$ spins.  They have
two possible chiral ground states.  Antiferromagnetic 
spin flip scattering with the \xxyy\ band occurs
in second-order where the \xxyy\ electron hops
onto the polaron and either an up or down spin polaron hops
back to the \xxyy\ band.  This leads to a Kondo effect.

The momentum dependence of the antiferromagnetic Kondo
spin exchange is largest for momenta near $(\pm\pi,0)$
and $(0,\pm\pi)$ since the polaron is comprised of A$_{1}$
orbitals.  The coupling is zero for momenta along
the diagonal.  The resistivity in
the plane is dominated by electrons with momentum near
the diagonals and transport out of the plane by
momenta near $(\pm\pi,0)$ and $(0,\pm\pi)$.
The resistivity upturn due to
Kondo scattering appears at a higher temperature for
out of plane transport.

There is no Kondo effect from Coulomb scattering with polarons
from $H_U+H_K$ defined in equations \ref{eqn:hc} and \ref{eqn:hk}
because the spin coupling is ferromagnetic.

\begin{figure}[tbp]
\centering  \includegraphics[width=\linewidth]{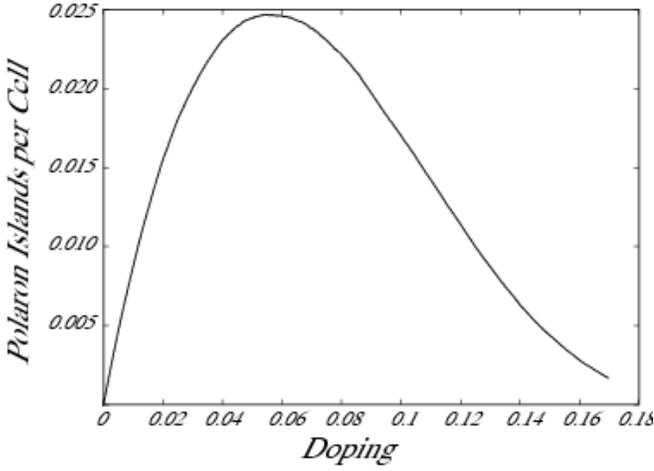}
\caption{Number of polaron islands per Cu as a function of
doping for a 2D Cu lattice.  An island is defined as a
four-site polaron having only undoped $d^9$ spins as
neighbors.
}
\label{fig:kondo}
\end{figure}

Figure \ref{fig:kondo} shows the number of islands per unit
cell as a function of doping for a 2D lattice with doped polarons.
The Kondo resistivity is expected to
disappear as the number of islands goes to zero.
From table \ref{table:percolation}, the polarons percolate in 2D
at $x\approx 0.15$.  This is approximately the doping where the
insulator to metal transition occurs for \lsco.\cite{boeb1}

The above suggestion has several caveats that could invalidate
the conclusion.  First, we do not know how many islands exist
with a small energy separation between the chiral polaron
time-reversed states.

Second, the g-factor for these
islands needs to be evaluated to determine the energy splitting
between the ``up" and ``down" states.
The Kondo effect is suppressed for temperatures less than
the splitting energy.  60 Tesla pulsed magnetic fields
were used to suppress the superconductivity in order to access
the low temperature normal state.\cite{boeb1}

The g-factor of an electron spin is two, $g_e=2$.  A 60 Tesla
field splits the up and down electron spin energies
by 80K.  For $g=0.1$, the splitting is 4K.  The Kondo effect
is active for $T>4$K.

An estimate of the polaron
g-factor, $g_p$, is made in section \ref{hall} by fitting to the
temperature dependent Hall effect for \lsco\ at $x=0.10$.  It is
argued that $g_p\sim 0.1$.  Also, for island polarons, figure
\ref{fig:4x4} shows the energy difference between a polaron and
time-reversed polaron is zero for large $J_{ch}$.
This order of magnitude reduction
in $g_p$ allows the Kondo effect to remain active at low
temperatures.

Third, the coupling to
the band electrons needs to be evaluated to determine if the
logarithmic resistivity is of the right magnitude.

\subsection{Hall Effect}
\label{hall}
Skew-scattering has been proposed\cite{varma1} to
explain the temperature dependence of the Hall effect
although the physical nature of the excitations causing the
skew-scattering was unclear.
In this section, we derive and estimate the magnitude and
temperature dependence of the skew-scattering of \xxyy\ band electrons
from chiral plaquette polarons.
The skew-scattering contribution to the Hall effect is
an extra term that is added to the ordinary Hall
effect.

Skew-scattering\cite{fert1, fert2, fert3, coleman1,skew_boltz1,
skew_boltz2} is a left-right scattering asymmetry occurring when
the scattering rate from $\mathbf{k\rightarrow k'}$ is not equal
to the scattering rate from $\mathbf{k'\rightarrow k}$,
$w(\mathbf{k\rightarrow k'})\ne w(\mathbf{k'\rightarrow k})$.
A combination of time-reversal and inversion symmetry
leads to no skew-scattering because
$w(\mathbf{k\rightarrow k'})=w(\mathbf{-k\rightarrow-k'})$ due
to inversion and
$w(\mathbf{-k\rightarrow-k'})=w(\mathbf{k'\rightarrow k})$
from time-reversal invariance.
An applied magnetic field breaks time-reversal invariance by
making the number of polarons of ``up" and ``down"
chirality different.

Skew-scattering first appears in third order for the scattering
Hamilitonian as can be seen in the simple example below
where we ignore the polarons and consider a band scattering matrix
element that is complex.

Let $H'$ be the electron scattering Hamiltonian,
\begin{equation}
\label{eqn:hall1}
H'=\sum_{\mathbf{k,k'}\sigma}V_{\mathbf{k'k}}
d^{\dagger}_{\mathbf{k'}\sigma}d^{}_{\mathbf{k}\sigma},
\end{equation}
where we ignore any spin dependence in the matrix element
$V_{\mathbf{kk'}}$.  Since $H'$ is Hermitean,
$V^{}_{\mathbf{k'k}}=V^\ast_{\mathbf{kk'}}$.
The scattering $T$ matrix is given by
\begin{equation}
\label{eqn:hall2}
T(E)=V+VG(E)V+\cdots,
\end{equation}
and the scattering rate is,
\begin{equation}
\label{eqn:hall3}
w(\mathbf{k\rightarrow k'})=\frac{2\pi}{\hbar}
|\langle\mathbf{k'}|T(\epsilon_k)|\mathbf{k}\rangle|^2
\delta(\epsilon_{k'}-\epsilon_{k}).
\end{equation}
Expanding the $T$ matrix to second-order,
\begin{equation}
\label{eqn:hall4}
\langle\mathbf{k'}|T(\epsilon_k)|\mathbf{k}\rangle=
V_{\mathbf{k'k}}+
\sum_{\mathbf{k''}}
\frac{V_{\mathbf{k'k''}}V_{\mathbf{k''k}}}
{\epsilon_k - \epsilon_{k'}+i\delta} + O(V^3).
\end{equation}
Substituting into equation \ref{eqn:hall3} and neglecting
terms of $O(V^4)$,
\begin{eqnarray}
\label{eqn:hall5}
w(\mathbf{k\rightarrow k'})=\frac{2\pi}{\hbar}
\left[
|V_{\mathbf{k'k}}|^2+ \sum_{\mathbf{k''}}
\frac{2\mathrm{Re}(V_{\mathbf{k'k}}V_{\mathbf{kk''}}
V_{\mathbf{k''k'}})}{\epsilon_k - \epsilon_{k''}}\right.
\nonumber \\
\left.
-2\pi\sum_{\mathbf{k''}}
\mathrm{Im}(V_{\mathbf{k'k}}V_{\mathbf{kk''}}
V_{\mathbf{k''k'}})
\delta(\epsilon_k-\epsilon_{k''})
\right]
\delta(\epsilon_k-\epsilon_{k'}).\ \ 
\end{eqnarray}
As a check, equation \ref{eqn:hall5} is invariant under any
redefinition of the $\mathbf{k}$ states,
$V_{\mathbf{k'k}}\rightarrow e^{i(\theta_{\mathbf{k'}}
-\theta_{\mathbf{k}})}V_{\mathbf{k'k}}$.

If $V$ is real, then there is no skew-scattering since
$w(\mathbf{k\rightarrow k'})= w(\mathbf{k'\rightarrow k})$.
If $V$ is complex, then interchanging $\mathbf{k}$ and
$\mathbf{k'}$ changes the sign of the third term.  This
is the lowest order skew-scattering term.
By interchanging $\mathbf{k'\leftrightarrow k''}$
in the third term of
the equation, the sum over $\mathbf{k'}$ of the skew term satisfies
$\sum_{k'}w_{skew}(\mathbf{k\rightarrow k'})=0$.

The applied magnetic field causes skew-scattering
by creating left-right asymmetries.
The relevant question is not if there is any skew
scattering, but whether
the scattering is large enough to account for the
experimental Hall effect and its temperature dependence.

We derive the skew-scattering from the Coulomb Hamiltonian
$H_U$ in equation \ref{eqn:hc}.  Including the exchange term
$H_K$ in equation \ref{eqn:hk} does not change the results
below because it leads to an average over $U$ and $U-K$.
Thus, we focus on $H_U$ and ignore the electron spin.

Consider a single polaron.  In the hole-doped cuprates, one
of the chiral polaron states, $P^{\pm}\equiv P_{x'}\pm iP_{y'}$
is initially
unoccupied and all the remaining polaron states shown in
figure \ref{fig:p1} are occupied.  The matrix elements for
scattering an \xxyy\ $\mathbf{k}$ electron and polaron electron
to a chiral polaron state
with a change in the \xxyy\ band electron momentum to
$\mathbf{k'}$ is,
\begin{equation}
\label{eqn:hall6}
\langle\mathbf{k'}P^{\pm}|H_U|\mathbf{k}S\rangle =
\left(\frac{U}{2N}\right)(-i)
\left(\sin q_{x'}\mp i\sin q_{y'}\right),
\end{equation}
\begin{equation}
\label{eqn:hall7}
\langle\mathbf{k'}P^{\pm}|H_U|\mathbf{k}D_{xy}\rangle =
\left(\frac{U}{2N}\right)(-i)
\left(\sin q_{x'}\pm i\sin q_{y'}\right),
\end{equation}
\begin{eqnarray}
\label{eqn:hall8a}
\langle\mathbf{k'}P^{\pm}|H_U|\mathbf{k}
P^{\pm}\rangle & = &
\left(\frac{U}{2N}\right)
\left(\cos q_{x'}+\cos q_{y'}\right), \\
\label{eqn:hall8}
 & = &
\left(\frac{U}{N}\right)
\cos\frac{1}{2}q_x\cos\frac{1}{2}q_y, 
\end{eqnarray}
\begin{eqnarray}
\label{eqn:hall9c}
\langle\mathbf{k'}P^{\pm}|H_U|\mathbf{k}
P^{\mp}\rangle & = &
\left(\frac{U}{2N}\right)
\left(\cos q_{x'}-\cos q_{y'}\right), \\
\label{eqn:hall9}
 & = &
(-)\left(\frac{U}{N}\right)
\sin\frac{1}{2}q_x\sin\frac{1}{2}q_y,
\end{eqnarray}
where $\mathbf{q=k'-k}$ and
\begin{eqnarray}
\label{eqn:hall9a}
q_{x'} & = & \frac{1}{2}\left(q_x + q_y\right),  \\
\label{eqn:hall9b}
q_{y'} & = & \frac{1}{2}\left(-q_x + q_y\right).
\end{eqnarray}
The chiral polaron to chiral polaron scattering terms in equations
\ref{eqn:hall8} and \ref{eqn:hall9} are real and lead
to no skew-scattering, as shown below.  Second-order mixing
with $S$ and $D_{xy}$ polaron symmetries leads to skew-scattering.

The Coulomb repulsion, $U$, in the above matrix elements is positive,
$U>0$.  The matrix elements corresponding to equations 
\ref{eqn:hall6}$-$\ref{eqn:hall9} for hole polarons
are obtained by the substitution $U\rightarrow -U$.
We use the expressions in equations
\ref{eqn:hall6}$-$\ref{eqn:hall9} and remember to take $U<0$ for
the hole-doped materials and $U>0$ for the electron-doped materials.

The $T$ matrix element is defined as,
\begin{equation}
\label{eqn:hall10}
T^{\pm}_{\mathbf{k'k}}=\langle\mathbf{k'}P^{\pm}
|T|\mathbf{k}P^{\pm}\rangle,
\end{equation}
where the hole resides in $P^{\pm}$ (the electron
is in $P^{\mp}$).  Expanding $T^{\pm}_{\mathbf{k'k}}$
to order $U^2$,
\begin{equation}
\label{eqn:hall11}
T^{\pm}_{\mathbf{k'k}}=V^{(1)}_{\mathbf{k'k}}+
\sum_{I}V^{(2)(\pm)}_{\mathbf{k'k}}(I)+O(U^3),
\end{equation}
where the sum is over all intermediate polaron states $I$.
$V^{(1)}_{\mathbf{k'k}}$ is defined as,
\begin{equation}
\label{eqn:hall12}
V^{(1)}_{\mathbf{k'k}}=
\langle\mathbf{k'}P^{\pm}|H_U|\mathbf{k}P^{\pm}\rangle,
\end{equation}
and is independent of the polaron chirality and real
from equation \ref{eqn:hall8}.  Thus,
$V^{(1)}_{\mathbf{k'k}}=V^{(1)}_{\mathbf{kk'}}$.
$V^{(2)}$ is,
\begin{equation}
\label{eqn:hall13}
V^{(2)(\pm)}_{\mathbf{k'k}}=
\sum_{\mathbf{p}} 
\frac{
\langle\mathbf{k'}P^{\pm}|H_U|
\mathbf{p}I\rangle
\langle\mathbf{p}I|H_U|
\mathbf{k}P^{\pm}\rangle
}{\epsilon_k-\epsilon_p-\left(E_I-E^{\pm}\right) +i\delta},
\end{equation}
where the sum is over all intermediate \xxyy\ momenta $\mathbf{p}$.
$E^{\pm}$ and $E_I$ are the chiral polaron
and the intermediate polaron energies respectively.  $\epsilon_p$
is the \xxyy\ band energy for momentum $\mathbf{p}$.

Since $V^{(1)}$ is real, $|T^{\pm}_{\mathbf{k'k}}|^2$ is,
\begin{equation}
\label{eqn:hall14}
|T^{\pm}_{\mathbf{k'k}}|^2=
\left(V^{(1)}_{\mathbf{k'k}}\right)^2 +
V^{(1)}_{\mathbf{k'k}}
\sum_{I}2\mathrm{Re} V^{(2)(\pm)}_{\mathbf{k'k}}(I)+O(U^3).
\end{equation}
The skew-scattering contribution
is antisymmetric under interchange, $\mathbf{k'\leftrightarrow k}$.
The first term in equation \ref{eqn:hall14} is symmetric and so is
$V^{(2)(\pm)}_{\mathbf{k'k}}(I)$ for $I=P^{\pm}$.  The lowest
order skew-scattering term is,
\begin{equation}
\label{eqn:hall15}
|T^{\pm}_{\mathbf{k'k}}|^2_{s}=
V^{(1)}_{\mathbf{k'k}}\cdot
2\mathrm{Re}\left[V^{(2)(\pm)}_{\mathbf{k'k}}(S)+
V^{(2)(\pm)}_{\mathbf{k'k}}(D_{xy})\right].
\end{equation}

In appendix \ref{ap:hall}, it is shown the antisymmetric terms
in $V^{(2)(\pm)}(S,D_{xy})$ are,
\begin{equation}
\label{eqn:hall16}
\mathrm{Re}V^{(2)(\pm)}_{\mathbf{k'k}}(S)=
\mp\left(\frac{U^2}{4N}\right)
\pi F_1\left[\epsilon_k-(E_S-E^\pm)\right]
A_{\mathbf{k'},\mathbf{k}},
\end{equation}
\begin{equation}
\label{eqn:hall17}
\mathrm{Re}V^{(2\pm)}_{\mathbf{k'k}}(D)=
\pm\left(\frac{U^2}{4N}\right)
\pi F_1\left[\epsilon_k-(E_D-E^\pm)\right]
A_{\mathbf{k'},\mathbf{k}},
\end{equation}
where the antisymmetric function, $A_{\mathbf{kk'}}=
-A_{\mathbf{k'k}}$, is,
\begin{equation}
\label{eqn:hall18}
A_{\mathbf{k'},\mathbf{k}}=
\sin k_{x'}\sin k'_{y'}-\sin k_{y'}\sin k'_{x'}.
\end{equation}
Equations \ref{eqn:hall9a} and \ref{eqn:hall9b}
define the $x'$ and $y'$ components of $\mathbf{k}$ and
$\mathbf{k'}$.  The function $F_1$ is the sum
over the Brillouin zone,
\begin{equation}
\label{eqn:hall19}
F_1(\omega)=\frac{1}{N}
\sum_{\mathbf{p}}\frac{1}{2}
\left(\cos p_x + \cos p_y\right)
\delta(\omega-\epsilon_p).
\end{equation}
The $D_{xy}$ contribution is negative the $S$ contribution with
$E_D$ substituted for $E_S$.

The skew-scattering term is,
\begin{eqnarray}
\label{eqn:hall20}
|T^{\pm}_{\mathbf{k'k}}|^2_{s} & = & \mp
\left(\frac{1}{N^2}\right)U^3\left(\frac{\pi}{2}\right)
\cos\frac{1}{2}q_x\cos\frac{1}{2}q_y \nonumber \\
 & \times &
\left\{F_1\left[\epsilon_k-\left(E_S-E^\pm\right)\right]
\right.- \nonumber \\
 & &
\left.
F_1\left[\epsilon_k-\left(E_D-E^\pm\right)\right]\right\}
A_{\mathbf{k'k}},
\end{eqnarray}
where $\mathbf{q=k'-k}$.

Equation \ref{eqn:hall20} is the skew-scattering for a single
polaron with polaron orbital energies $E^\pm$, $E_S$, and $E_D$.
These energies have probability distributions
$\rho^\pm(E^\pm)$, $\rho_S(E_S)$, and $\rho_D(E_D)$.
The mean value of equation \ref{eqn:hall20} is,
\begin{eqnarray}
\label{eqn:hall21}
\langle|T^{\pm}_{\mathbf{k'k}}|^2_{s}\rangle=
\int dE^{\pm}dE_S dE_D\rho^\pm(E^\pm)\rho_S(E_S)\rho_D(E_D)
\nonumber \\
\times
\ |T^{\pm}_{\mathbf{k'k}}|^2_{s}
(E^\pm,E_S,E_D).\ \ 
\end{eqnarray}
If polarons with $S$ and $D_{xy}$ symmetry have identical
energy distributions, then the $\pm$ components
of the skew-scattering are always zero.

The range of energies
that $E^+$ and $E^-$ span is $\sim J_{dd}$ and is much smaller than
the hopping matrix element scale of $E_S$ and $E_D\sim t_{hop}$.
Therefore, we set $E^\pm = 0$ in
$F_1$ and evaluate the mean over $E_S$ and $E_D$,
\begin{eqnarray}
\label{eqn:hall22}
\langle F_1\left[\epsilon_f-E_S\right]\rangle & = &
\int_0^{\Delta^S_{max}}d\omega\rho^S(\omega)
F_1(\epsilon_f-\omega), \nonumber \\
 & = & \frac{\lambda^S}{\Delta^S_{max}},
\end{eqnarray}
where 
\begin{equation}
\label{eqn:hall23}
\lambda^S=\left(\frac{1}{N}\right)\sum_{\epsilon_f-\Delta^S_{max}
\le\epsilon_p\le\epsilon_f}
\frac{1}{2}(\cos p_x+\cos p_y),
\end{equation}
and $\rho^S=1/\Delta^S_{max}$ in the interval
$0\le\Delta\le\Delta^S_{max}$ and zero outside.
There is a similar expression for the mean over $E_D$.

If $\epsilon_f-\Delta^{S,D}_{max}$ is less than the bottom of the
\xxyy\ band, then $\lambda^S=\lambda^D=\lambda$ where $\lambda$ is
the sum over occupied states,
\begin{equation}
\label{eqn:hall24}
\lambda=\left(\frac{1}{N}\right)\sum_{occ}
\frac{1}{2}(\cos p_x+\cos p_y).
\end{equation}
In general, $-1<\lambda<1$.
Cuprate band structures are occupied at $(0,0)$ and unoccupied
at $(\pi,\pi)$ leading to $0<\lambda<1$.  For \lsco\ at
$x=0.10$, $\lambda=0.19$.
Substituting the mean of $F_1$ into equation \ref{eqn:hall20},
\begin{eqnarray}
\label{eqn:hall25}
\langle|T^{\pm}_{\mathbf{k'k}}|^2_{s}\rangle & = & \mp
\left(\frac{1}{N^2}\right)U^3\left(\frac{\pi}{2}\right)
\cos\frac{1}{2}q_x\cos\frac{1}{2}q_y \nonumber \\
 & \times & \lambda\left[
\frac{1}{\Delta^S_{max}}- \frac{1}{\Delta^D_{max}}\right]
A_{\mathbf{k'k}},
\end{eqnarray}
\begin{equation}
\label{eqn:hall26}
w_{skew}^\pm(\mathbf{k\rightarrow k'})=
\frac{2\pi}{\hbar}
\langle|T^{\pm}_{\mathbf{k'k}}|^2_{s}\rangle
\delta(\epsilon_{k'}-\epsilon_k).
\end{equation}
The total skew-scattering is,
\begin{equation}
\label{eqn:hall27}
w_{skew}(\mathbf{k\rightarrow k'})=N_p\langle n^- - n^+\rangle
w_{skew}^+(\mathbf{k\rightarrow k'}),
\end{equation}
where $N_p$ is the total number of polarons and
$\langle n^- - n^+\rangle$ is the mean number of $P^-$ polarons
minus the number of $P^+$ polarons.  This difference is non-zero
in a magnetic field and is proportional to the field $B$.

To evaluate $\langle n^- - n^+\rangle$, consider a polaron
with energy difference between $P^-$ and $P^+$ of $\Delta$,
$E^+ - E^-=\Delta$.
The field changes the energies to
$E^\pm = E^\pm \mp g_p\mu_B B$,
where
$g_p$ is the polaron g-factor and $\mu_B$ is the Bohr magneton.
The occupations become,
\begin{equation}
\label{eqn:hall28}
n^-=f(-\Delta) + 2g_p\mu_B B\left(-\frac{\partial f}
{\partial\epsilon}\right)_{\epsilon=-\Delta}B+O(B^2),
\end{equation}
\begin{equation}
\label{eqn:hall29}
n^+=f(\Delta) - 2g_p\mu_B B\left(-\frac{\partial f}
{\partial\epsilon}\right)_{\epsilon=\Delta}B+O(B^2),
\end{equation}
where $f(\epsilon)$ is the Fermi-Dirac function defined in
equation \ref{eqn:fd}.
Taking the probability density of $\Delta$ to be uniform
over the range $\Delta_{min}\le|\Delta|\le\Delta_{max}$ makes
the density $\rho_0=1/2(\Delta_{max}-\Delta_{min})$.  The
mean polaron difference is,
\begin{equation}
\label{eqn:hall30}
\langle n^- - n^+\rangle =
4g_p\mu_B B(2\rho_0)
\int_{\Delta_{min}}^{\Delta_{max}}d\Delta
\left(-\frac{\partial f}{\partial\epsilon}\right), 
\end{equation}
\begin{equation}
\label{eqn:hall31}
\langle n^- - n^+\rangle =
4g_p\mu_B\frac{\left[f(\Delta_{min})-f(\Delta_{max})\right]}
{\Delta_{max}-\Delta_{min}}B.
\end{equation}

All of the temperature dependence of the Hall effect is contained
in equation \ref{eqn:hall31}.  $\Delta_{max}$ is on the order
of $J_{dd}\sim 0.1$ eV.  $\Delta_{min}$ and $\Delta_{max}$
are larger for small doping since there are more
undoped $d^9$ spins to split the energy between
time-reversed polarons.  The $\Delta_{max}$ in
equation \ref{eqn:hall31} is smaller than the $\Delta_{max}$
in equation \ref{eqn:e5} since the latter includes the splittings
of the $S$, $D_{xy}$, and bonding combinations of
$P_{x'}\pm iP_{y'}$. 
From equation \ref{eqn:hall31}, the difference 
$\langle n^- - n^+\rangle$ is zero for $T\ll\Delta_{min}$ and
rises to a maximum for some $T$ between $\Delta_{min}$
and $\Delta_{max}$.  When
$T\gg\Delta_{max}$, the difference decreases to zero as $1/T$.
The temperature dependence from skew-scattering is added
to the ordinary band contribution to the Hall
effect.  The ordinary term is hole-like and positive for the
cuprates.  The coefficient that multiplies equation
\ref{eqn:hall31} in $\rho^{skew}_{xy}$ can be either positive
or negative depending on the details of the polaron energy
distributions and the sign of the Coulomb repulsion $U$.

\begin{figure}[tbp]
\centering  \includegraphics[width=\linewidth]{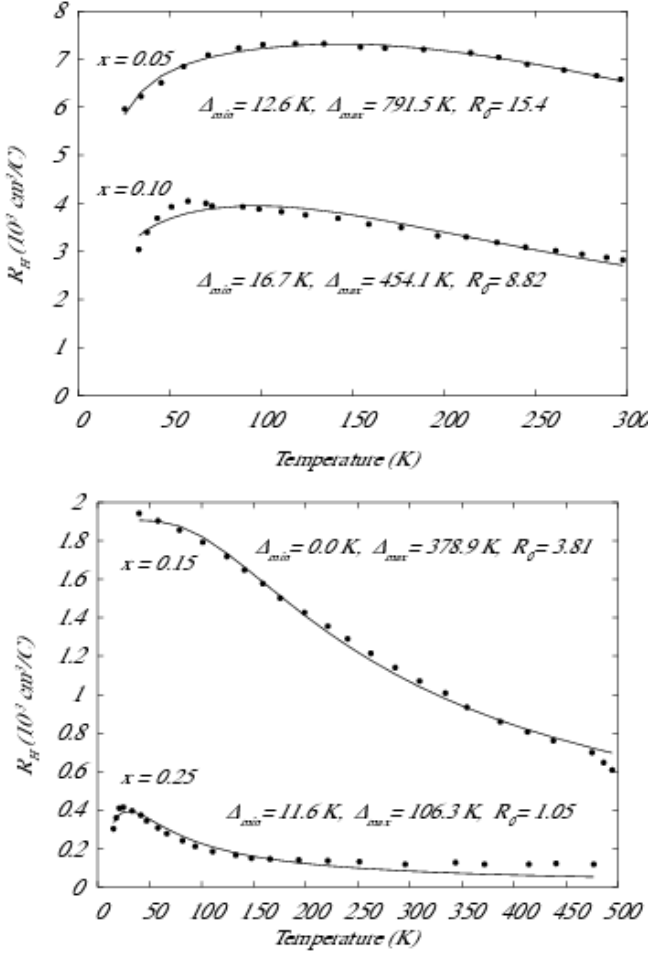}
\caption{Fit to Hall effect data\cite{cava_hall} for \lsco\ using
the expression in equation \ref{eqn:hall32}.
For $x=0.05$, the parameters are
$\Delta_{min}=12.6$ K, $\Delta_{max}=791.5$ K, and
$R_0=15.4\times 10^{-3}$\ cm$^3$/C.
For $x=0.10$,
$\Delta_{min}=16.7$ K, $\Delta_{max}=454.1$ K, and $R_0=8.82$.
For $x=0.15$,
$\Delta_{min}=0.0$ K, $\Delta_{max}=387.9$ K, and $R_0=3.81$.
For $x=0.25$,
$\Delta_{min}=11.6$ K, $\Delta_{max}=106.3$ K, and $R_0=1.05$.}
\label{fig:hall}
\end{figure}

The observed temperature and doping dependence
of \lsco\cite{cava_hall} is consistent with chiral plaquette polaron
skew-scattering as shown in figure \ref{fig:hall}.   We fit
the experimental data to the expression,
\begin{equation}
\label{eqn:hall32}
R_H=R_0\left[f(\Delta_{min})-f(\Delta_{max})\right],
\end{equation}
where $f(\epsilon)$ is the Fermi-Dirac function defined in
equation \ref{eqn:fd}.

There is sufficient structure in the
temperature dependence derived above to account for
the various temperature behaviors of the electron-doped
materials, too.\cite{kontani1}

The additive contribution to the
Hall resistivity arising from skew-scattering is derived in
appendix \ref{ap:hall} and is,
\begin{eqnarray}
\label{eqn:hall33}
\rho^{skew}_{xy} & = & 2\left(\frac{m^2}{n^2e^2}\right)
\left(\frac{1}{\Omega}\right)(-)\sum_{\mathbf{k'k}}
\left(-\frac{\partial f}{\partial\epsilon_k}\right) \nonumber \\
 & \times &
\frac{1}{2}\left(v_{kx}v_{k'y}-v_{ky}{v_{k'x}}\right)
w_{skew}(\mathbf{k\rightarrow k'}).\ 
\end{eqnarray}

To determine if the magnitude of the
chiral polaron skew-scattering is large
enough to be compatible with experimental data,
we computed $R_H c=\rho^{skew}_{xy}/B$ for \lsco\ using
the band structure in equation \ref{eqn:lsco} at $x=0.10$
and estimates of various parameters.

We take $m$ to be one electron mass and $n=4x/\Omega_{cell}$ to
be the number of charge carriers per volume (since each polaron
adds four \xxyy\ band electrons in our percolating model).
The polaron g-factor is taken to be one, $g_p=1$.
$\Delta_{max}=454.1$ K and $\Delta_{min}=16.7$ K from
the fit in figure \ref{fig:hall}.  Finally, we choose
$\Delta^S_{max}=1.0$ eV and $\Delta^{D}_{max}=1.5$ eV for the
energy distribution widths of the $S$ and $D_{xy}$ polaron energies.  

The reason we chose $\Delta^{D}_{max}>\Delta^{S}_{max}$ is because
the energy separation of
the $D_{xy}$ states from the $P^\pm$ chiral polarons is larger
than $S$ in table \ref{table:polaron_energies}.
This does not prove $\Delta^{D}_{max}>\Delta^{S}_{max}$ is
always the correct choice, but it is suggestive of the expected
energy splitting.  Also, our values for
the average skew-scattering $T$ matrix in equation \ref{eqn:hall25}
assume a uniform energy density $\rho^{S}$ and $\rho^{D}$
and this may not be correct.  The idea here is to get a
reasonable order of magnitude estimate for the size of the 
skew-scattering.

The final parameter to choose is the Coulomb repulsion matrix
element energy $U$.  $U\sim1.0$ eV is the correct energy scale
because
$U=(1/\Omega_{cell})4\pi e^2/q_{D}^2$ where $q_{D}=4\pi e^2N(0)$ is
the Debye screening length arising from the many-body response
of the band electrons to a perturbing potential.  $N(0)$ is
the band density of states.  A typical metallic density of states
is one state per eV per unit cell leading to $U\sim 1.0$ eV.
Since the density of charge carriers is reduced to $4x=0.40$ per unit
cell, the band electrons are less effective at screening the
Coulomb repulsion.  This increases $U$ by a factor of $1/0.4=2.5$.
We take $U=2.5$ eV.

$U$ is negative for the hole-doped cuprates because
the band structure arises from fully occupied polaron states
and a hole in a chiral polaron state amounts to subtracting
a Coulomb matrix element.
This was discussed previously following equations 
\ref{eqn:hall6}$-$\ref{eqn:hall9}.
For the electron-doped systems, $U$ is positive.  Since $U^3$
appears in equation \ref{eqn:hall25}, the sign difference in
the temperature dependence of the Hall effect between hole-doped
and electron-doped cuprates is obtained by assuming all
the other parameters discussed above do not change.  We
take $U=-2.5$ eV for \lsco.

Calculating $\rho^{skew}_{xy}$ in equation \ref{eqn:hall33}, we obtain
$R_H c=5.39\times 10^{-3}\mathrm{cm}^3/\mathrm{C}$
at 100K or a charge density of $0.116$ carriers per unit cell
with the chosen parameters.  The experimental
value\cite{cava_hall} for $x=0.10$ at $100$ K is approximately
$4\times 10^{-3}\mathrm{cm}^3/\mathrm{C}$ as can be seen in
figure \ref{fig:hall}.
Our result is a factor of 1.35 too large.

The largest errors in our calculation come from
the estimate of the Coulomb repulsion $U$, 
the effective $(n/m)_{eff}$, and the value of polaron
g-factor, $g_p=1$.  We used one electron
mass for $m$ and chose $n$ to be $4x=0.4$ carriers per cell
to obtain $(n/m)_{eff}$.  This is likely an overestimate
of $(n/m)_{eff}$ since the carriers must traverse a percolating
swath through the crystal and the scattering rate $1/\tau$ for
the ordinary conductivity is not expected to be large.
Since $(n/m)_{eff}^{-2}$ and $U^3$ appear in $\rho_{xy}$, small
changes lead to a large enhancement of the computed
$\rho_{xy}$.  We conclude the polaron
g-factor is overestimated and may be an order of magnitude
smaller, leading to $g_p\sim 0.1$.  A reduced $g_p$ allows
the Kondo resistivity discussed in section \ref{resist} to remain
active in the pulsed $60$ Tesla magnetic fields\cite{boeb1} used
to suppress superconductivity and measure the low temperature
resistivity.

The analysis of this section can be done for the real polaron
states $P_{x'}$ and $P_{y'}$ instead of the chiral states.
In this case, the magnetic field creates the Van-Vleck
paramagnetic states $P_{x'}+iBP_{y'}$ and $P_{y'}+iBP_{x'}$.
The $B$ in the equations is a constant times the applied field.
The skew-scattering contribution from a polaron where the
energy of $P_{y'}$ is greater than $P_{x'}$ by $\Delta$
exactly cancels the contribution from a polaron
where $P_{x'}$ is greater than $P_{y'}$ by
$\Delta$.  Thus, real polaron states cannot lead to skew
scattering.

For a planar
magnetic field, the g-factor is substantially smaller
because the angular momentum of the chiral polarons is normal
to the plane.  Therefore, for planar magnetic fields, the Hall
effect should be temperature independent, as observed.

\section{ARPES}
\label{sec:arpes}

\subsection{Pseudogap}
\label{sec:pseudo}
The undoped $d^9$ spins have antiferromagnetic fluctuations with
wavevector $\mathbf{q}\approx(\pi,\pi)$
that couple an \xxyy\ band state with momentum $\mathbf{k}$ to
$\mathbf{k\pm q}$.  In this section, we show
that this coupling leads to a pseudogap at the Fermi energy
with magnitude proportional to the square
of the coupling.

At fixed temperature,
the coupling is larger for low dopings because there
are more undoped $d^9$ spins.  The coupling
decreases as the doping increases.
This leads to a larger pseudogap for low dopings, as observed.
\cite{norman1, loeser1, marshall1, ding1}

At fixed doping, the strength of the antiferromagnetic $d^9$
fluctuations decreases with increasing temperature.  Thus, the
coupling to the \xxyy\ band should decrease with increasing
temperature.  This leads to the closing of the pseudogap with
temperature.

The momentum dependence of the pseudogap is determined by
the shape of the Fermi surface.  If
the state with momentum $\mathbf{k\pm q}$ is
close to the Fermi surface, then mixing with the $\mathbf{k}$
state leads to a reduction in the spectral function
and a pseudogap.
Such nesting occurs for states near $(\pm\pi,0)$ and $(0,\pm\pi)$.
Along the Brillouin zone diagonals, there is no nesting.
This leads to a pseudogap that is zero along the diagonal and
increases as one moves toward $(\pi,0)$.
The model does not guarantee zero pseudogap along the diagonal
although the magnitude of the gap is expected to be small.

ARPES measures the spectral function 
$A(\mathbf{k},\omega)=(-1/\pi)\mathrm{Im}\ G(\mathbf{k},\omega)$.
Evaluating $A(\mathbf{k},\omega)$ for 
$\mathbf{k}$ vectors on the Fermi surface, $\epsilon_k=\epsilon_f$,
and at the Fermi energy, $\omega=\epsilon_f$, determines the
density of states suppression at the Fermi level and its
temperature dependence.  The relative change in the spectral
function is,
\begin{equation}
\label{eqn:arp1}
r(\mathbf{k},\omega,V)=
\frac{A(\mathbf{k},\omega)-A_0(\mathbf{k},\omega)}
{A_0(\mathbf{k},\omega)},
\end{equation}
where $A_0(\mathbf{k},\omega)$ is the $V=0$ spectral function
and $A(\mathbf{k},\omega)$ is the spectral function including the
$d^9$ interaction $V$.
The pseudogap is zero when $V=0$ at high temperature.
Using equation \ref{eqn:g3},
\begin{equation}
\label{eqn:arp2}
r=
V^2\cdot
\frac{\mathrm{Im}\ G_0^2(\mathbf{k},\omega)
\left[G_0(\mathbf{k+q},\omega)+
G_0(\mathbf{k-q},\omega)\right]}
{\mathrm{Im}\ G_0(\mathbf{k},\omega)}.
\end{equation}
From equation \ref{eqn:g6},
\begin{equation}
\label{eqn:arp3}
G_0(\mathbf{k},\epsilon_k)=\frac{1}{i\Gamma_k}.
\end{equation}
The relative spectral function change at the Fermi level,
$\omega=\epsilon_f$, is,
\begin{eqnarray}
\label{eqn:arp4}
r & = & -V^2\left(\frac{1}{\Gamma_k}\right)
\left[
\frac{\Gamma_{k+q}}
{\left(\epsilon_f-\epsilon_{k+q}\right)^2+\Gamma_{k+q}^2}\right.
 \nonumber \\
 &   &
+\left.
\frac{\Gamma_{k-q}}
{\left(\epsilon_f-\epsilon_{k-q}\right)^2+\Gamma_{k-q}^2}\right].
\end{eqnarray}

$r$ is always negative, $r<0$, and $r\rightarrow 0$
as $V\rightarrow 0$ at high temperatures.  The
density of states decreases at the Fermi energy leading to
a temperature dependent pseudogap that closes as the
temperature increases.  The linewidths $\Gamma_k$ and
$\Gamma_{k\pm q}$ are of the order $\sim 0.01$ eV due
to the broadening of a $\mathbf{k}$ state arising from
the non-uniform percolating path in underdoped systems.
The coupling of the $d^9$ spins to the \xxyy\ band must
be less than or of the order of $J_{dd}\sim 0.1$ eV.
The relative density of states change is on the order of
$r\sim(V/\Gamma_k)^2$.  Thus, the change in the ARPES
spectra arising from
coupling between the undoped $d^9$ spins and \xxyy\ is
large enough to qualitatively account for the observations.

\subsection{2212 Bi-Layer Splitting}
\label{sec:split}
There is an $\sim 0.1$ eV splitting
between the bonding and antibonding \xxyy\ bands in
bilayer Bi$_2$Sr$_2$CaCu$_2$O$_{8+\delta}$ for $\mathbf{k}$ vectors
near $(\pi,0)$ arising from coupling through out-of-plane orbitals.
The splitting along the diagonal is
very small and not experimentally resolvable.
The $(\pi,0)$ splitting has been observed\cite{bilayer1} by
ARPES for overdoped systems.
No splitting has been observed for underdoped systems.
In addition, the bilayer splitting disappears
as the temperature is increased with the temperature separating
the coherent (splitting) and incoherent (no splitting) regimes
increasing with further overdoping.\cite{bilayer1}

XAFS\cite{haskel_Sr1, haskel_Sr2} and first principles calculations
\cite{ub3lyp_dope} find the distance between an apical O and
planar Cu decreases $\approx0.2-0.3$ \AA\ due to Sr doping in \lsco.
Coherent hopping from one CuO$_2$
layer to another requires coupling of \xxyy\ with an out-of-plane
orbital.  The \xxyy\ band state is delocalized over
the percolating polaron swath.
Hopping onto an out-of-plane polaron orbital induces a local
structural deformation.  The matrix element 
is reduced due to the large structural change as in the small
polaron problem.\cite{mahan1}  The \xxyy\ band
electronic response to the changed potential of the doubly
occupied polaron suppresses the matrix element due
to the orthogonality catastrophe.\cite{mahan1}

The second-order coherent bilayer hopping through
an out-of-plane polaron orbital is not completely suppressed by the
orthogonality catastrophe because the initial and final polaron states
have the same occupation.  Bilayer splitting of
the \xxyy\ band is induced by this coupling.
The splitting is strongly temperature dependent
and its magnitude decreases with increasing temperature.

In section \ref{resist}, a similar second-order hopping
leading to d-like anisotropic scattering was argued to be weakly
temperature dependent.  The difference between the previous
intra-layer and current inter-layer process is that the matrix
element for the latter is very sensitive to any out-of-plane
atomic displacement leading to a strong temperature dependence.

Qualitatively, for underdoped 2212, the effective inter-layer
hopping matrix element is small due to the reduced number
of Cu pairs between layers that are both part of the percolating
band swath.  In this scenario, the bilayer splitting is
small and may not have been resolved by ARPES at low temperatures.

As the doping is increased, adjacent Cu pairs between layers
become available for coherent hopping and
bilayer splitting becomes resolvable.
This splitting decreases with increasing
temperature in complete analogy to the small polaron problem.
At high temperatures, the splitting disappears.

For highly overdoped materials, the splitting is large at
low temperatures.  As the polaron plaquettes become more
crowded, any atomic distortion from hopping
to or from the out-of-plane orbitals is reduced.  This leads
to a higher temperature before bilayer splitting becomes too
small to be resolvable by ARPES.

\section{NMR}
\label{sec:nmr}
A potential problem with our chiral plaquette polaron theory
of cuprate superconductivity involves the NMR experiments by
Takigawa et al\cite{takigawa1} on underdoped
YBa$_{2}$Cu$_{3}$O$_{6.63}$ that shows the same
temperature dependent Knight
shift for both the planar Cu and O.  The most reasonable conclusion
from this experiment is that only one electronic component is
involved in the cuprate NMR.  Similar conclusions were reached for
\lsco.\cite{johnston1, nakano1, walstedt2}
This would appear to contradict our assertion of an out-of-plane
polaron because one would expect the polaron to have different
hyperfine couplings to the planar Cu and O atoms than the \xxyy\ band,
leading to a total Knight Shift temperature dependence (sum of
\xxyy\ band plus polaron) that is different for Cu and O.

On the other hand,  small-tip angle spin-echo double resonance
(SEDOR) measurements\cite{pennington1}
find the electron mediated nuclear-nuclear couplings between
neighboring Cu-O and Cu-Cu are incompatible with one-component
cuprate theories.  In addition, the high-temperature Y spin
relaxation and Knight Shifts\cite{nandor1}
are incompatible with one-component cuprate theories.
Indeed, recent phenomenological fits\cite{uldry1}
imply there are two contributions to the Redfield correlation time.

Clearly the effect of the out-of-plane orbitals in our model cannot
be ignored in interpreting the NMR data.  In our model, the polarons
lead to a weakly temperature dependent Knight shift of the form in
equation \ref{eqn:hall31}
for the magnetic susceptibility due to polarons and
shown in figure \ref{fig:hall}
for the temperature dependence of the Hall effect.
The polaron hyperfine coupling to the planar Cu and O sites is weaker
than the \xxyy\ band because the hole character is primarily in the
out-of-plane O \pz\ orbital.  This weak temperature dependence is
seen in fully doped YBa$_{2}$Cu$_{3}$O$_{7}$
because the \xxyy\ band electrons
Knight shift contribution is constant with temperature.\cite{nandor1}
For underdoped samples, the temperature dependence of the polaron
shift is masked by the larger temperature dependence of the
Knight shift from the \xxyy\ band electrons.  The latter shift arises
from the decrease in the density of states of the \xxyy\ band due
to the coupling with the undoped Cu $d^9$ spins as shown in equation
\ref{eqn:arp4}
for the ARPES pseudogap.  We intend to develop the methodology
to estimate the size of these effects and their detailed
temperature dependence in order to compare the results with
experimental data. However, at the present time the NMR must be
considered a possible Achilles Heel for our chiral plaquette
polaron theory of cuprate superconductivity.

\section{Conclusions}
\label{sec:conclusions}
Ab-initio quantum mechanical calculations find localized holes
in out-of-plane orbitals in contrast to the t-J model.
Based on the results of calculations with explicity doped Sr in \lsco,
we postulated that chiral plaquette polarons are created by doping
and that a delocalized \xxyy/\psigma\ band is formed when the
plaquette polarons percolate through the crystal.

The D-wave superconductivity, temperature dependent Hall effect,
neutron $\omega/T$ scaling, neutron incommensurability, resistivity,
the doping value for the insulator-metal phase transition,
optical absorption, low-temperature $\log(T)$ resistivity, ARPES
pseudogap, and the evolution of bi-layer splitting in Bi-2212 are
explained by chiral plaquette polarons.

We have not shown ab-initio that chiral plaquette polarons are formed
with doping.  For \lsco, ab-initio evidence for holes
in out-of-plane orbitals has been demonstrated.
Out-of-plane hole orbitals
are plausible for cuprates where the
doping arises from interstitial O atoms.
YBa$_2$Cu$_3$O$_{7}$, where the
Cu$-$O chain is completely full, is most intriguing.
We believe there are local structural
deformations of the apical O atoms favoring localized
out-of-plane orbitals.

Detailed calculations of the spectral function for the \xxyy\ band
measured in ARPES are necessary in order to study the evolution of
the background and linewidth as a function of doping and
temperature.  For underdoped systems, the percolating path
is not uniform
through the crystal and leads to broadening of the
$\mathbf{k}$ state spectral function.  Mixing with the
A$_1$ orbitals of the polarons increases the broadening
near $(\pi,0)$.  

A calculation using the Coherent Potential Approximation
(CPA)\cite{cpa1, cpa2}
may be the way to obtain quantitative estimates
of the ARPES.  CPA calculations would lead to an accurate
determination of the magnitude of the resistivity and would
allow us to determine the correct value of
$(n/m)_{eff}$ to use in the results for the magnitude of
the skew-scattering contribution to the Hall effect.

Since neutron spin scattering
couples to the electron spin and transport couples to its charge,
we regard our results to be strong evidence for
the existence of chiral plaquette polarons.

\begin{acknowledgments}
\label{sec:acknowledgments}
The authors thank Dr. Jason K. Perry for
innumerable discussions and insight into the problem.
The facilities of the
MSC are supported by ONR-DURIP, ARO-DURIP, NSF-MURI, and the Beckman
Institute.  This research was supported in part by DARPA-PROM
(ONR N00014-06-1-0938) and by NSF (CCF-0524490).
\end{acknowledgments}

\appendix

\section{Perturbed Band Energy}
\label{ap:neut}
In this appendix, we derive the expressions used in section
\ref{subsec:ke}
to compute the contribution of the \xxyy\ band energy for
different incommensurate neutron q-vectors.

Using equations \ref{eqn:g5}, \ref{eqn:g6}, and \ref{eqn:g11},
\begin{equation}
\label{a1_1}
n(\mathbf{k},\epsilon) = n_0(\mathbf{k},\epsilon) +
V^2 \int_{-\infty}^{\epsilon}d\omega
\ P(\mathbf{k},\omega),
\end{equation}
where we have defined,
\begin{equation}
\label{a1_2}
P(\mathbf{k},\omega) = R(\mathbf{k},\mathbf{k+q},\omega) +
R(\mathbf{k},\mathbf{k-q},\omega).
\end{equation}
The total number of electrons is
$n_{tot}=\sum_{\mathbf{k}}n(\mathbf{k},\epsilon_f+\delta\epsilon_f)$.
From equation \ref{a1_1}, 
\begin{equation}
\label{a1_3}
n_{tot}=
\sum_{\mathbf{k}}
n_0(\mathbf{k},\epsilon_f+\delta\epsilon_f) +
V^2 \int_{-\infty}^{\epsilon_f+\delta\epsilon_f}d\omega
\sum_{\mathbf{k}}P(\mathbf{k},\omega).
\end{equation}
The first term on the right-hand side can be expanded in a power
series in $\delta\epsilon_f$,
\begin{eqnarray}
\label{a1_4}
\sum_{\mathbf{k}} n_0(\mathbf{k},\epsilon_f) +
\delta\epsilon_f
\frac{\partial}{\partial\epsilon}
\left[
\sum_{\mathbf{k}} n_0(\mathbf{k},\epsilon_f)
\right] + O(\delta\epsilon_f^2) \nonumber \\
\label{a1_5}
=n_{tot} + \delta\epsilon_f N_0(\epsilon_f) + O(\delta\epsilon_f^2),
\end{eqnarray}
where we have used equations \ref{eqn:g5a} and \ref{eqn:g8}.
$N_0(\epsilon_f)$ is the unperturbed density of states per spin.
The second term in equation \ref{a1_3} becomes,
\begin{equation}
\label{a1:6}
V^2 \int_{-\infty}^{\epsilon_f}d\omega
\sum_{\mathbf{k}}P(\mathbf{k},\omega) +
O(V^2\delta\epsilon_f).
\end{equation}
Since $\delta\epsilon_f\sim O(V^2)$, the second term is of order
$\delta\epsilon_f^2\sim O(V^4)$ and can be neglected to lowest
order.  Substituting into \ref{a1_3} leads to,
\begin{equation}
\label{a1_7}
\delta\epsilon_f N_0(\epsilon_f) + V^2
\int_{-\infty}^{\epsilon_f}d\omega\sum_{\mathbf{k}}
P(\mathbf{k},\omega) = 0.
\end{equation}
From equation \ref{a1_2}, \ref{a1_7} is identical equation
\ref{eqn:g10}.

Substituting \ref{a1_1} into equation \ref{eqn:g9} for the
total energy, using equation \ref{eqn:g5a},
and keeping terms up to $O(V^2)$,
\begin{eqnarray}
\label{a1_8}
E_{tot}(\mathbf{q},V) & = & E_G+
\epsilon_f\cdot
\delta\epsilon_f N_0(\epsilon_f) \nonumber \\
 &  & + V^2 \int_{-\infty}^{\epsilon_f}d\omega\ \omega
\sum_{\mathbf{k}}P(\mathbf{k},\omega).
\end{eqnarray}
The second term proportional to $\delta\epsilon_f$ can be
eliminated by using equation \ref{a1_7} leading to,
\begin{equation}
\label{a1_9}
E_{tot}(\mathbf{q},V)=E_G + V^2
\int_{-\infty}^{\epsilon_f}d\omega (\omega-\epsilon_f)
\sum_{\mathbf{k}}
P(\mathbf{k},\omega).
\end{equation}
$E_G$ is the unperturbed $(V=0)$ ground state energy.
Using \ref{a1_2}, this is the result shown in equation \ref{eqn:g11}.

To compute the change in energy due to incommensurability,
we need to evaluate integrals of the form,
\begin{equation}
\label{a1_10}
I(E,E',\epsilon_f)\equiv
\int_{-\infty}^{\epsilon_f}\frac{d\omega}
{\left(\omega-E\right)^2\left(\omega-E'\right)},
\end{equation}
\begin{equation}
\label{a1_11}
J(E,E',\epsilon_f)\equiv
\int_{-\infty}^{\epsilon_f}d\omega
\frac{\left(\omega-\epsilon_f\right)}
{\left(\omega-E\right)^2\left(\omega-E'\right)},
\end{equation}
where
\begin{equation}
\label{a1_11a}
E=\epsilon-i\Gamma,\ E'=\epsilon'-i\Gamma'\ \ \mathrm{and}
\ \Gamma,\ \Gamma'>0.
\end{equation}
From equation \ref{eqn:g12},
\begin{eqnarray}
\label{a1_12}
\int_{-\infty}^{\epsilon_f}R(\mathbf{k},\mathbf{p},\omega)d\omega=
\ \ \ \ \ \ \ \ \ \ \ \nonumber \\
\left(-\frac{1}{\pi}\right)\mathrm{Im}
\ I(\epsilon_k-i\Gamma_k,\epsilon_p-i\Gamma_p,\epsilon_f).
\end{eqnarray}
The imaginary part of $J(E,E',\epsilon_f)$ appears in
\ref{a1_9}.

The integrand of \ref{a1_10} can be expanded in partial fractions,
\begin{eqnarray}
\label{a1_13}
\frac{1}{\left(\omega-E\right)^2\left(\omega-E'\right)} & = &
\delta E^{-2}\left(
\frac{1}{\omega-E'}-\frac{1}{\omega-E} \right) \nonumber \\
 & - & \delta E^{-1}
\frac{1}{\left(\omega-E\right)^2},
\end{eqnarray}
where $\delta E=E'-E$.
The second term on the right-hand side can be integrated,
\begin{equation}
\label{a1_14}
\int_{-\infty}^{\epsilon_f}d\omega
\frac{(-)}{\left(\omega-E\right)^2}=
\frac{1}{\epsilon_f-E}.
\end{equation}
Each individual integral in the first term is infinite.  The
integral is performed by substituting $-\Lambda$ for $-\infty$ and
taking the limit $\Lambda\rightarrow+\infty$.
\begin{eqnarray}
\label{a1_15a}
\mathrm{Re}\int_{-\Lambda}^{\epsilon_f}d\omega\left[
\frac{1}{\omega-E'}-\frac{1}{\omega-E}
\right] = \ \ \ \ \ \ \ \nonumber \\
\int_{-\Lambda}^{\epsilon_f}d\omega\left[
\frac{\left(\omega-\epsilon'\right)}
{\left(\omega-\epsilon'\right)^2+\Gamma'^2}-
\frac{\left(\omega-\epsilon\right)}
{\left(\omega-\epsilon\right)^2+\Gamma^2}
\right],
\end{eqnarray}
\begin{eqnarray}
\label{a1_15b}
\mathrm{Im}\int_{-\Lambda}^{\epsilon_f}d\omega\left[
\frac{1}{\omega-E'}-\frac{1}{\omega-E}
\right] = \ \ \ \ \ \ \ \nonumber \\
\int_{-\Lambda}^{\epsilon_f}d\omega\left[
\frac{\Gamma}{\left(\omega-\epsilon\right)^2+\Gamma^2}-
\frac{\Gamma'}{\left(\omega-\epsilon'\right)^2+\Gamma'^2}
\right].
\end{eqnarray}
The real part can be solved,
\begin{eqnarray}
\label{a1_16}
\int_{(-\Lambda-\epsilon')/\Gamma'}^
{(\epsilon_f-\epsilon')/\Gamma'} \frac{xdx}{x^2+1} -
\int_{(-\Lambda-\epsilon)/\Gamma}^
{(\epsilon_f-\epsilon)/\Gamma} \frac{xdx}{x^2+1}= \ \ \ \ 
\nonumber \\
\frac{1}{2}\ln\left[
\frac{\left(\epsilon_f-\epsilon'\right)^2/\Gamma'^2+1}
{\left(\epsilon_f-\epsilon\right)^2/\Gamma^2+1}\right]
\nonumber \ \ \ \ \ \ \ \ \ \ \\
+ \ \ \frac{1}{2}\ln\left[
\frac{\left(-\Lambda-\epsilon\right)^2/\Gamma^2+1}
{\left(-\Lambda-\epsilon'\right)^2/\Gamma'^2+1}
\right].  \ \ \ \ \\
\end{eqnarray}
Taking the limit $\Lambda\rightarrow+\infty$, the real part
becomes,
\begin{equation}
\label{a1_17}
\mathrm{Re}\int_{-\infty}^{\epsilon_f}d\omega\left[
\frac{1}{\omega-E'}-\frac{1}{\omega-E}
\right]= 
\frac{1}{2}\ln\left[
\frac{\Gamma'^2\left(x'^2+1\right)}
{\Gamma^2\left(x^2+1\right)}
\right].
\end{equation}
The imaginary part integrates to,
\begin{eqnarray}
\label{a1_18}
\mathrm{Im}\int_{-\infty}^{\epsilon_f}d\omega\left[
\frac{1}{\omega-E'}-\frac{1}{\omega-E}
\right] & = & \tan^{-1}x  \nonumber \\
 & -  & \tan^{-1}x', 
\end{eqnarray}
where $x=(\epsilon_f-\epsilon)/\Gamma$ and
$x'=(\epsilon_f-\epsilon')/\Gamma'$.  Substituting into equation
\ref{a1_10},
\begin{eqnarray}
\label{a1_19}
I(E,E',\epsilon_f) & = &
\delta E^{-2}\left\{
\frac{1}{2}\ln\left[
\frac{\Gamma'^2\left(x'^2+1\right)}
{\Gamma^2\left(x^2+1\right)}\right]\right. \nonumber \\
 & + & \left.i\left(\tan^{-1}x-\tan^{-1}x'\right)
\right\} \nonumber \\
 & + &
\delta E^{-1}\cdot\frac{1}{\epsilon_f-E},
\end{eqnarray}
where $E$ and $E'$ are defined in equation \ref{a1_11a} and
$\delta E=E'-E$.

Writing $\omega-\epsilon_f=(\omega-E')+(E'-\epsilon_f)$ in
equation \ref{a1_11} for $J(E,E',\epsilon_f)$,
\begin{equation}
\label{a1_20}
J(E,E',\epsilon_f) =
\left(E'-\epsilon_f\right)I(E,E',\epsilon_f)
-\frac{1}{\left(\epsilon_f-E\right)},
\end{equation}
leading to,
\begin{eqnarray}
\label{a1_21}
J(E,E',\epsilon_f) & = &
\frac{\left(E'-\epsilon_f\right)}{\left(E'-E\right)^2}\left\{
\frac{1}{2}\ln\left[
\frac{\Gamma'^2\left(x'^2+1\right)}
{\Gamma^2\left(x^2+1\right)}\right]\right. \nonumber \\
 & + & \left.i\left(\tan^{-1}x-\tan^{-1}x'\right)
\right\} \nonumber \\
 & + &
\frac{1}{E-E'}.
\end{eqnarray}

\section{Hall Effect}
\label{ap:hall}
In this appendix, we derive equations
\ref{eqn:hall16}-\ref{eqn:hall19} and \ref{eqn:hall33}
in section \ref{hall}.  Equation \ref{eqn:hall13} is
\begin{equation}
\label{a2_1}
V^{(2)(\pm)}_{\mathbf{k'k}}=
\sum_{\mathbf{p}} 
\frac{
\langle\mathbf{k'}P^{\pm}|H_U|
\mathbf{p}I\rangle
\langle\mathbf{p}I|H_U|
\mathbf{k}P^{\pm}\rangle
}{\epsilon_k-\epsilon_p-\left(E_I-E^{\pm}\right) +i\delta}.
\end{equation}
Expanding for $I=S$ using equation \ref{eqn:hall6},
\begin{widetext}
\begin{equation}
\label{a2_2}
V^{(2)(\pm)}_{\mathbf{k'k}}(S)=
\left(\frac{U}{2N}\right)^2\sum_{\mathbf{p}}
\frac{
\left[
\sin\left(k'-p\right)_{x'}\mp i \sin\left(k'-p\right)_{y'}
\right]
\left[
\sin\left(k-p\right)_{x'}\mp i \sin\left(k-p\right)_{y'}
\right]^\ast}
{\epsilon_k-\epsilon_p-\left(E_S-E^\pm\right)+i\delta}.
\end{equation}
Expanding the numerator inside the summation into products
$\sin k'_{x'}\cos p_{x'}$ etc,
\begin{eqnarray}
\label{a2_3}
\mathrm{Num} & = &
\left[
\left(\sin k'_{x'}\cos p_{x'} - \cos k'_{x'}\sin p_{x'}\right)
\right.
 \mp
\left.
i\left(\sin k'_{y'}\cos p_{y'} - \cos k'_{y'}\sin p_{y'}\right)
\right]  \nonumber \\
 & \times &
\left[
\left(\sin k_{x'}\cos p_{x'} - \cos k_{x'}\sin p_{x'}\right)
\right.
 \mp
\left.
i\left(\sin k_{y'}\cos p_{y'} - \cos k_{y'}\sin p_{y'}\right)
\right]^\ast.
\end{eqnarray}
\end{widetext}
This can be further expanded into terms with products of the form
$\cos^2 p_{x'}$, $\cos p_{x'}\cos p_{y'}$, $\cos p_{x'}\sin p_{y'}$,
etc.  

The denominator in equation \ref{a2_2} only depends on
$\mathbf{p}$ through the energy $\epsilon_p$ and has
$D_{4h}$ crystal symmetry.  Thus, for each $\mathbf{p}$,
the sum includes terms with $p_{x'}\rightarrow -p_{x'}$
and $p_{y'}\rightarrow -p_{y'}$.  The only non-zero terms in the
numerator have products $\cos^2 p_{x'}$, $\cos^2 p_{y'}$,
$\cos p_{x'}\cos p_{y'}$, $\sin^2 p_{x'}$, and $\sin^2 p_{y'}$.

The real part of the numerator is,
\begin{eqnarray}
\label{a2_4}
\mathrm{Re(Num)} & = &
\left[ \left(
\sin k'_{x'}\sin k_{x'}\cos^2 p_{x'} \right.\right. \nonumber \\
 & + & \left.\cos k'_{x'}\cos k_{x'}\sin^2 p_{x'}\right) \nonumber \\
 & + &
\left(\sin k'_{y'}\sin k_{y'}\cos^2 p_{y'}\right. \nonumber \\
 & + & \left.\left.
\cos k'_{y'}\cos k_{y'}\sin^2 p_{y'} \right)\right].
\end{eqnarray}
Using $\cos^2 p_{x'}=1/2(1+\cos 2p_{x'})$ and
$\sin^2 p_{x'}=1/2(1-\cos 2p_{x'})$ with similar equations for
$p_{y'}$,
\begin{eqnarray}
\label{a2_5}
\mathrm{Re(Num)}=
\frac{1}{2}
\left[
\cos (k'-k)_{x'} + \cos (k'-k)_{y'}\right] \nonumber \\
-\frac{1}{2}\left[\cos (k'+k)_{x'}
+\cos (k'+k)_{y'}\right]\cos 2p_{x'},
\end{eqnarray}
where we have used $\cos 2p_{y'}=\cos 2p_{x'}$ due to symmetry
in the summation.  The definitions of the primed coordinates
in equations \ref{eqn:hall9a} and \ref{eqn:hall9b} lead to,
\begin{eqnarray}
\label{a2_6}
\mathrm{Re(Num)}=\cos\frac{1}{2}\left(k'_x-k_x\right)
\cos\frac{1}{2}\left(k'_y-k_y\right) \nonumber \\
- \cos 2p_{x'} \cos\frac{1}{2}\left(k'_x+k_x\right)
\cos\frac{1}{2}\left(k'_y+k_y\right).
\end{eqnarray}
A similar expansion for the imaginary part leads to,
\begin{eqnarray}
\label{a2_7}
\mathrm{Im(Num)} & = & \left(\mp \right)
\left(\sin k_{x'}\sin k'_{y'}\right. \nonumber \\
 & - & \left.\sin k_{y'}\sin k'_{x'}\right)
\cos p_{x'}\cos p_{y'} \nonumber \\
 & = & A_{\mathbf{k'k}} \cos p_{x'}\cos p_{y'},
\end{eqnarray}
with $A_{\mathbf{k'k}}$ defined in equation \ref{eqn:hall18}.
Collecting the real and imaginary terms,
\begin{eqnarray}
\label{a2_8}
V^{(2)(\pm)}_{\mathbf{k'k}}(S) & = & \left(\frac{U}{4N}\right)^2
\nonumber \\
 & \times & \left\{\tilde F_0(\omega_0)
\cos\frac{1}{2}\left(k'_x-k_x\right)
\cos\frac{1}{2}\left(k'_y-k_y\right)\right. \nonumber \\
 & - & \tilde F_2(\omega_0)
\cos\frac{1}{2}\left(k'_x+k_x\right)
\cos\frac{1}{2}\left(k'_y+k_y\right) \nonumber \\
 & \mp & \left.i\tilde F_1(\omega_0) A_{\mathbf{k'k}} \right\},
\end{eqnarray}
where $\omega_0=\epsilon_k-(E_S - E^\pm)$ and the $\tilde F$
functions are defined as,
\begin{equation}
\label{a2_9}
\tilde F_0(\omega)=\left(\frac{1}{N}\right)\sum_{\mathbf{p}}
\frac{1}
{\omega-\epsilon_p+i\delta},
\end{equation}
\begin{equation}
\label{a2_10}
\tilde F_1(\omega)=\left(\frac{1}{N}\right)\sum_{\mathbf{p}}
\frac{\cos p_{x'}\cos p_{y'}}
{\omega-\epsilon_p+i\delta},
\end{equation}
\begin{equation}
\label{a2_11}
\tilde F_2(\omega)=\left(\frac{1}{N}\right)\sum_{\mathbf{p}}
\frac{\cos 2p_{x'}}
{\omega-\epsilon_p+i\delta}.
\end{equation}
Expanding the cosines in \ref{a2_10} and \ref{a2_11},
\begin{equation}
\label{a2_12}
\cos p_{x'}\cos p_{y'}=
\frac{1}{2}\left(\cos p_x + \cos p_y\right),
\end{equation}
\begin{eqnarray}
\label{a2_13}
\cos 2p_{x'}=\cos(p_x+p_y) & = & \cos p_x\cos p_y \nonumber \\
 & - & \sin p_x\sin p_y.
\end{eqnarray}
The second term in equation \ref{a2_13} averages to zero when summed in
$\tilde F_2$.  $\tilde F_1$ and $\tilde F_2$ become
\begin{equation}
\label{a2_14}
\tilde F_1(\omega)=\left(\frac{1}{N}\right)\sum_{\mathbf{p}}
\frac{\frac{1}{2}\left(\cos p_x + \cos p_y\right)}
{\omega-\epsilon_p+i\delta},
\end{equation}
\begin{equation}
\label{a2_15}
\tilde F_2(\omega)=\left(\frac{1}{N}\right)\sum_{\mathbf{p}}
\frac{\cos p_x\cos p_y}
{\omega-\epsilon_p+i\delta}.
\end{equation}
The equation for $V^{(2)(\pm)}_{\mathbf{k'k}}(D)$ is the same
as equation \ref{a2_8} with $E_S\rightarrow E_D$ and
$\mp(i)\rightarrow\pm(i)$.

From equation \ref{eqn:hall15},
the real part of $V^{(2)(\pm)}_{\mathbf{k'k}}(S)$
is necessary for the lowest order skew-scattering.  In
addition, only terms antisymmetric under interchange of
$\mathbf{k'}$ and $\mathbf{k}$ contribute to 
$\rho^{skew}_{xy}$.  The last term in equation
\ref{a2_8} is antisymmetric and only the imaginary part of
$\tilde F_1$ contributes to the skew-scattering.

Defining $F_1(\omega)=(-1/\pi)\mathrm{Im}\tilde F_1(\omega)$,
\begin{equation}
\label{a2_16}
F_1(\omega)=\left(\frac{1}{N}\right)\sum_{\mathbf{p}}
\frac{1}{2}\left(\cos p_x+\cos p_y\right)
\delta(\omega-\epsilon_p),
\end{equation}
leads to equations \ref{eqn:hall16} and \ref{eqn:hall17}.

Equation \ref{eqn:hall33} for the skew-scattering is obtained by
solving the linearized Boltzmann transport equation for
skew-scattering,\cite{varma1, skew_boltz1, skew_boltz2}
\begin{equation}
\label{a2_17}
\left(\frac{\partial f_k}{\partial t}\right)_{field}+
\left(\frac{\partial f_k}{\partial t}\right)_{scatt}=0,
\end{equation}
\begin{eqnarray}
\label{a2_18}
\left(\frac{\partial f_k}{\partial t}\right)_{field} & = &
\left(\frac{\partial f_0}{\partial\epsilon_k}\right)
(-e)\left(\mathbf{v_k\cdot E}\right) \nonumber \\
 & + & \left(\frac{e}{\hbar c}\right)
\frac{\partial g_k}{\partial\mathbf{k}}\cdot
\left(\mathbf{v_k\wedge B}\right),
\end{eqnarray}
\begin{eqnarray}
\label{a2_19}
\left(\frac{\partial f_k}{\partial t}\right)_{scatt} & = &
-\frac{g_k}{\tau_k} + \sum_{k'}\left[
-w_{skew}\left(k\rightarrow k'\right)g_k\right. \nonumber \\
 & + & \left. w_{skew}\left(k'\rightarrow k\right)g_{k'}
\right],
\end{eqnarray}
where $e>0$ and $f_k=f_0 + g_k$ with $f_0$ the Fermi-Dirac function
at energy $\epsilon_k$.  $1/\tau_k$ is the ordinary scattering
rate.  This is given by equation \ref{eqn:e7} in our model.
From equations
\ref{eqn:hall25}$-$\ref{eqn:hall27},
$\sum_{k'}w_{skew}(k\rightarrow k')=0$ making the second term
on the right hand side of equation \ref{a2_19} vanish.

Substituting,
\begin{equation}
\label{a2_20}
g_k= \left(-\frac{\partial f_0}{\partial\epsilon_k}\right)\Phi_k,
\end{equation}
the transport equation becomes,
\begin{eqnarray}
\label{a2_21}
-e\left(\mathbf{v_k\cdot E}\right)+
\left(\frac{e}{\hbar c}\right)
\frac{\partial \Phi_k}{\partial\mathbf{k}}\cdot
\left(\mathbf{v_k\wedge B}\right) \nonumber \\
-\frac{\Phi_k}{\tau_k}
-\sum_{k'}w_{skew}(k\rightarrow k')\Phi_{k'}=0.
\end{eqnarray}

$\Phi_k$ can be expanded in the series,
\begin{equation}
\label{a2_22}
\Phi_k=\Phi^{(0)}_k + \Phi^{(1)(ord)}_k + \Phi^{(1)(skew)}_k + \ldots,
\end{equation}
where $\Phi^{(0)}\sim O(E)$ and $\Phi^{(1)}\sim O(EB)$.
$\Phi^{(1)(ord)}_k$ is the ordinary band contribution
to the Hall effect.  We do
not write down the expression for this term.  The second
$O(EB)$ term, $\Phi^{(1)(skew)}_k$, is the additional contribution
arising from the skew-scattering term, $w_{skew}$.  Substituting
\ref{a2_22} into \ref{a2_21} leads to,
\begin{equation}
\label{a2_23}
\Phi^{(0)}=-e\left(\tau_k\mathbf{v_k\cdot E}\right),
\end{equation}
\begin{equation}
\label{a2_24}
\Phi^{(1)(skew)}=
\sum_{k'}\tau_k\tau_{k'}e\left(\mathbf{v_k\cdot E}\right)
w_{skew}(k\rightarrow k').
\end{equation}

The current density per spin is,
\begin{equation}
\label{a2_25}
\mathbf{J}=\left(\frac{1}{\Omega}\right)
\sum_k (-e)g_k\mathbf{v_k},
\end{equation}
where $\Omega$ is the total volume.  The $O(EB)$ contribution
to the conductivity per spin
$\sigma_{yx}^{skew}$ from skew-scattering 
satisfies $J_y^{(1)(skew)}=\sigma_{yx}^{skew}E_x$,
\begin{eqnarray}
\label{a2_26}
\sigma_{yx}^{skew} & = &
(-e^2)\left(\frac{1}{\Omega}\right)\sum_{kk'}
\left(-\frac{\partial f_0}{\partial\epsilon_k}\right)
\tau_k\tau_{k'} \nonumber \\
 & \times & v_{ky}v_{k'x}w_{skew}(k\rightarrow k').
\end{eqnarray}
Using $\sigma_{yx}=-\sigma_{xy}$ from crystal symmetry and
$\rho_{xy}=\sigma_{xy}/\sigma^2$ for the Hall resistivity,
\begin{eqnarray}
\label{a2_27}
\rho_{xy}^{skew}\sigma^2 & = & (2e^2)
\left(\frac{1}{\Omega}\right)\sum_{kk'}
\left(-\frac{\partial f_0}{\partial\epsilon_k}\right)
\tau_k\tau_{k'} \nonumber \\
 & \times & v_{ky}v_{k'x}w_{skew}(k\rightarrow k'),
\end{eqnarray}
with the additional factor of $2$ on the right hand side to
include spin.  The conductivity $\sigma$ is,
\begin{equation}
\label{a2_28}
\sigma=e^2\tau\left(\frac{n}{m}\right)_{eff},
\end{equation}
where $(n/m)_{eff}$ is the effective $n/m$ and we have taken
$\tau_k$ to be independent of k and equal to $\tau$.
Since $\rho_{yx}=-\rho_{xy}$ by symmetry, we symmetrize
the expression for $\rho_{xy}$ by setting
$\rho_{xy}=(1/2)(\rho_{xy}-\rho_{yx})$, leading to equation
\ref{eqn:hall33}.

\end{document}